\documentclass[a4paper,fleqn,usenatbib]{mnras}

\usepackage{newtxtext,newtxmath}
\usepackage[T1]{fontenc}
\usepackage{ae,aecompl}

\usepackage{graphicx}	% Including figure files
\usepackage{amsmath}	% Advanced maths commands
\usepackage{amssymb}	% Extra maths symbols

\title[Understanding probability weights]{Confronting missing observations with probability weights: Fourier space and generalised formalism}

\author[D. Bianchi and L. Verde.]{
Davide Bianchi,$^{1}$\thanks{E-mail: davide.bianchi@icc.ub.edu}
and Licia Verde$^{1}$
\\
$^{1}$ICC, University of Barcelona, IEEC-UB, Mart\'{ı} i Franqu\'{e}s, 1, E-08028 Barcelona, Spain
}

\date{Accepted XXX. Received YYY; in original form ZZZ}

\pubyear{2018}

\begin{document}
\label{firstpage}
\pagerange{\pageref{firstpage}--\pageref{lastpage}}
\maketitle

\begin{abstract}
%This is a simple template for authors to write new MNRAS papers.
%The abstract should briefly describe the aims, methods, and main results of the paper.
%It should be a single paragraph not more than 250 words (200 words for Letters).
%No references should appear in the abstract.
Due to instrumental limitations, the nature of which vary from case to case, spectroscopic galaxy redshift surveys usually do not collect redshifts for all galaxies in the population of potential targets.
Especially problematic is the entanglement between this incompleteness and the true cosmological signal, arising from the fact that the proportion of successful observations is typically lower in regions with higher density of galaxies.
%If not properly corrected, the resulting suppression of the observed galaxy clustering can have dramatic impact on cosmological constraints.  
%The result is a fictitious suppression of the galaxy clustering that can severely affect the inference of cosmological parameter.
The result is a fictitious suppression of the galaxy clustering that, if not properly corrected, can impact severely on cosmological-parameter inference.
Recent developments in the field have shown that an unbiased estimate of the 2-point correlation function in the presence of missing observations can be obtained by weighting each pair by its inverse probability of being targeted. 
In this work we expand on the concept of probability weights by developing a more mature statistical formalism, which provides us with a deeper understanding of their fundamental properties.
We take advantage of this novel perspective to handle the problem of estimating the inverse probability, specifically, we discuss how to efficiently determine the weights from a finite set of realisations of the targeting and how to model exactly the resulting sampling effects.
%We take advantage of this novel perspective to tackle the complexity of estimating the inverse probability, specifically, we discuss how to efficiently determine the weights from a finite set of realisations of the targeting and how to model exactly the resulting sampling effects.
This allows us to derive an inverse-probability-based estimator for the power spectrum, which is the main result of this work, but also to improve robustness and computational efficiency of the already existing configuration-space estimator.
Finally, we propose a strategy to further extend the concept of inverse probability, providing examples of how traditional missing-observation countermeasures can be included in this more general picture.
The effectiveness of the different models and weighting schemes discussed in this work is demonstrated using realisations of an idealised simple survey strategy.
\end{abstract}

\begin{keywords}
cosmology: large-scale structure of Universe -- methods: statistical
\end{keywords}

%%%%%%%%%%%%%%%%%%%%%%%%%%%%%%%%%%%%%%%%%%%%%%%%%%

%%%%%%%%%%%%%%%%% BODY OF PAPER %%%%%%%%%%%%%%%%%%

\section{Introduction}

Spectroscopic redshift surveys give us access to a wealth of cosmological information encoded in the 3-dimensional galaxy clustering.
Being able to identify and remove the galaxy-density fluctuations that arise from non-cosmological sources is of fundamental importance for a systematic-free inference of cosmological parameters.
One of the dominant sources for these spurious fluctuations is that of missing observations.
Due to intrinsic limitations of the instrument, spectroscopic surveys typically collect redshifts for just a fraction of the full set of potential targets.
%A typical example of these limitations is the impossibility of allocating two optical fibres close enough in the focal plane to observe simultaneously a close pair of galaxies.
A typical example of this limitations is the impossibility of  positioning two optical fibres on the focal plane closer to each other than a minimal distance, making it therefore impossible to observe close pairs of galaxies. 
This results in a complicated pattern of missing observations, unevenly distributed across the survey footprint, mostly localised in the high density regions.
The net effect is a non-trivial scale-dependent suppression of the amplitude of the observed galaxy clustering, which, if not properly corrected, can have dramatic impact on cosmological constraints.  

For the updated Sloan telescope \citep{gunn2006}, used by the Baryon Oscillation Spectroscopic Survey \citep[BOSS;][]{dawson2013} , two optical fibres cannot be allocated at a separation smaller than 62$\arcsec$ on the focal plane.
As a consequence of these fibre-collisions, in the final Data Release 12 \citep{alam2015} of BOSS, approximately 5\% of the targets  were not observed \citep{reid2016}.
The resulting density-dependent completeness of the sample has a strong impact on the small-scale clustering measurements, see e.g. \citet{hahn2017}.
%The range of scales potentially affected by fibre collisions grows with the depth of the survey, as is the case for the extended BOSS \citep{dawson2016}.
For deeper surveys, such as the extended BOSS \citep{dawson2016}, the range of scales potentially affected by fibre collisions gets even larger.
The collision does not necessarily have to be between fibres, e.g. the VIPERS survey \citep{guzzo2014} presents a very similar issue, but for the allocation of slits \citep{mohammad2018}.

The Dark Energy Spectroscopic Instrument \citep[DESI;][]{amir2016a, amir2016b} will construct its main survey, covering  an area of about 14000\,deg$^2$, by making approximately 10000 observations, with 5000 spectra in each 7.5\,deg$^2$ field of view.
Different observations overlap in such a way that each patch in the survey footprint will be observed 5 times on average.
The instrument consists of an array of robotic fibres, in which the freedom of movement of each fibre is limited to a patrol radius of 1.48$\arcmin$.
Compared to, e.g., BOSS, this yields a conceptually more complex fibre-allocation issue that goes beyond the simple idea of fibre collisions, but the net effect is a similar: in regions of high target density, and for targets of low priority in the ranking of different target classes, there will be missing observations with a consequent scale-dependent distortion of the observed clustering \citep{pinol2017,burden2017,bianchi2018,smith2019}.

Different missing-observation countermeasures have been proposed in the literature, often focusing on a specific survey of interest, e.g. \citet{hawkins2003, anderson2012, guo2012, reid2014, burden2017, pinol2017, hahn2017}. 
We refer the reader to \citet{bianchi2017} for a concise description of the different approaches and discussion about their advantages and downsides.

The first unbiased-by-construction missing-observation correction was introduced by \citet{bianchi2017}.
It that work it was shown that an unbiased estimate of the 2-point correlation function can be obtained by weighting each pair of galaxies by its inverse probability of being observed (pairwise inverse probability, PIP), provided that there are no zero-probability pairs in the parent sample of potential targets.
The idea was then successfully applied to specific scenarios in a few subsequent works: \citet{bianchi2018} and \citet{smith2019} tested the effectiveness of the method on simulated catalogues of the DESI survey (emission-line and bright galaxy sample, respectively); \citet{mohammad2018} used it to extract robust constrain on the growth rate of the cosmological structures from the VIPERS survey.

This paper builds on the original \citet{bianchi2017} idea %by providing its natural extension to the power spectrum and discussing possible generalisations of the PIP weights.  
by providing new insights and not-tied-to-any-survey generalisations to the general concept of probability weights, the most relevant of which is probably its extension to Fourier space.
Also important is the extensive discussion of how in practice the inverse probability can be estimated from a finite number of realisations of the targeting algorithm and how the resulting sampling effects can be modelled. 

The layout of our paper is as follows:
in Sec.~\ref{sec:IP} we show how the inverse-probability prescription works and how it can be formalised in a universal statistical language;
in Sec.~\ref{sec:simulations} we describe the simulations that we use for testing;
in Sec.~\ref{sec:sampling} we discuss the problem of selection-probability estimation;
in Sec.~\ref{sec:Fourier} we introduce a PIP-based estimator for the power spectrum;
in Sec.~\ref{sec:generalising} we discuss general extensions to the inverse-probability approach and how the variance is influenced by the choice of different weighting scemes;
we conclude summarising our results in Sec.~\ref{sec:conclusions}. 
In order to make the paper more readable and focused, we decided to relegate to the appendix the derivation of important formulae (Apps. \ref{app:evaluation_expect}, \ref{app:distributions}, \ref{app:PIPco_derivation}) and relevant discussions/examples regarding the concept of generalised weights (Apps \ref{app:toy}, \ref{app:norm}, \ref{app:generalised_comparison}, \ref{app:xipe}).

\section{Inverse probability weights}\label{sec:IP}

The inverse probability weights introduced by \citet{bianchi2017} are based on the idea that if a pair appears $c$ times out of $K$ equally-likely realisations of the targeting, then by assigning a weight $f=K/c$ to that pair, it will be counted $K$ times when summing over all the realisations, i.e. once on average. 
This can be expressed as
\begin{equation}\label{eq:invprob}
\frac{1}{K} \sum_{n=1}^K f b_\eta = f \frac{1}{K} \sum_{\eta=1}^K b_\eta  = 1 \ ,
\end{equation} 
where $b$ is a bitwise (i.e. logical) array such that
\begin{equation}
b_\eta =
\begin{cases}
0 \text{ (false)} & \text{pair discarded in the $\eta$-th realisation} \\
1 \text{ (true)} & \text{pair selected in the $\eta$-th realisation} \\
\end{cases} \ .
\end{equation}
%(throughout the whole paper we will use greek letter for indices that refers to realisations, to avoid confusion).
Trivially, $\sum_{\eta=1}^K b_\eta = c$.
Since the average count of each individual pair is unbiased, the resulting correlation function estimator, which is a linear function of all pair counts, is unbiased as well. 
Purpose of this work is to provide a rigorous formalism for this simple idea and explore possible generalisations.

Clearly, the quantity of interest is the binary random variable $b \in \{0,1\}$, which is the appropriate statistical counterpart of the array $b_\eta$ discussed above ($b_\eta$ can be seen as a finite set of realisations of $b$).
The variable $b$ follows a binomial distribution,
\begin{equation}
\mathcal{P}(b) =  p^b (1-p)^{1-b} \ ,
\end{equation}
where $p$ is the selection probability.
Since $\langle b \rangle = p$, it follows immediately that, for $p > 0$, $\langle b/p \rangle = 1$.
This proves the intrinsic unbiasedness of the estimator in a well-defined statistical environment, i.e. in terms of expectation values. 

Pairs with $p=0$ are by definition unknowable.
Irregardless of the statistical approach adopted, any meaningful consideration about such objects involves assumptions whose nature and validity varies from case to case (see e.g. discussion in \citealt{bianchi2018}). 
For the rest of this work we assume that there are no zero-probability objects, unless otherwise stated.

Interestingly, $b/p$ is not the only possible function of $b$ and $p$ with expectation value equal to one.
For example, another simple solution is given by $b + 1-p$, and more complex expressions can be obtained via combinations, such as $a b/p + (1-a) (b+1-p)$, with $a \in R$.
Additive expressions like $b + 1-p$ are appealing because of their intrinsic stability for small values of $p$, but they assign non-zero weight to objects that are not observed.
%This is exactly what we do not want to do for the missing-obsevation problem, since it would require to guess the redshift of the galaxies we did not observe.
In the context of missing observations in a redshift survey, this implies guessing the redshift of the galaxies we did not observe, thus forcing the correction to be model dependent.
What makes the inverse probability special is that $b/p = 0$ when $b=0$. 
%This basic property is what is needed when dealing with missing observations. 
 
It is also interesting to note that the whole inverse-probability description is founded on the simple equality $\langle b \rangle = p$.
This suggest a very natural extension of the concept to generic (non binary) random variables $\phi$, in which $b/p$ is replaced by $\phi / \langle \phi \rangle$.
We discuss explicitly this idea later, in Sec. \ref{sec:generalising}.
 
The true selection probability and its inverse are, in general, unknown. 
In practice, we have to estimate them, e.g., by rerunning several times a targeting algorithm\footnote{If, as it is often the case, the targeting algorithm is a stochastic process initialised by random seeds, it suffices to change the random seeds. If it is deterministic, as e.g. for the VIPERS survey, we can take advantage of the isotropy of the Universe and shift the tiling of the survey from one realisation to one another. The two approaches can be also combined. See \citet{bianchi2018,mohammad2018,smith2019} for discussions/examples.}, thus potentially creating sampling effects that need to be understood.
In Sec. \ref{sec:sampling} we provide a well defined formalism to address this issue.

One simple but important point to make is that, so far we focused on pairs just for finiteness, but the formalism is completely general.
All the above considerations hold for any $n$-plet and its corresponding $n$-point correlation function.
Beside the growing interest in 3- and even 4-point functions and the consequent need for adequate missing-observation countermeasures, it is interesting to note that the reasoning also works for $n=1$.
In this case, following the nomenclature introduced by \citet{bianchi2017}, we talk about individual inverse probability (IIP).
The IIP allows us to recover an unbiased estimate of the full spatial distribution of the galaxies, which can be used, e.g., as an input for reconstruction algorithms (e.g. \citealt{eisenstein2007}).
Note however that the unbiasedness is guaranteed only for quantities that depend linearly on the $n$-plet under examination.
For example, IIP weights return unbiased counts in cell at any position in the sky but the corresponding 2-point function is not necessarily unbiased, as extensively discussed in \citet{bianchi2017}, because it depends quadratically on the distribution of particles.
On the other hand, the 2-point function can be expressed as a linear combination of pairs and, therefore, we need pairwise weights for the prescrition to work.   
Since reconstruction, in principle, depends on all higher order statistics or, in other words, it is not a linear operation on the individual 1-point counts, it is not guaranteed that the reconstructed field obtained via observed sample plus IIP will coincide with what we would obtain if we could apply the reconstruction algorithm directly to the parent sample.    
Nonetheless, being reconstruction a large-scale process, which involves smoothing of the scales $\lesssim 15h^{-1}$Mpc, and being the IIP-PIP discrepancy typically confined to separations smaller than this smoothing range\footnote{Note, however, that the smoothing is a 3-dimensional process, whereas the IIP-PIP discrepancy in expressed in terms of angles. It is not rigorously true that this discrepancy is confined to small 3-dimensional scales, but it is certainly true that its impact on the 3-dimensional clustering is stronger on such scales.} (see Sec. \ref{sec:sampling}), it seems likely that IIP will provide accurately reconstructed fields in the vast majority of the scenarios of interest.
We will address this topic and the analysis of 3-point functions in future works.

%From an even larger perspective, we can interpret the inverse-probability weights as a special case of a general statistical formalism in which, rather than focusing on the probability distribution of a generic field $\mathcal{P}(\varphi)$, we consider weight functions $\mathcal{W}$, conceptually not too dissimilar from the wave functions of quantum mechanics.
%These weight functions are associated to tracers of the field (in this sense they have a ``Lagrangian nature'') and contain the same information as $\mathcal{P}(\varphi)$.
%They do not have a specific meaning on their own, in the sense that there is no natural order with respect to $\eta$ (i.e. it does not matter how the realisations are sorted), but the $n$-pt statistics can be obtained via scalar products.
%$\sum_\eta \mathcal{W}_{i\eta}$ gives the 1-pt weight,  $\sum_\eta \mathcal{W}_{i\eta} \mathcal{W}_{j\eta}$ the 2-pt weight, $\sum_\eta \mathcal{W}_{i\eta} \mathcal{W}_{j\eta} \mathcal{W}_{k\eta}$ the 3-pt weight, etc.
%TO BE FINSHED, MAYBE WORTHED TO HAVE A SMALL SECTION ON THIS TOPIC OR DISCUSS IT IN THE CONCLUSIONS 

\subsection{Notation: operating instructions}

There is one further important introductory comment to be made, which concerns the notation and terminology adopted.
This work spans a relatively wide variety of topics and keeping the notation fully rigorous and self consistent throughout the whole text would result in overcomplicated mathematical expressions.
Specifically, we warn the reader that, despite our goal is to provide a formalism for summary statistics such as, e.g., the correlation function of a large set of objects,
it is actually often convenient to focus at first on the behaviour of a single isolated object. This is what we have done for the most part of this section and what we will do often for the rest of the paper.    
In this respect, a more general notation for the binary (selected/discarded) variable $b$ would be $b_{i\eta}$, where the index $i$ indicates that $b_i$ is the binary random variable associated to the $i$-th object in the sample.
Furthermore, we have to consider different realisations of these random variables, which are encoded in the index $\eta$. 
For simplicity, both indices, $i$ and $\eta$, will be suppressed throughout the rest of this work, unless there is obvious ambiguity.
To avoid confusion, when we need to make the dependence on the realisation explicit we will always use greek letters, e.g. $\eta$.  

Another potential source of confusion for the reader comes from the fact that, so far, in the literature the discussion about probability weights has been limited to the dichotomy individual versus pairwise weights, in the context of galaxy pair counts.
In such scenario each galaxy has it own binary variable $b_i$ and, as a consequence, the binary random variable associated to each pair is $b_{ij} = b_i b_j$, where $b_i$ and $b_j$ are, in general, not independent.
The fact that the selection properties of a pair (or a $n$-plet) can be inferred from those of individual objects is irrelevant for many of the problems considered in this work.
As discussed above, if we are dealing with $n$-point functions, the irreducible building blocks are $n$-plets.
For this reason we will often talk about objects rather than galaxies, pairs, triplet, etc.
When we do so it is intended that the object is the appropriate irreducible entity for the statistics under examination, e.g. a pair if we are dealing with 2-point correlation functions.    
For the same reason we will often use the simple notation $b$, rather than, e.g., $b_{ij}$, unless there is obvious ambiguity (as e.g. for the random variable $\phi$ in the discussion on generalised nearest neighbour weights, Sec. \ref{sec:generalising}).

\section{Simulations}\label{sec:simulations}

Many of the results presented in this paper are of theoretical nature and, strictly speaking, would not require validation via simulations.
The main reason why we want to perform tests with simulated data, beside providing a proof of concept, is that the true distribution of the selection probabilities is a free input in our modelling.
More explicitly, we can model the statistical properties of, e.g., a pair of galaxies with selection probability $p$, but different pairs will have different probabilities and the statistical properties of the overall clustering will depend on how these probabilities are distributed across the whole sample.
It is therefore desirable to the test the performance of different models and estimators in the presence of realistic clustering and selection algorithm or, in other words, realistic distribution of the selection probabilities.

For this purpose we use the same simulation and targeting algorithm adopted in \citealt{bianchi2017}.
Specifically, we use the data from the MultiDark MDR1 run \citep{prada2012}, which adopts WMAP
cosmology, $\{\Omega_m, \Omega_{\Lambda}, \Omega_b, \sigma_8, n_s\} = \{0.27,
0.73, 0.047, 0.82, 0.95\}$, to trace the evolution of $2048^3$ particles over a ${(1000 h^{-1}{\rm Mpc})}^3$ cubical volume.
We create a parent sample of potential targets by applying a $0.005\%$ dilution factor to the $z=0.5$ snapshot. 
The resulting catalogue consists of $\sim 4.3 \times 10^5$ dark matter particles, with a number density of $\sim 4.3 \times 10^{-4} h^3 \text{Mpc}^{-3}$, compatible with that of a modern galaxy survey.
In the following we sometimes refer to these particles as galaxies.

We then obtain a collided catalogue by running two passes of the of the maximum-randomness (MR) targeting algorithm introduced in \citet{bianchi2017}.
Different realisations of the targeting, i.e. different collided catalogues, are obtained by changing the random seed of the algorithm.
We refer the reader to \citet{bianchi2017}, specifically observing strategy OS2, for a detailed description of the process.
In brief, the MR algorithm works by randomly picking a pair among those with separation smaller than a given angular scale and randomly discarding one of the two galaxies; the procedure is repeated until there are no collided pairs left.
Since we are dealing with a box, the collision scale is not an angle but rather a length, which we arbitrarily set to $1 h^{-1}$Mpc.  
The so obtained fraction of observed galaxy is $\sim 0.84$, fully compatible with, e.g., the completeness expected for the final emission-line-galaxy sample of the DESI survey. 

Performing two passes of the algorithm is the simplest way to ensure that there are no zero-probability pairs in the sample, thus allowing us to isolate any sampling effect coming from the finiteness of the of the number or realisations from zero-probability-related features, which have already been discussed in previous works \citep{bianchi2017,bianchi2018,mohammad2018,smith2019}.
Having multiple passes is indeed very common in modern surveys, but the way these passes are performed is a survey-specific issue.
Each survey has its own observing strategy, simulating the different tiling patterns adopted to cover the different survey footprints would go against the all-embracing nature of this paper.

\section{Sampling the inverse probability}\label{sec:sampling}

As discussed in Sec. \ref{sec:IP}, the inverse probability weights yield unbiased counts in cells for any object with non-zero probability of being observed.
The quantity of interested $1/p$ is unknown but can be estimated by running the targeting algorithm $K$ times over a given parent sample and counting how many times $c$ each object gets selected over the $K$ independent realisations.
To avoid confusion between different kind of counts we will refer to $c$ as the {\it recurrence}.
Clearly, the number of realisations needed for a fair sampling of the probabilities depends on the probabilities themselves.
%In this section we discuss what kind of estimator gives a faster convergence and which kind of tests are best suited to evaluate such convergence. 
In this section we discuss the convergence properties (with respect to $K$) of different estimators and we show how to model subtle but relevant sampling effects. 

We first consider a scenario in which, given a parent sample, the inverse probability is evaluated from a set of $K$ realisations of the targeting and then used to correct the clustering obtained from a new independent realisation of the targeting.      
The recurrence $c \in \{0,\dots,K\}$ is a random variable, distributed as
\begin{equation}
\mathcal{P}(c) = \binom{K}{c} p^c (1-p)^{K-c} \ .
\end{equation}
For a given $K$, the expectation value of any function $w_K=w_K(c,b)$ can be evaluated as
\begin{equation}
\langle w_K \rangle = \sum_{c=0}^K \sum_{b=0}^1 w_K(c,b) \mathcal{P}(c) \mathcal{P}(b) \ .
\end{equation}
%where we made the implicit assumption that $b$ and $c$ are independent (a different approach is discussed Sec. \ref{sec:previous}).
Since we want to assign zero weight to objects that are not selected (see Sec. \ref{sec:IP}), we are interested in functions of the form $w_K(c,b) = b f_K(c)$.
Explicitly, we have 
\begin{equation}
\langle w_K \rangle = \sum_{c=0}^K \sum_{b=0}^1 b f_K(c) \binom{K}{c} p^{c+b} (1-p)^{K-c+(1-b)} \ .
\end{equation}
By performing the trivial summation over $b$, we obtain
%\begin{equation}
%\langle w_K \rangle = \sum_{c=0}^K f_K(c) \binom{K}{c} p^{c+1} (1-p)^{K+1- (c+1)} \ ,
%\end{equation}
%\begin{align}
%\langle w_{pK} \rangle = \sum_{c=0}^{K} f_K(c) \binom{K}{c} \left(p^{1+\frac{1}{c}}\right)^c (1-p)^{K-c}
%\end{align
\begin{equation}\label{eq:expectation}
\langle w_K \rangle = p \sum_{c=0}^K f_K(c) \binom{K}{c} p^c (1-p)^{K-c} \ ,
\end{equation}
which makes clear that what we are looking for is a function $f_K$ of the binomial variable $c$ whose expectation value is equal to $1/p$.
Trivially, $f_K = 1/p$ satisfies this requirement, but this is of no practical use since we are looking for weights that depend only on $c$ (and the parameter $K$) and not on $p$, which is unknown.

\subsection{Inverse-count estimator}

The most obvious ansatz, which we will refer to as {\it inverse-count}~(subscript ic), is $f_K = K/c$ plus an appropriate prescription for handling the $c=0$ divergence:
\begin{equation}
f_{K} =
\begin{cases}
w_0 & c=0  \\
\frac{K}{c} & c>0 \\
\end{cases} \ , \quad \text{i.e.} \quad
w_{\rm ic} =
\begin{cases}
b w_0 & c=0  \\
b \frac{K}{c} & c>0 \\
\end{cases} \ .
\end{equation}
The expectation value is obtained by substituting in Eq. (\ref{eq:expectation}),
%\footnote{A natural generalisation is given by
%\begin{equation}
%w_{K} =
%\begin{cases}
%b w_0 & c=0  \\
%b \frac{K}{c} & c>0
%\end{cases} \ , \nonumber
%\end{equation} 
%where $w_0$ is an arbitrary weight, e.g. $w_0=K$.
%This contributes an additive term $w_0 p (1-p)^K$ to the expectation value.},
\begin{equation}
\langle w_{\rm ic} \rangle = w_0 p (1-p)^K + \sum_{c=1}^K \frac{K}{c} \binom{K}{c} p^{c+1} (1-p)^{K-c} \ .
\end{equation}
Reasonable choices for $w_0$ include $w_0=0$ and $w_0=K$.
%This definition of $w_K$ is the one adopted in all previous works\footnote{FIGURE IN BIANCHI2017} \citep{bianchi2017,bianchi2018,smith2019,mohammad2018}, or, more precisely, its most obvious extension to include the $c=0$ case.
%We will come back to the interpretation of such extension later on.

\subsection{Efficient estimator}

Here we consider an alternative approach, denoted as {\it efficient}~(subscript e), which consists of setting $f_K(c) = \frac{K+1}{c+1}$ or, equivalently,
\begin{equation}
w_{\rm e} =  b \frac{K+1}{c+1} \ .
\end{equation}
By substituting in Eq. (\ref{eq:expectation}), it is easy to see (App. \ref{app:evaluation_expect}) that
\begin{equation}\label{eq:eff_expect}
\langle w_{\rm e} \rangle = 1 - {(1-p)}^{K+1} \ .
\end{equation}
Thanks to the fact that we have a simple analytic expression for $\langle w_{\rm e} \rangle$, it is possible to define a completely unbiased estimator in a compact form,
\begin{equation}
w_{\rm deb} =  b \frac{K+1}{c+1} \frac{1}{1 - {(1-p)}^{K+1}}\ ,
\end{equation}
but, similarly to the $1/p$ case just discussed, it is just an academic exercise, since the knowledge of the true $p$ is formally required.

\subsection{Zero-trucated estimator}\label{sec:zt_estimator}

We now discuss the scenario in which the realisation under examination is included in the set of realisations used to evaluate the probabilities.
In essence, this is the strategy implicitly adopted in \citet{bianchi2017,bianchi2018,mohammad2018,smith2019}.
In this scenario, when performing counts in cells it is impossible to have $c=0$: by construction all the observed objects have been selected at least once.
The most natural estimator, which we refer to as {\it zero-truncated} (subscript zt), is
\begin{equation}
w_{\rm zt} =
\begin{cases}
0 & b=0  \\
\frac{K}{c} & b=1 \\
\end{cases} \ ,
\end{equation}
where $c \ge 1$ if $b=1$.
Despite the definition of the zero-truncated estimator resembling that of its inverse-count counterpart, as we show below, its behaviour is actually more closely related to that of the efficient estimator.
In the zero-truncated scenario the random variable $c$ is distributed as

%$c = d +1$ with $d \in \{0,\dots,K-1\}$, which means $c \in \{1,\dots,K\}$
%\begin{equation}
%\mathcal{P}(d) = \binom{K-1}{d} p^d (1-p)^{K-1-d} \ .
%\end{equation}
%The expectation value of $w_{con}$ is equivalent to ...
%$c \in \{1,\dots,K\}$

\begin{equation}
\mathcal{P}(c) = \binom{K-1}{c-1} p^{c-1} (1-p)^{K-c} \ ,
\end{equation}
with $c \in \{1,\dots,K\}$. 
Following the same reasoning used for the previous two estimators, we get
\begin{equation}
\langle w_K \rangle = p \sum_{c=1}^K f_K(c) \binom{K-1}{c-1} p^{c-1} (1-p)^{K-c} \ ,
\end{equation}
which for $f_K(c) = K/c$, or, equivalently,
\begin{equation}
w_{\rm zt} = b \frac{K}{c} \ ,
\end{equation}
yields (see App. \ref{app:evaluation_expect})
\begin{equation}\label{eq:zt_expect}
\langle w_{\rm zt} \rangle = 1 - (1-p)^K \ .
\end{equation}
If we consider that by including the observed sample in the set used to evaluate the weights we have effectively increased the number of realisations, $K \rightsquigarrow K+1$, by comparison with Eq.(\ref{eq:eff_expect}), it becomes clear that the efficient and zero-truncated estimator have identical convergence properties.

\subsection{Comparison of the estimators}

\begin{figure}
 \begin{center}
   \includegraphics[width=9cm]{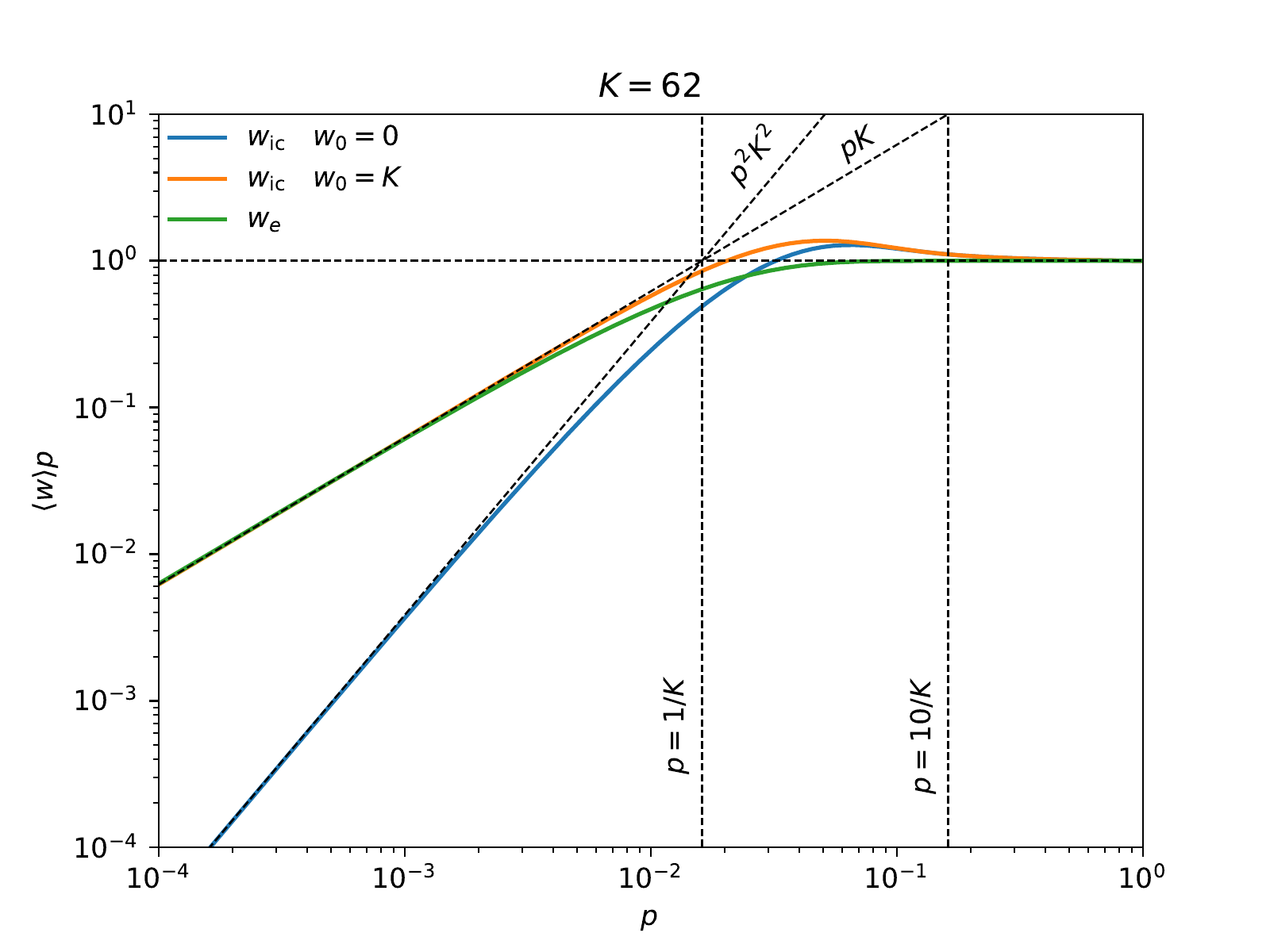}
   \includegraphics[width=9cm]{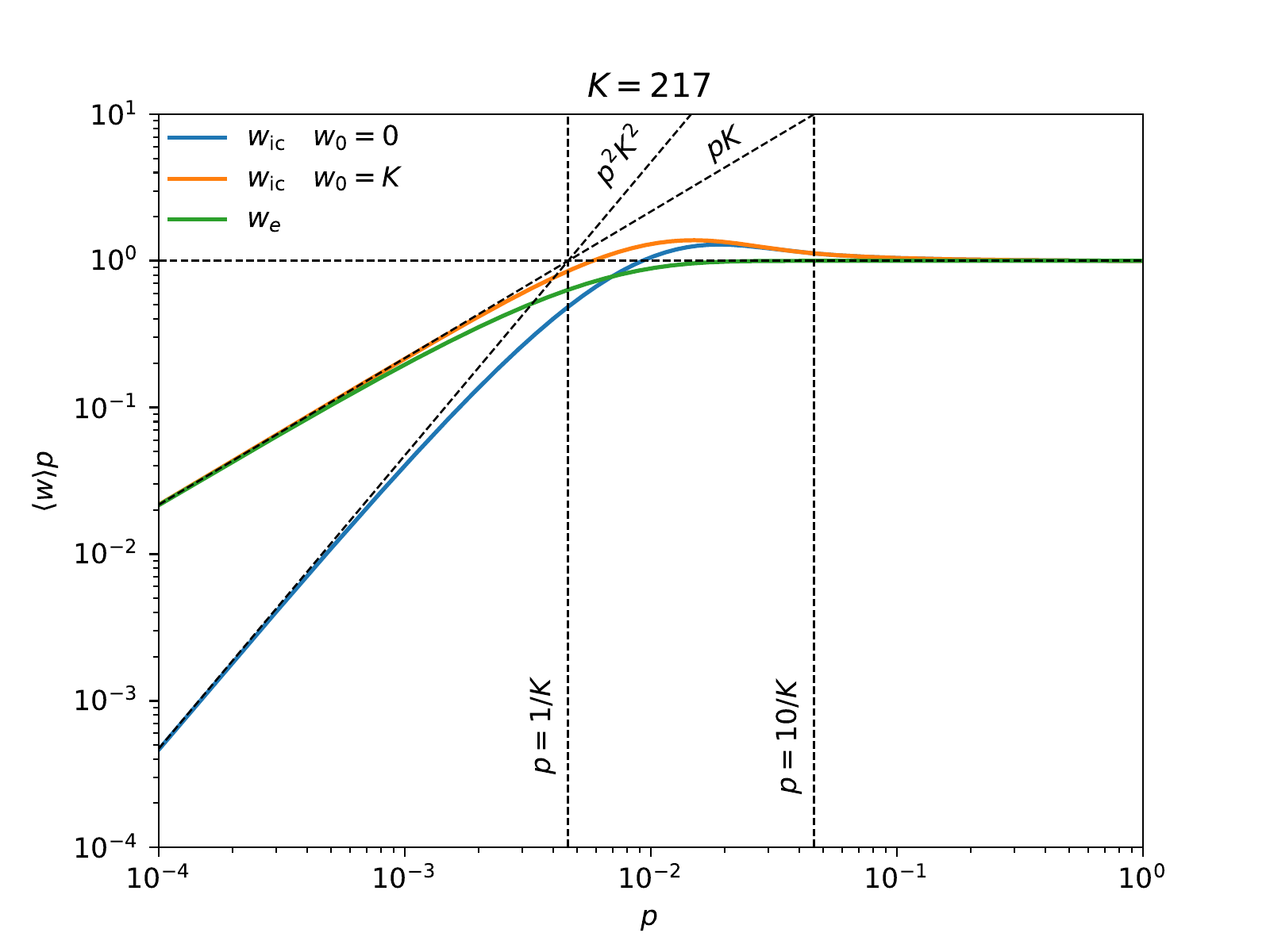}   
   \includegraphics[width=9cm]{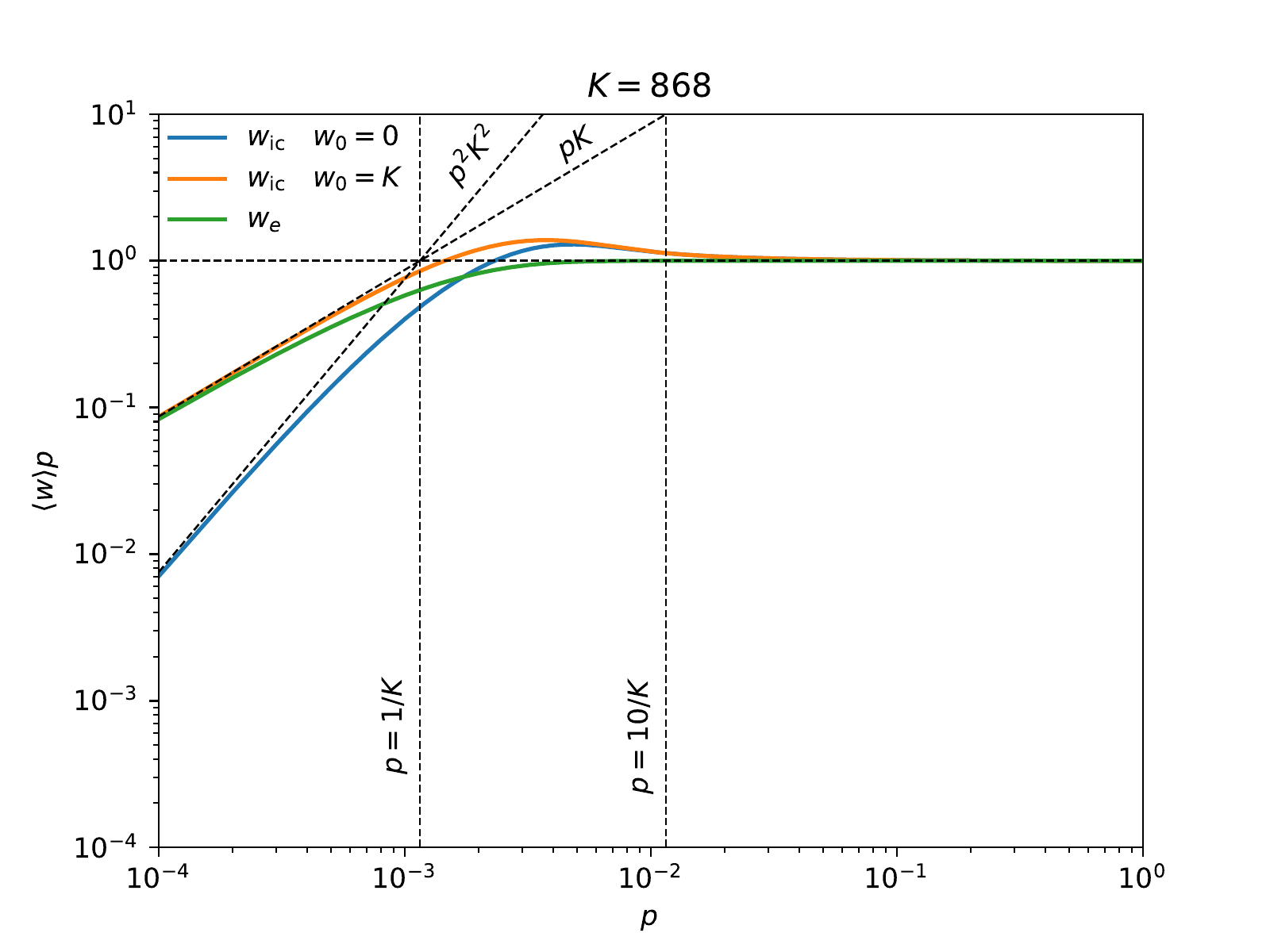}   
   \caption{Ratio between the expectation value of the weight $\langle w \rangle$ and the true value of the quantity that we are estimating, i.e. the inverse probability $1/p$, for three different number of realisations of the selection process $K$ (top of each panel). Three different definitions of the weights are considered: inverse count with $w_0=0$ (solid blue), inverse count with $w_0=K$ (solid orange), efficient estimator (solid green).}
   \label{fig:w_comp}
 \end{center}
\end{figure}

In Fig.~\ref{fig:w_comp} we compare the performance of the different definitions of $w_K$ as a function of the true probability $p$, for different number of realisations $K$. 
Note that the curves in the figure are not obtained via simulations, but rather directly from Eq. \ref{eq:expectation}.
Specifically, we consider two versions of the inverse-count estimator $w_{\rm ic}$, corresponding to $w_0=0$ (blue) and $w_0=K$ (orange), and the efficient estimator $w_{\rm e}$ (green).
As discussed above, for this test, the performance of the zero-truncated estimator $w_{\rm zt}$ are identical by construction to those of the efficient estimator.
Ideally, one would wish to recover the reference amplitude $\langle w_K \rangle / p^{-1}= 1$ for any value of the variable $p$, but this is in practice unfeasible with a finite number of realisations.
As expected, a regular improvement of the performance of all the estimators can clearly be seen as $K$ grows. 
It is also clear that, for any fixed $K$, the efficient estimator returns a significantly better estimate of $1/p$ compared to its inverse-count counterpart.
From a quantitative point of view, $w_{\rm e}$ yields an error of $\mathcal{O}(10^{-5})$ at $p=10/K$, which drops very quickly below numerical precision for larger values of $p$.
For comparison, at the same abscissa $p=10/K$ the accuracy of $w_{\rm ic}$ is about four order of magnitude worse.
Perhaps even more importantly, the overall behaviour the inverse-count estimator is non monotonic with respect to $p$, for both values of $w_0$.
This may cause uncontrolled cancellations when testing its convergence on real datasets.
When dealing with galaxy surveys, these non monotonic features could even induce scale-dependent effects on the measured clustering, at least in theory. 

\begin{figure}
 \begin{center}
    \vspace{-0.5cm}
   \includegraphics[width=7.4cm]{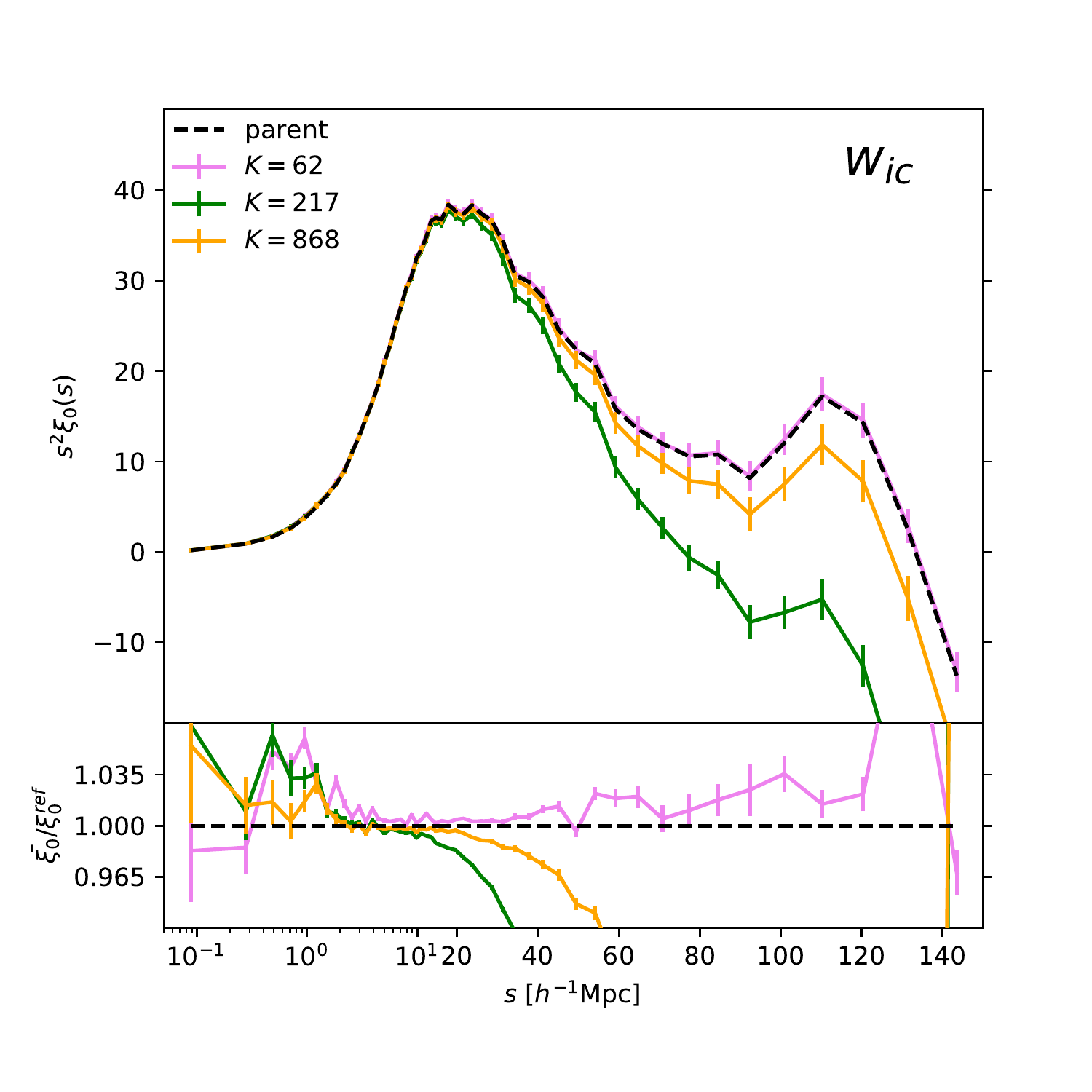}\\
   \vspace{-0.6cm}
   \includegraphics[width=7.4cm]{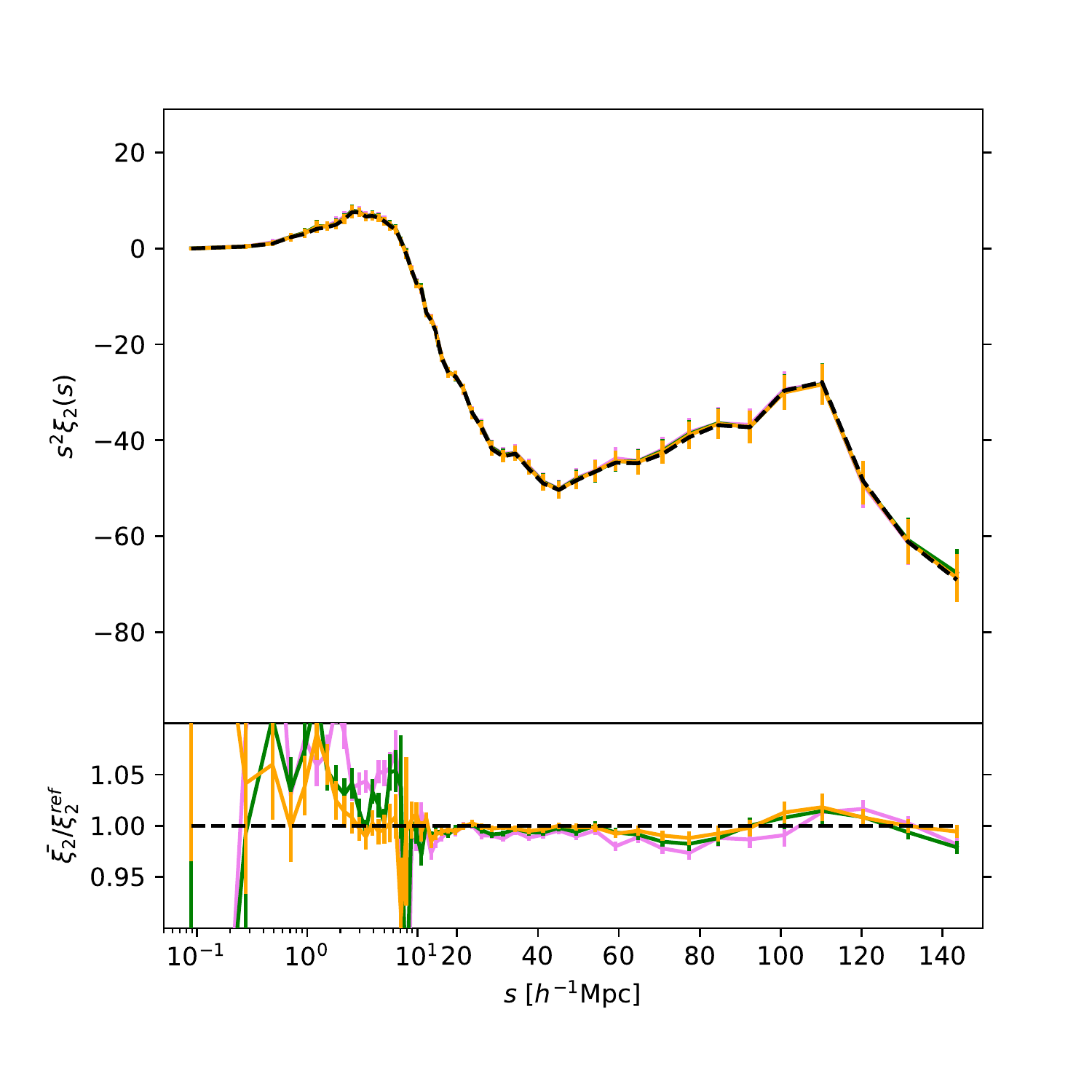}\\ 
   \vspace{-0.6cm}  
   \includegraphics[width=7.4cm]{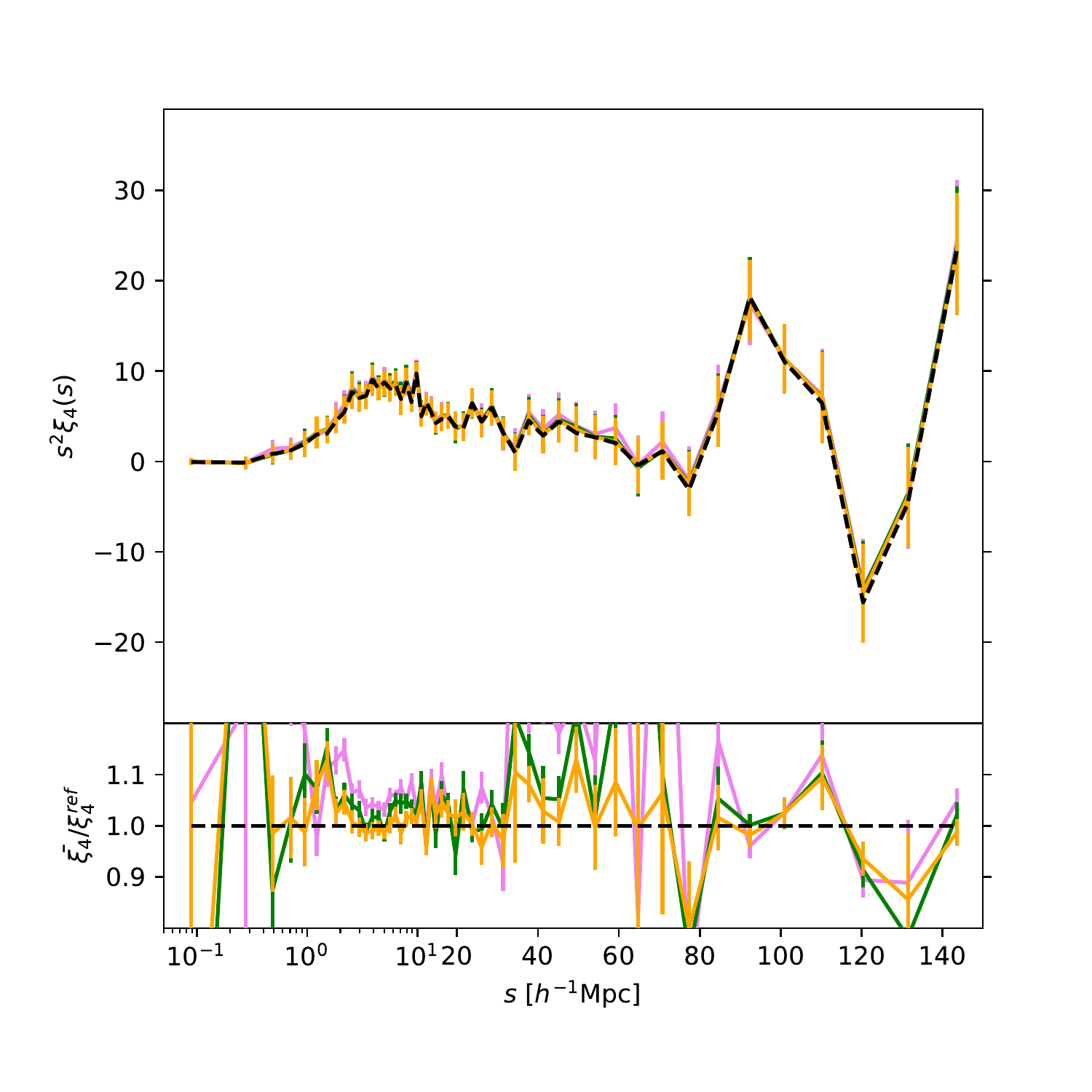}\\
   \vspace{-0.6cm}        
   \caption{Legendre multipoles of the redshift-space correlation $\xi$ (multiplied by the square of the separation $s^2$ for visualisation purposes)  measured from the simulation via the inverse-count estimator compared to their reference values obtained directly from the parent sample (black dashed). When deriving the PIP weights, three different number of realisations of the targeting have been considered: $K=62$ (solid purple), $K=217$ (solid green), $K=868$ (solid orange). In the top panel of each plot we show the mean obtained from 124 realisations of the targeting (independent from those used to compute the weights) with error bars corresponding to their standard deviation. In the lower panels we show the ratio between mean and reference, with error bars of the mean.}
   \label{fig:xi_comp_std}
 \end{center}
\end{figure}

\begin{figure}
 \begin{center}
   \vspace{-0.5cm} 
   \includegraphics[width=7.4cm]{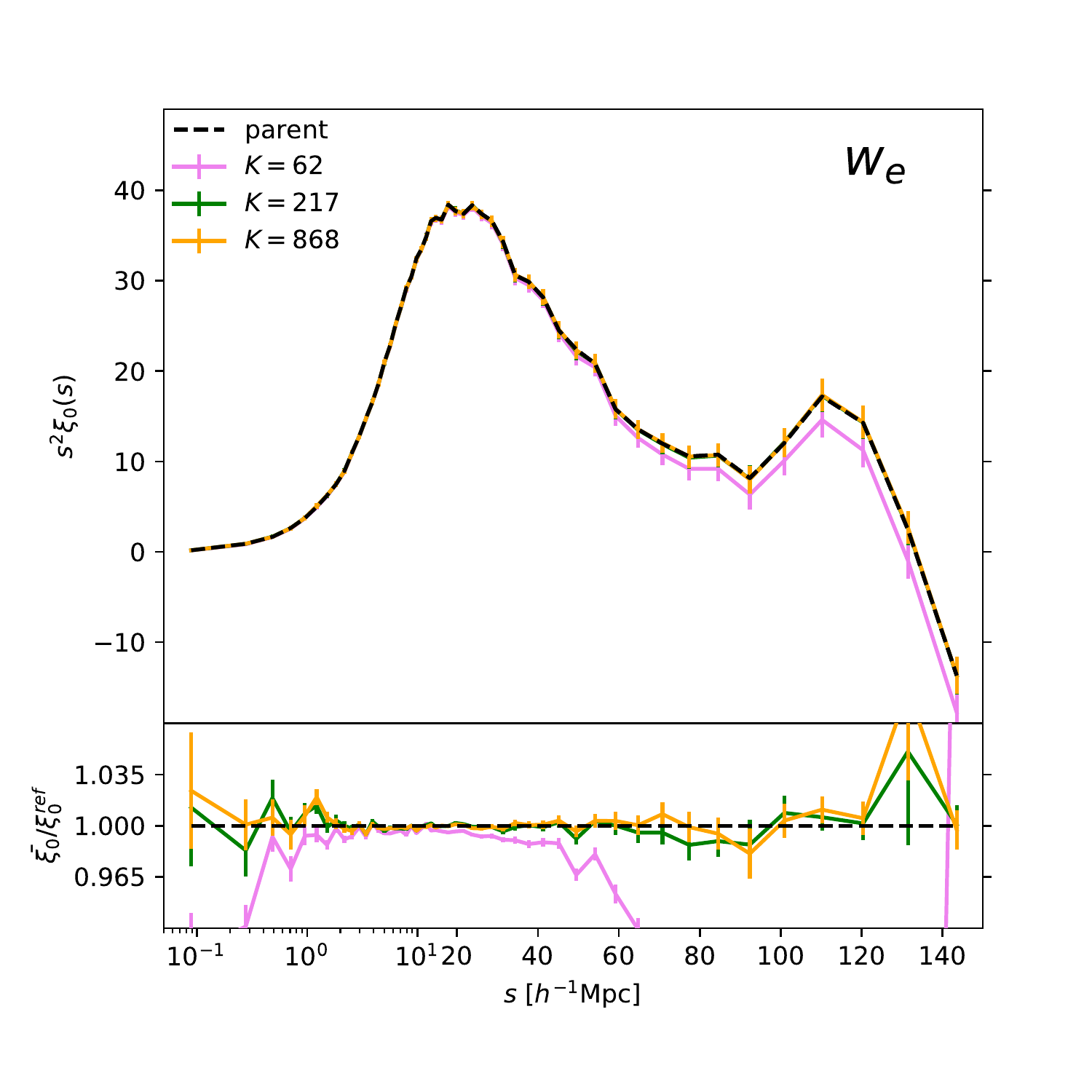}\\
   \vspace{-0.6cm}
   \includegraphics[width=7.4cm]{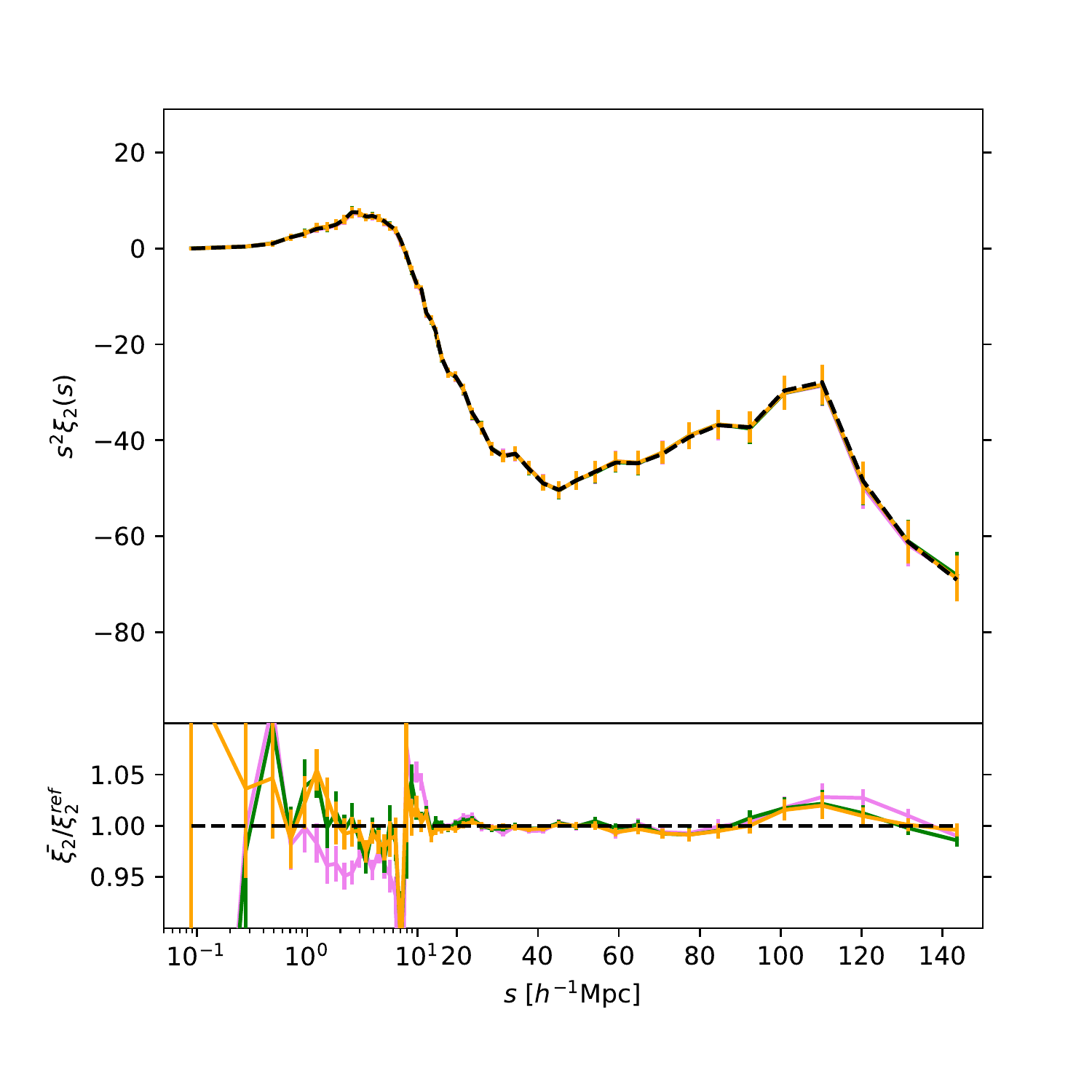}\\ 
   \vspace{-0.6cm}4
   \includegraphics[width=7.4cm]{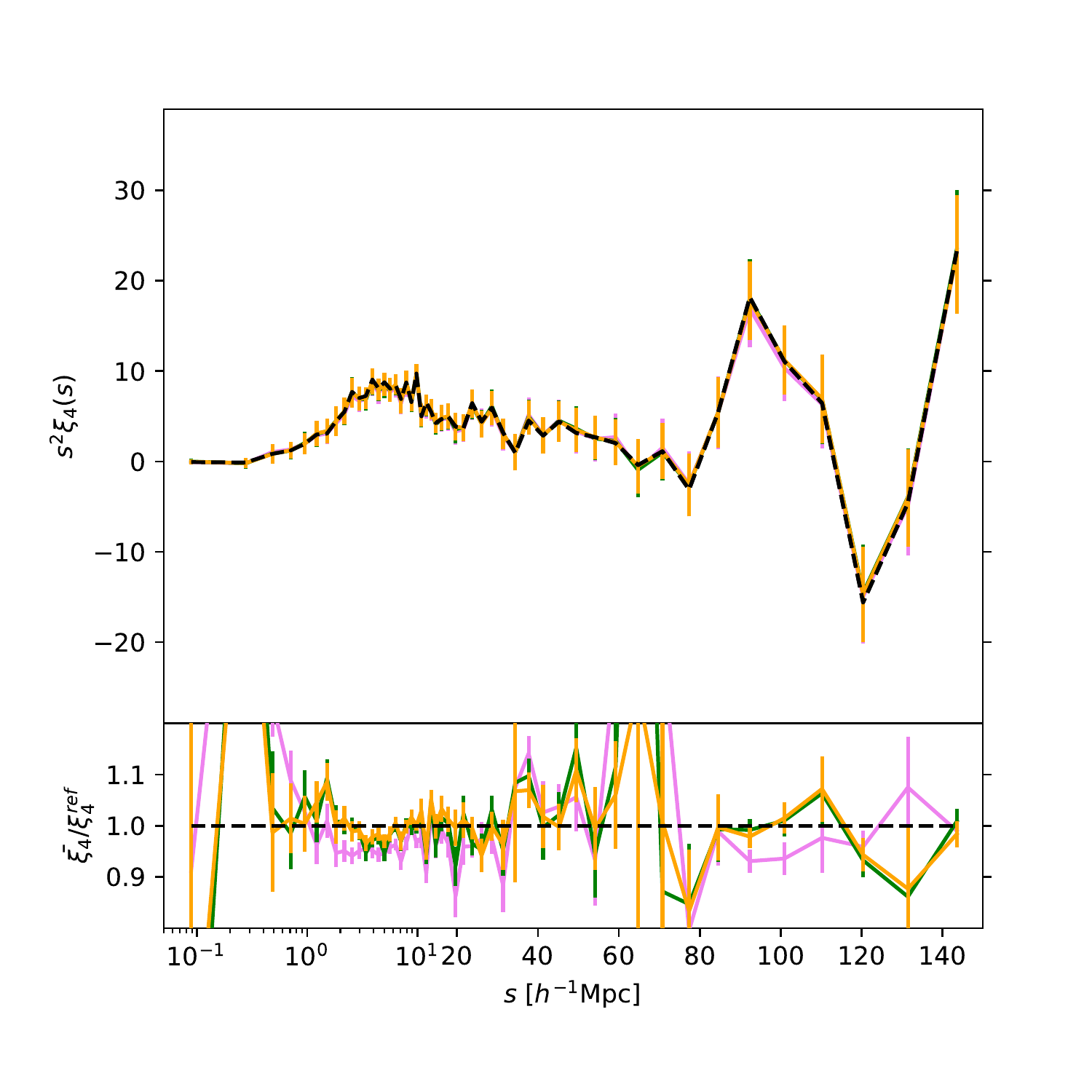}   
   \caption{Same as Fig. \ref{fig:xi_comp_std} but for the efficient estimator.}
   \label{fig:xi_comp_eff}
 \end{center}
\end{figure}

%To understand how correlation-function measurements respond to the different choices of $K$ and estimator just discussed, we need to imagine of integrating the curves in Fig. \ref{fig:w_comp} weighted by the distributions of the selection probabilities of the pairs a the various separation scales.
To understand how correlation-function measurements respond to the different choices of the estimator and how quickly they converge by increasing $K$, we need to imagine of integrating the curves in Fig.~\ref{fig:w_comp} weighted by the distributions of the selection probabilities of the pairs a the various separation scales.
%The overall effect on the correlation function is essentially an integral of the curves in Fig. \ref{fig:w_comp} weighted by the distribution of the selection probabilities across the pairs at a given separation.
Since any global effect on the amplitude gets absorbed by the pair-counts normalisation, it is the scale dependence of these distributions that can create deviations from the true correlation.
%For this reason we suggest not to infer a minimum reasonable value for $K$ directly from Fig. \ref{fig:w_comp} (e.g. $K_{min} \sim 10/p_{min}$ for the efficient and zero-truncated estimators, where $p_{min}$ is the probability of the most unlikely pairs), but rather to perform convergence tests.
For this reason trying to infer a minimum reasonable value for $K$ directly from Fig.~\ref{fig:w_comp}, e.g. $K_{min} \sim 10/p_{min}$ for the efficient and zero-truncated estimators, where $p_{min}$ is the probability of the most unlikely pairs, would realistically lead to overly conservative conclusion.
We rather suggest to perform convergence tests.
In Fig.~\ref{fig:xi_comp_std} we show the performance of the inverse-count estimator on measurements of the (Legendre) monopole $\xi_0$ (top panel), quadrupole $\xi_2$ (central panel), and hexadecapole $\xi_4$ (lower panel), from our simulation (Sec. \ref{sec:simulations}).
The PIP weights are computed for three different number of realisation $K=62,217,868$, violet, green and orange, respectively, and they are used to correct the clustering measurements from 124 independent realisation of the targeting.
The pair counts are normalised by the total number of weighted pairs in each sample.    
Not surprisingly, the convergence to the true value, measured directly from the parent sample, does not seem to be monotonic.
Especially for the monopole, it seems clear that the $K=62$ measurement agrees more with the true value than the (in principle) more accurate $K=217$ and $K=868$ measurements.
This is clearly a chance coincidence, explained by the non monotonic behaviour, with respect to $p$, of the inverse-count estimator and confirmed by the more regular convergence, for increasing $K$, of the other two measurements.
Furthermore, we note that even the largest value of $K$ considered in this work leads to a non negligible underestimate of the true monopole.
In Fig.~\ref{fig:xi_comp_eff} we show the results for the efficient estimator.
Clearly all the issue encountered with the inverse-count estimator are removed by this new approach.
The convergence with $K$ is regular and much faster.
Even with just $K=217$ the measurements are in good agreement with the true value. 

In light of the above, we recommend to use either the efficient or the zero-truncated estimator, which share the same convergence properties, instead of the inverse-count estimator.
Choosing between efficient and zero-truncated appears to be a practice-related issue.
If we have a parent sample with $K$ realisations of the selection process, it is convenient to use all the realisations for both deriving the weights and testing (e.g. by taking the mean of the weighted samples).  
This can be done with the zero-truncated estimator.
When adopting this strategy it is important to be aware that we are enforcing the exact cancellation of the fluctuations around the mean value, which, formally, should be reached only for $K \rightarrow \infty$.  
In other words, we will always find an apparent inconsistency between the behaviour of the mean and its error bars, the former matching the expected value better than what suggested by latter.   
Other than this visualisation artefact, we did not find any obvious reason against this strategy and, as consequence, we decided to adopt it for the rest of this work.

\subsection{Sampling individual and pair probabilities}\label{sec:sampling_ind_vs_pair}

In this section we discuss sampling effects that arise when comparing individual (IIP) with pairwise inverse probabilities (PIP).
There are, at least, two common circumstances in which these effects become relevant and require appropriate modelling: (i) when computing the normalisation of the correlation function; (ii) when evaluating the characteristic length $s_\perp^{(c)} \in [0,\infty]$ above which the probability of pairs can be well approximated by the product of individual probabilities\footnote{With a more realistic survey geometry, beyond plane parallel approximation, the selection-correlation length obviously becomes a separation angle $\theta_c \in [0,2\pi]$.}.

We consider the common situation of having $K$ realisations of the targeting from which we want evaluate both individual and pair weights.
Given a pair formed by galaxies 1 and 2, we denote with $c_{12}$, $c_1$, $c_2$ the corresponding recurrences and $p_{12}$, $p_1$, $p_2$ the true probabilities.
Via combinatory calculus, it can be shown (App. \ref{app:distributions}) that if $p_{12}=p_1 p_2$, i.e. the galaxies are selected independently, the probability distribution of $c_{12}$ given $c_1$ and $c_2$ is given by
\begin{equation}\label{eq:dist_ind_vs_pair}
\mathcal{P}(c_{12} | c_1, c_2) = \binom{c_1}{c_{12}} \binom{K - c_1}{c_2 -c_{12}} {\binom{K}{c_2}}^{-1}  \ ,
\end{equation}
where\footnote{
Despite the appearances, Eq. (\ref{eq:dist_ind_vs_pair}) is symmetric with respect to $c_1$~and~$c_2$.
This can be seen, e.g., by expressing it in terms of factorials, 
\begin{equation}
\mathcal{P}(c_{12} | c_1, c_2)  =  \frac{c_1! (K - c_1)! c_2! (K - c_2)!}{c_{12}! (c_1 - c_{12})! (c_2 - c_{12})! (K - c_1 - c_2 + c_{12})!  K!} \ . \nonumber
\end{equation}
} $0 \le c_{12} \le \min(c_1,c_2)$.

\begin{figure}
 \begin{center}
   \includegraphics[width=9cm]{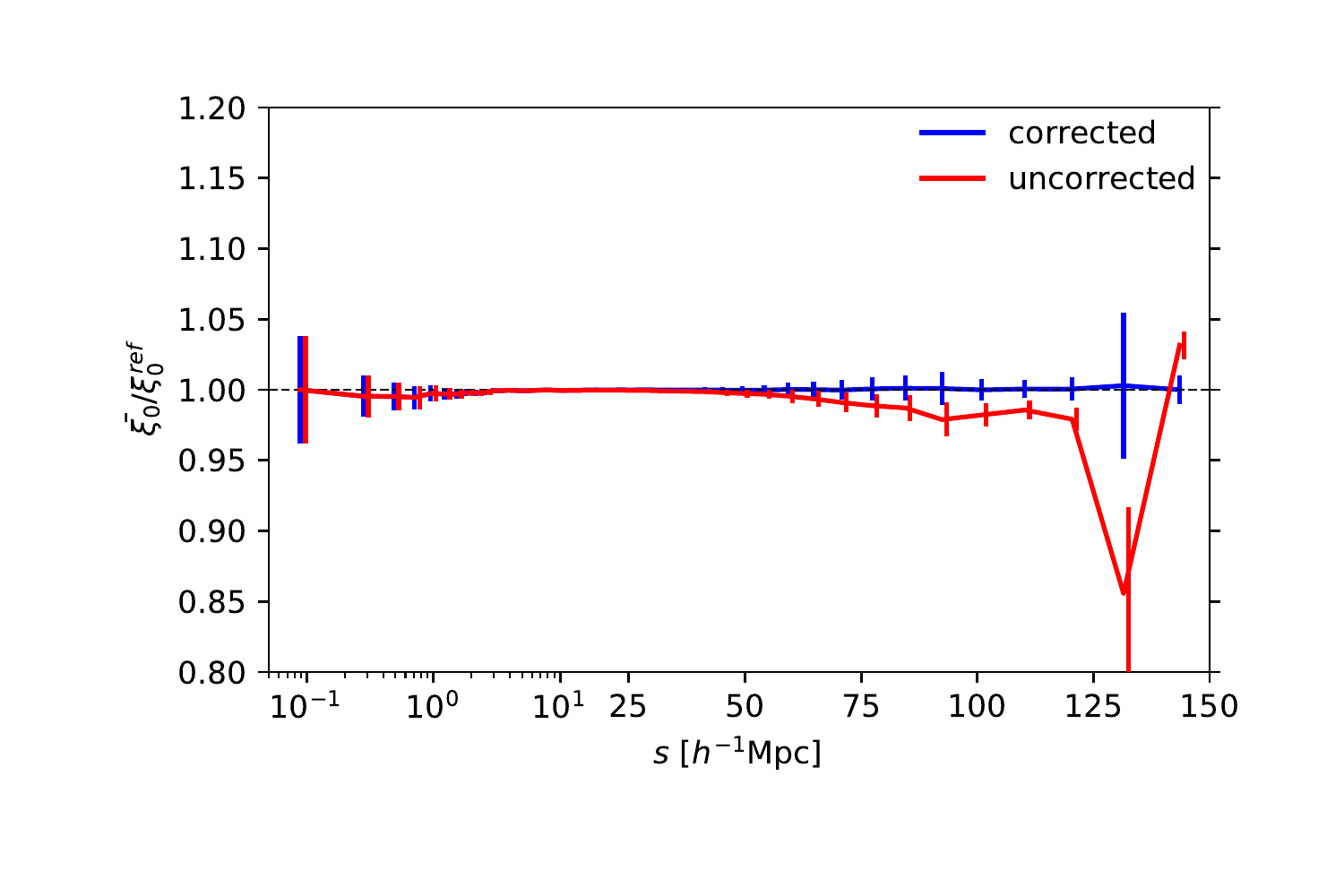} 
   \caption{Impact on the Legendre monopole of the correlation function $\xi_0$ of ignoring the sampling correction in the pair-count normalisation $n_{DD}$. Specifically, we compare the mean of the monopole obtained via PIP weights (zero-truncated estimator) from $K=217$ realisations of the targeting with $n_{DD} = [ \left( \sum w_{\rm IIP} \right)^2 - \sum w_{\rm IIP}^2 ] / 2$ (solid red) and $n_{DD}$ obtained via Eq.~\ref{eq:norm} (solid blue). Both measurements are divided by the reference value measured directly from the parent sample. The error bars are the standard deviation of the mean, slightly shifted for visualisation purposes.}
   \label{fig:normPIP_vs_normIIP}
 \end{center}
\end{figure}

\begin{figure}
 \begin{center}
   \includegraphics[width=7.5cm]{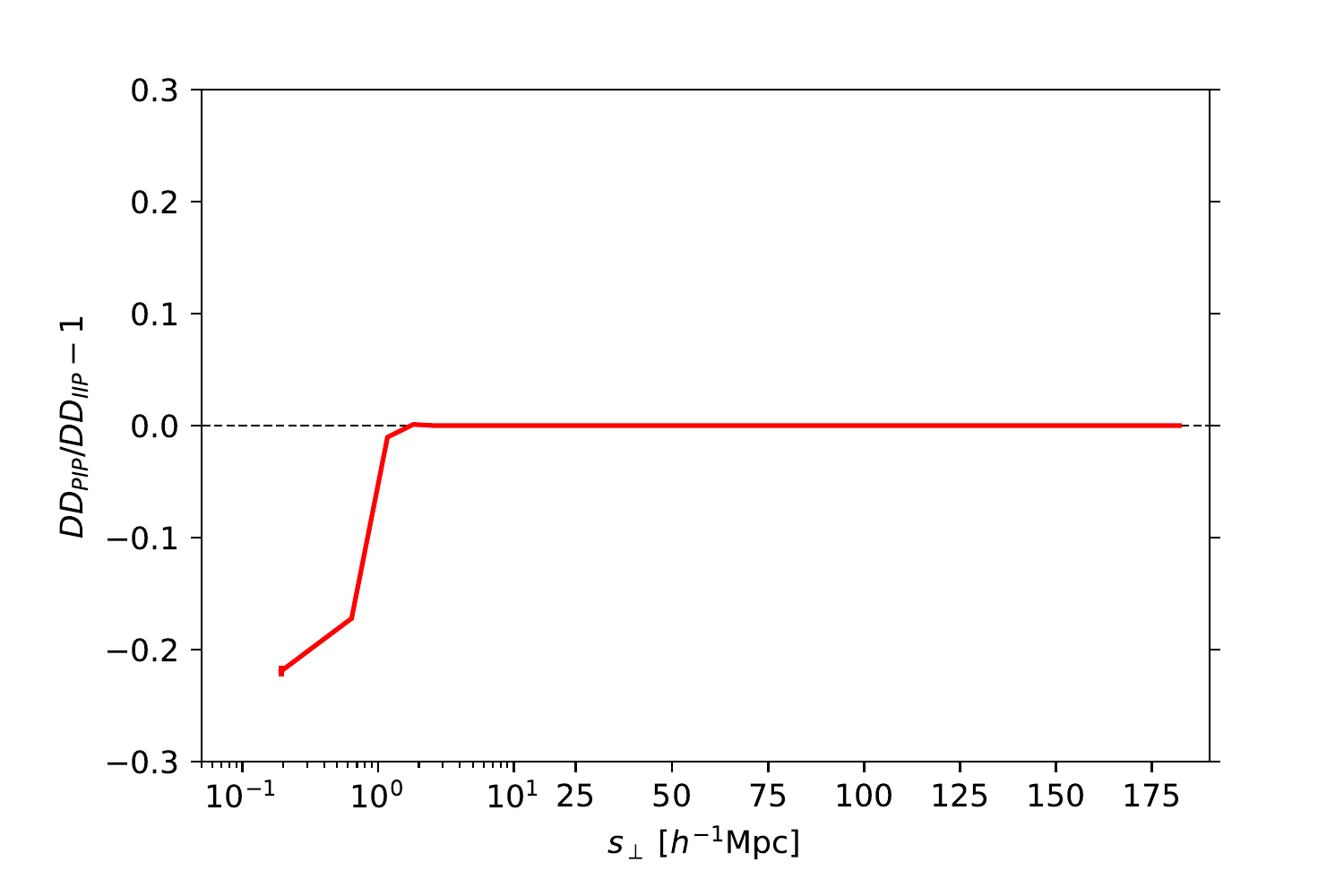}\\
    \hspace{-0.9cm}
    \includegraphics[width=7.5cm]{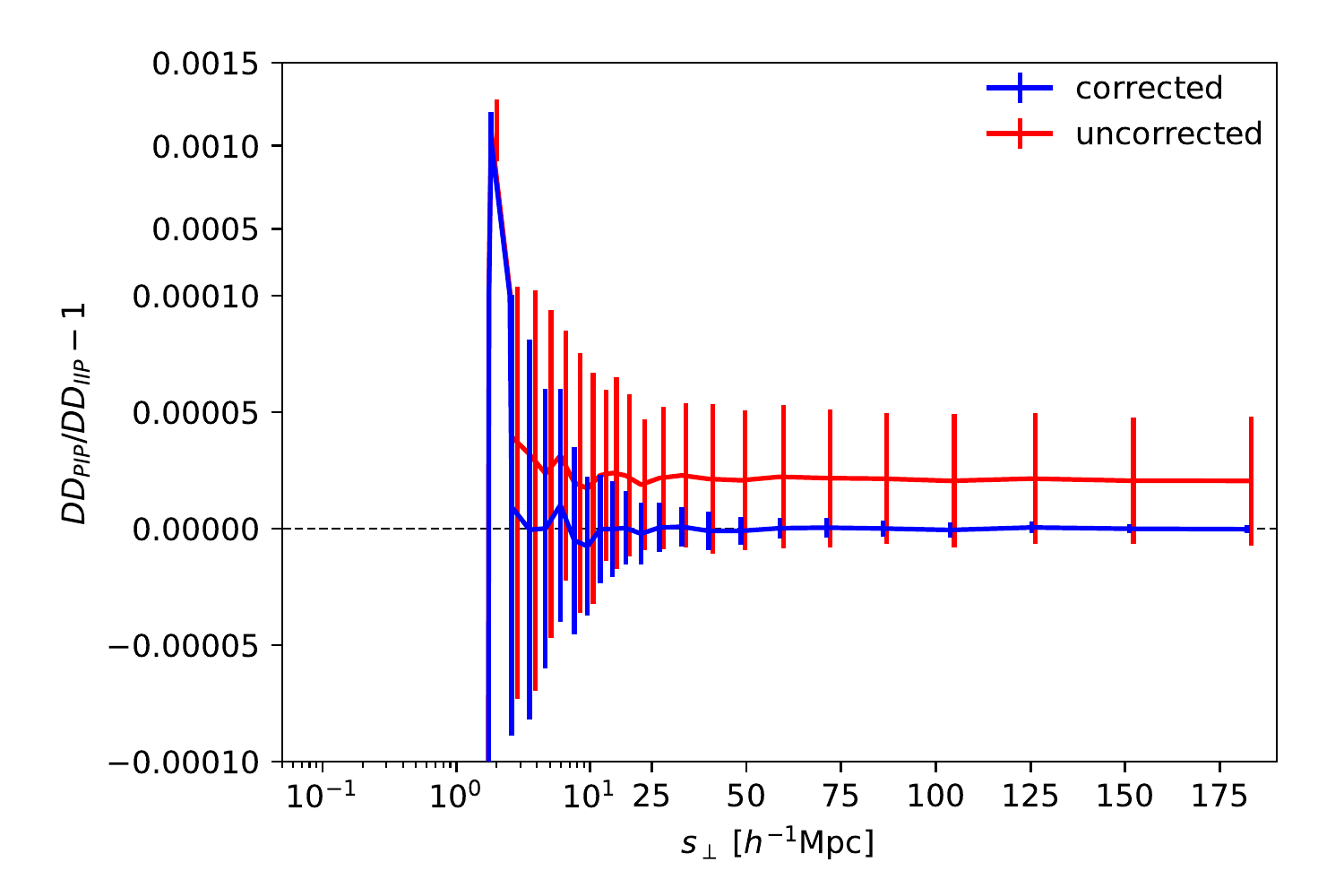}   \\
   \includegraphics[width=7.5cm]{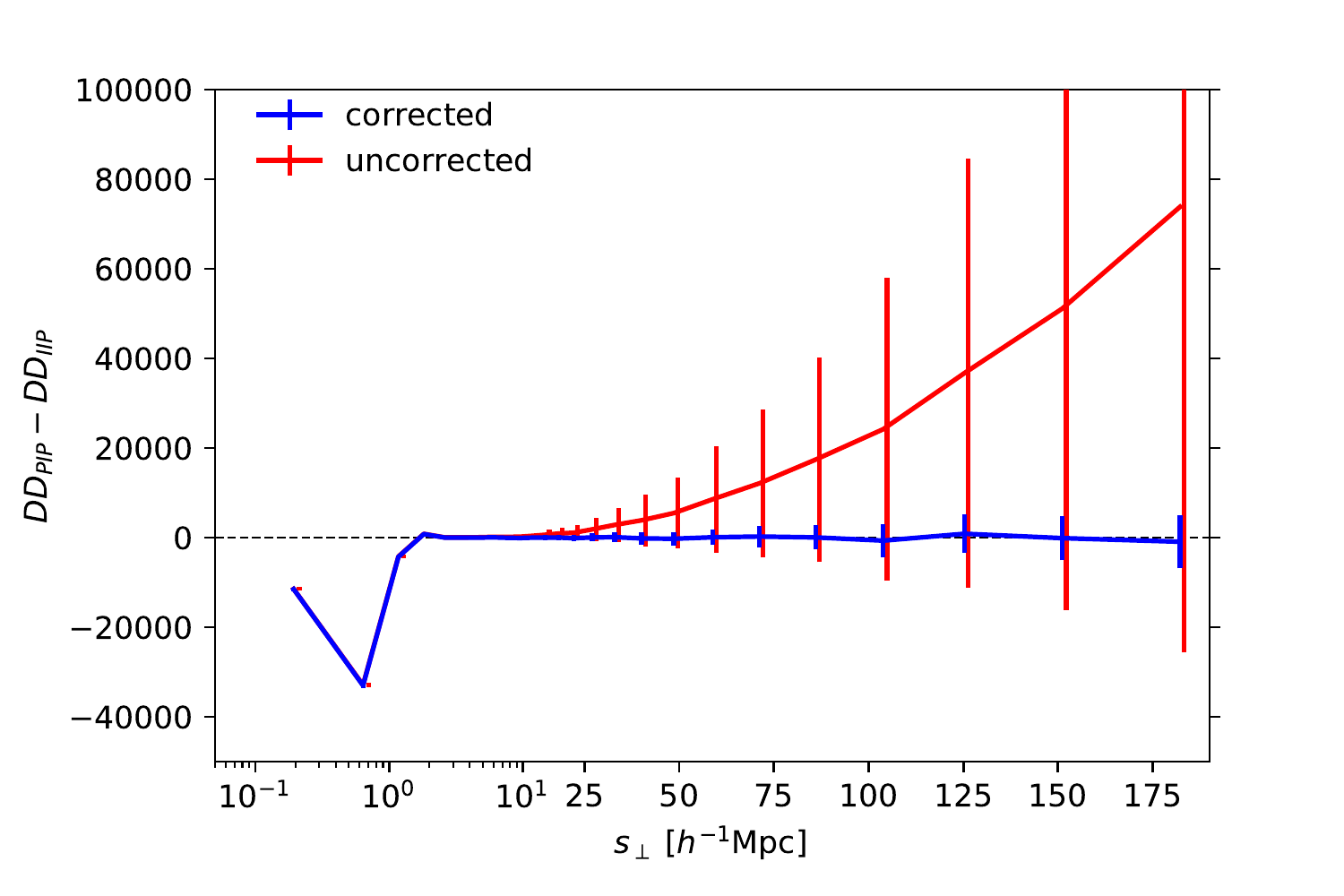}
   \caption{Determination of the selection-correlation length. Top panel: selection-correlation coefficient $DD_{\rm PIP}/DD_{\rm IIP}-1$ as a function of the perpendicular separation $s_\perp$, where $DD_{\rm PIP}$ and $DD_{\rm IIP}$ are the mean of the pair counts obtained via PIP and IIP weights, respectively, from $K=217$ realisations of the targeting. Central panel: same as top panel but with a finer scale for the $y$ axis. Two different measurements are reported, corresponding to different definitions of $DD_{\rm IIP}$, with (solid blue) and without (solid red) sampling correction, as defined in the text. Note that, in order to accommodate in the figure the sharp (positive) correlation peak at $\sim 2 h^{-1}$Mpc, the $y$-axis scale actually gets coarser in the upper part of the panel. Lower panel: same as central panel but for the quantity $DD_{\rm PIP}-DD_{\rm IIP}$. The error bars (slightly shifted for visualisation purposes) are the standard deviation of the mean.}
   \label{fig:corr_eff}
 \end{center}
\end{figure}

As a consequence, the (conditional) expected value for $f_K(c_{12})$ becomes
\begin{equation}\label{eq:effpc}
g_{K}(c_1,c_2) =  \sum_{c_{12}=0}^{\min(c_1,c_2)} f_K (c_{12}) \mathcal{P}(c_{12} | c_1, c_2) \ne f_K(c_1) f_K(c_2) \ .
\end{equation}
Similarly, for the zero-truncated scenario we have 
\begin{equation}\label{eq:dist_ind_vs_pair_zt}
\mathcal{P}_{\rm zt}(c_{12} | c_1, c_2) = \binom{c_1 - 1}{c_{12} - 1} \binom{K - c_1}{c_2 -c_{12}} {\binom{K - 1}{c_2 - 1}}^{-1}  \ ,
\end{equation}
where $1 \le c_{12} \le \min(c_1,c_2)$, which leads to the same conclusion, %$g_{K}(c_1,c_2) \ne f_K(c_1) f_K(c_2)$, as before .
\begin{equation}\label{eq:ztpc}
g_{K}(c_1,c_2) =  \sum_{c_{12}=1}^{\min(c_1,c_2)} f_K (c_{12}) \mathcal{P}_{{\rm zt}}(c_{12} | c_1, c_2) \ne f_K(c_1) f_K(c_2) \ ,
\end{equation}
%\begin{equation}
%g_{K}(c_1,c_2) =  \sum_{c_{12}=0}^{\min(c_1,c_2)} f_K (c_{12}) \binom{c_1}{c_{12}} \binom{K - c_1}{c_2 - c_{12}} {\binom{K}{c_2}}^{-1} \ne f_K(c_1) f_K(c_2) \ ,
%\end{equation}
as before.
This ultimately shows that, when we try to assess the selection correlation length, e.g. by taking the ratio of PIP and IIP pair counts or their difference, we measure an apparent correlation even when the true one is zero.
As anticipated there are two relevant consequences, which we discuss in the following. 
\begin{enumerate}
\item
The exact normalisation $n_{DD}$ of the galaxy pair counts $DD$ requires computational demanding counts over all the pairs in the sample.
Compared to $DD$ counts, the direct evaluation of $n_{DD}$ is order of magnitudes more time consuming since the process cannot be speeded up through linked lists or similar computational shortcuts.  
If the selection-correlation length is much smaller than the size of the survey, $s_c \ll L$, it should be possible to obtain the normalisation from the individual weights as $n_{DD} = \frac{1}{2}\left[ \left( \sum w_{\rm IIP} \right)^2 - \sum w_{\rm IIP}^2 \right]$, where the sums are performed over all the observed galaxies.
Unfortunately, as shown in Fig. \ref{fig:normPIP_vs_normIIP}, this approach tends to fail.
Against the intuition, in the vast majority of  the cases, such behaviour does not mean that the selection-correlation has a large impact on the normalisation factor, but rather it is a direct manifestation of the sampling effects just discussed.
%The correct normalisation can be obtained via Eqs. (\ref{eq:dist_ind_vs_pair}) or (\ref{eq:dist_ind_vs_pair_zt}) as
%\begin{equation}\label{eq:norm}
%n_{DD} = \frac{1}{2} \left[ \sum_{i=1}^{K} \sum_{j=1}^{K} g_K(\tilde{c}_i,\tilde{c}_j) \ h(\tilde{c}_i) \ h(\tilde{c}_j)  -  \mathcal{A} \right] \ ,
%\end{equation}
%where $h(\tilde{c}_i) = \sum_{n=1}^{N_{gal}} \delta^{(K)}_{c_n,\tilde{c}_i}$, with $\tilde{c}_i \in \{1,\dots,K\}$ and $\delta^{(K)}$ is the Kronecker delta, which is just a formal way to say that $h$ is the unnormalised distribution of $c$ measured from the galaxies in the sample.
The correct normalisation can be obtained as
\begin{equation}\label{eq:norm}
n_{DD} = \frac{1}{2} \left[ \sum_{i}^{K} \sum_{j}^{K} g_K(i,j) \ h_i \ h_j  -  \mathcal{A} \right] \ ,
\end{equation}
with $h_i = \sum_{n=1}^{N_{gal}} \delta^\mathcal{K}_{c_n,i}$, where $\delta^\mathcal{K}$ is the Kronecker delta, which is just a formal way to say that $h_i$ is the unnormalised distribution, measured from all the galaxies in the sample, of their recurrence $c_n$.
For the inverse-count and efficient estimator $g_K$ is obtained from Eq. \ref{eq:effpc} and the sums are performed from $i,j=0$, whereas for the zero-truncated estimator, $g_K$ is obtained from Eq. \ref{eq:ztpc} and the sums are performed from $i,j=1$.  
The $\mathcal{A}$ term on the righthand side of the equation represents the autocorrelation of the galaxies with themselves.
This term is typically subdominant but there is no point in neglecting it since it can be easily computed as  
\begin{equation}
\mathcal{A} = \sum_{n=1}^{N_{gal}} g_K(c_n,c_n) \ .
\end{equation}
%Note that here the sum is over the number of galaxies $N_{gal}$ and $c_n$ is the actual statistical count of the $n$-th galaxy.
We stress that, for any realistic value of $K$, the evaluation of Eq.\ref{eq:norm} is computationally costless\footnote{Since the evaluation of $g_K$ involves lengthy factorial calculations we strongly recommend to tabulate once for ever the correspondent 2-dimensional array, of effective size $K (K+1)/2$ (the array is symmetric with respect to the diagonal), for the $K$ of interest and simply recall it when computing normalisations and, especially, when performing pair counts.}.

\item 
When dealing with missing observations, it is useful to know if it exists a characteristic length $s_\perp^{(c)}$ above which the selection-correlation is negligible or, in other words, above which we are allowed to use IIP instead of PIP weights.
One advantage of using IIP weights is that they can be straightforwardly assigned to a grid and, consequently, used as an input for fast Fourier transform (FFT) algorithms (see Sec. \ref{sec:Fourier} for a more exhaustive discussion).      
A practical way to infer $s_\perp^{(c)}$ is to take the ratio of the pair counts obtained via PIP with those obtained via IIP. 
In Fig. \ref{fig:corr_eff} is reported the selection-correlation coefficient $DD_{\rm PIP}/DD_{\rm IIP}-1$, as a function of the perpendicular separation\footnote{Since we are dealing with a periodic box, $s_\perp$ can be interpreted as an angle.} $s_\perp$,  measured from the simulation via the zero-truncated estimator.
The inverse-count and efficient estimators yield similar results.
The upper panel shows the overall behaviour of the coefficient, with a y-axis scale large enough to follow its largest fluctuations.
Not surprisingly, the selection correlation becomes negative below about $1 h^{-1}$Mpc, which corresponds to the collision scale.
To understand what happens at larger scales we need to zoom in on the y-axis, central panel.
Two different measurements are reported, the red curve is the same as the upper panel, the blue curve is obtained via Eq. \ref{eq:ztpc}.
Specifically, $DD_{\rm PIP} = \sum w^{(12)}_{\rm zt}$ is the same for both measurements, whereas $DD_{\rm IIP} = \sum w^{(1)}_{\rm zt} w^{(2)}_{\rm zt}$ for the uncorrected case and $DD_{\rm IIP} = \sum w^{(12)}_{\rm ind}$ for the corrected case, where we defined $w^{(12)}_{\rm ind} = b g_K(c_1,c_2)$. As usual, $b$ is the binary random variable encoding whether the pair is selected or not and the sums are performed in bins of the separation.
The uncorrected measurements show a correlation tail that extends to the largest separations, which completely disappears for the corrected estimator.

The same behaviour is observed for $DD_{\rm PIP} - DD_{\rm IIP}$, which is the quantity of interest for the power spectrum, as discussed in Sec.~\ref{sec:Fourier}.
In lower panel of Fig.~\ref{fig:corr_eff} we show this quantity measured with (solid blue) and without (solid red) the sampling correction. 
As expected we see that without a proper treatment of the sampling effects, Eq.~\ref{eq:dist_ind_vs_pair_zt}, we would be misled by the presence of an apparent large-scale selection correlation, even more evident than for the $DD_{\rm PIP}/DD_{\rm IIP}$ case. 

To summarise, we have assumed that the selection process is independent, $p_{12}=p_1p_2$, derived a theoretical prediction for the sampling-induced selection correlation and shown that it exactly matches what observed in our simulation on scales larger than about $3 h^{-1}$Mpc.
We conclude that, on those scales, the selection probabilities are actually independent.     
%Note that this also implies that the estimates of the pair probabilities are not fully converged for the number of realisations $K$ adopted, which is consistent with fact that for the figure we use $K=62$.
%If they where converged we should have measured negligible selection correlation on the scales at which the true correlation is zero (we verified this behaviour for larger $K$).
%
%For practical purposes, we recommend to evaluate the selection-correlation scale $s_{c}$ via Eq. \ref{eq:}, i.e. via\footnote{precompute} $w^{(12)}_{\rm ind}$.
%This approach disentangles the determination of such scale from sampling issues.
%If $s_c$ exists, for $s > s_c$ one is allowed to use individual probabilities (see Sec. \ref{sec:}).
%Also, once $s_c$ is determined, comparing IIP and PIP for $s > s_c$ can help understanding the convergence properties of the weights as function of $K$, i.e. how many realisations it is reasonable to use.    

Finally, we note an inversion of the sign of the selection correlation, with a sharp peak at about $2 h^{-1}$Mpc, which is barely noticeable in the upper and lower panel of Fig. \ref{fig:corr_eff}, but becomes quite apparent when we zoom in on the $y$-axis (central panel, not to be confused with the error bars).
%This counterintuitive behaviour seems to be universal, e.g., it can also be seen in figure 9 of \citet{smith2019}.
%Despite its impact on the measurements presented in this work being completely negligible,
Although small, this feature is interesting as it represents empirical evidence of the fact that, in addition to the expected anticorrelation below the collision length, fibre collisions also create a, more counterintuitive, positive correlation just above such length.  
\end{enumerate}

%\begin{figure*}
% \begin{center}
%   \includegraphics[width=8cm]{{/Users/davide/qua/corfu_pawe/plt/MDR1_62_0.00005_shift_spasst_ms1_w_bw62_IIPreg_norm_PIPvsIIP_reg_ang}.pdf} 
%   \includegraphics[width=8cm]{{/Users/davide/qua/corfu_pawe/plt/MDR1_62_0.00005_shift_spasst_ms1_w_bw62_IIPregpc_norm_PIPvsIIP_reg_ang_zoom_test_comp}.pdf}
%   \caption{???}
%   \label{fig:corr_eff}
% \end{center}
%\end{figure*}
%
%\begin{figure}
% \begin{center}
%   \includegraphics[width=7.5cm]{{/Users/davide/qua/corfu_pawe/plt/MDR1_62_0.00005_shift_spasst_ms1_w_bw217_IIPauwpc_PIPvsIIP_reg_ang_comp_diff}.pdf}
%   \caption{???}
%   \label{fig:sampling_P}
% \end{center}
%\end{figure}

%\begin{align}
%g_{K}(c_1,c_2) & =  \nonumber \\
%=  \sum_{c_{12}=1}^{c_1} & \frac{K}{c_{12}}\binom{c_1-1}{c_{12}-1}\frac{\prod_{i=1}^{c_{12} - 1} (c_2-i) \prod_{j=1}^{c_1 - c_{12}} (K - c_2 + 1 - j)}{\prod_{k=1}^{c_1 - 1} (K - k)} 
%\end{align}
%etc\footnote{
%Eq. (\ref{}) can be also expressed in terms of factorials,
%\begin{equation}
%g_{K}(c_1,c_2) =  \sum_{c_{12}=1}^{c_1} \frac{K}{c_{12}}\binom{c_1-1}{c_{12}-1}\frac{(c_2-1)! (K-c_2)! (K - c_1)!}{(c_2 - c_{12})! (K - c_2 - c_1 + c_{12})!  (K - 1)!} \ . \nonumber
%\end{equation}
%}

%WHAT HAPPENS IF I USE W REG?
%\begin{align}
%g_{K}(c_1,c_2) & =  \nonumber \\
%=  \sum_{c_{12}=0}^{c_1} & f_K (c_{12}) \binom{c_1}{c_{12}}\frac{\prod_{i=0}^{c_{12}-1 } (c_2-i) \prod_{j=0}^{c_1 - c_{12}-1} (K - c_2 - j)}{\prod_{k=0}^{c_1-1} (K - k)} 
%\end{align}
%
%\begin{align}
%g_{K}(c_1,c_2) & =  \nonumber \\
%=  \sum_{c_{12}=0}^{c_1} & f_K (c_{12}) \frac{\prod_{i=0}^{c_{12}-1 } [(c_1-i) (c_2-i)] \prod_{j=0}^{c_1 - c_{12}-1} (K - c_2 - j)}{c_{12}! \prod_{k=0}^{c_1-1} (K - k)} 
%\end{align} 

\section{Fourier space}\label{sec:Fourier}

In large-scale structure analysis it is customary to pair correlation-function measurements with measurements of its Fourier-space counterpart, the power spectrum $P({\bf k})$, which suffer of comparable missing-observation issues. 
Since the targeting is intrinsically a configuration space process, the PIP correction for the power spectrum is less straightforward but can be obtained through a similar reasoning.
The main concept is that, analogously to what happens in configuration space, where the correlation function is a sum over pairs that are either observed or not, the power spectrum can be seen as a sum over modes in $k$-space, which are either observed or not.
It follows that, by weighting each of this modes for its inverse probability of being observed we obtain unbiased estimates of the power spectrum.
Note that by modes in $k$-space we mean $\exp[i {\bf s} \cdot {\bf k}]$, where ${\bf s}$ is a wave number (corresponding to a pair separation) and ${\bf k}$ is the variable.
The problem with the usual configuration space modes, for which ${\bf k}$ is the wave number, is that they do not have the same true/false nature of their $k$-space counterpart, in other words, an individual configuration-space mode can be damped but not cancelled by missing observations.       
%Unfortunately, incorporating the full PIP machinery is possible but leads to prohibitively time consuming algorithms.
%More precisely, it is straightforward to include PIP weights in the more general version of the so called Yamamoto estimator \citep{yamamoto2006}, in which pair counts are performed over the whole set of galaxies.

%We start by defining the density fluctuation \citep{feldman1994}, 
%%\begin{equation}
%%F({\bf r})=\frac{w_{FKP}({\bf r})}{I^{1/2}} [n({\bf r}) - \alpha n_s({\bf r})],
%%\label{eq:FKP_factor}
%%\end{equation}
%\begin{equation}\label{eq:FKP_factor}
%F ({\bf r})= \frac{n({\bf r}) - \alpha n_s({\bf r})}{I} \ ,
%\end{equation}

Following common practice, we introduce the density fluctuation as $F ({\bf r})= n({\bf r}) - \alpha n_s({\bf r})$, where $n$ and $n_s$ are, respectively, the number density of galaxies and of a random catalog covering the same volume of the parent sample, but $1/\alpha$ times denser and fibre-assignment free.
% and $I$ normalises the amplitude of the observed power in accordance with its definition in a universe with no survey selection, $I = \int d^3 r {\bar{n}}^2({\bf r})$.
In order not to overcomplicate the notation, we do not consider standard FKP weights  \citep{feldman1994}, but their inclusion is straightforward.
%\begin{equation}
%\label{eq:I2def}
%I\equiv \int d{\bf r}\,w^2{\bar{n}}^2({\bf r}).
%\end{equation}
We define the multipole power spectrum estimator as \citep{feldman1994, yamamoto2006},
\begin{align}\label{eq:P_estimator} 
\nonumber P_\ell({ k})=& \ \frac{(2\ell+1)}{I}\int \frac{d\Omega_k}{4\pi}\, \left[ \int d^3 r_1\,\int d^3 r_2\,F({\bf r}_1)F({\bf r}_2)\right.\\
&\times \left.e^{i{\bf k}\cdot({\bf r}_1-{\bf r}_2)}\mathcal{L}_\ell(\hat{\bf k}\cdot\hat{\bf \eta})-S({\bf k})\right] \ ,
\end{align}
where $\hat{\bf \eta} = \hat{\bf \eta}({\bf r}_1, {\bf r}_2)$ denotes the line of sight (LOS) of the pair formed by galaxies $1$  and $2$, $d\Omega_k$ the solid-angle element in $k$-space, $\mathcal{L}_\ell$ the $\ell-${th} order Legendre polynomial and $S$ the shot noise term. %given by, 
%%$S(k) = (1+\alpha)\int d{\bf r}\, \bar{ n} ({\bf r})w^2({\bf r})\mathcal{L}_\ell (\hat{\bf k}\cdot\hat{\bf r})$. 
%\begin{equation}
%S({\bf k}) = (1+\alpha)\int d^3 r \ \bar{n} ({\bf r})w^2({\bf r})\mathcal{L}_\ell (\hat{\bf k}\cdot\hat{\bf r}) \ .
%\end{equation}
%For multipoles of order $\ell>0$, $S\ll \hat{P_\ell}$, and consequently the shot noise correction is negligible.
For the normalisation factor we adopt the standard $I = \int d^3 r \ {\bar{n}}^2({\bf r})$, where $\bar{n}$ is the expected mean space density of galaxies, which guaranties that the no-window-function limit is recovered.

It is well known that the evaluation of Eq. (\ref{eq:P_estimator}) can be speeded up by several order of magnitudes in two steps: (i) defining the LOS of a pair as the LOS of one of the two galaxies \citep{yamamoto2006}; (ii) properly expanding the integrand in a form that is compatible with the use of FFTs (\citealt{bianchi2015b,scoccimarro2015}, se also \citealt{hand2017}).
The resulting computationally-efficient algorithm is not pair based and, as a consequence, the incorporation of PIP weights is not straightforward. 
On the other hand, IIP weights are fully compatible with this modern version of the estimator.
A straightforward way to see it consists of expressing $n$ as an explicit sum over Dirac deltas, $n = \sum_i \delta_D({\bf r} - {\bf r}_i)$, and then replacing it with its weighted counterpart, $n_{\rm IIP} = \sum_i w^{\rm IIP}_i \delta_D({\bf r} - {\bf r}_i)$.
%\begin{equation}
%F_{\rm IIP} = w_{\rm IIP} ({\bf r}) n({\bf r}) + \alpha n_s({\bf r})
%\end{equation}
%\begin{equation}
%P^{\rm IIP}_\ell = ...
%\end{equation}
%where $S{noise}_0 = V \left(\sum w_{\rm IIP}\right)^2 / \sum w_{\rm IIP}^2$.

\begin{figure}
 \begin{center}
    \vspace{-0.5cm}
    \includegraphics[width=7.3cm]{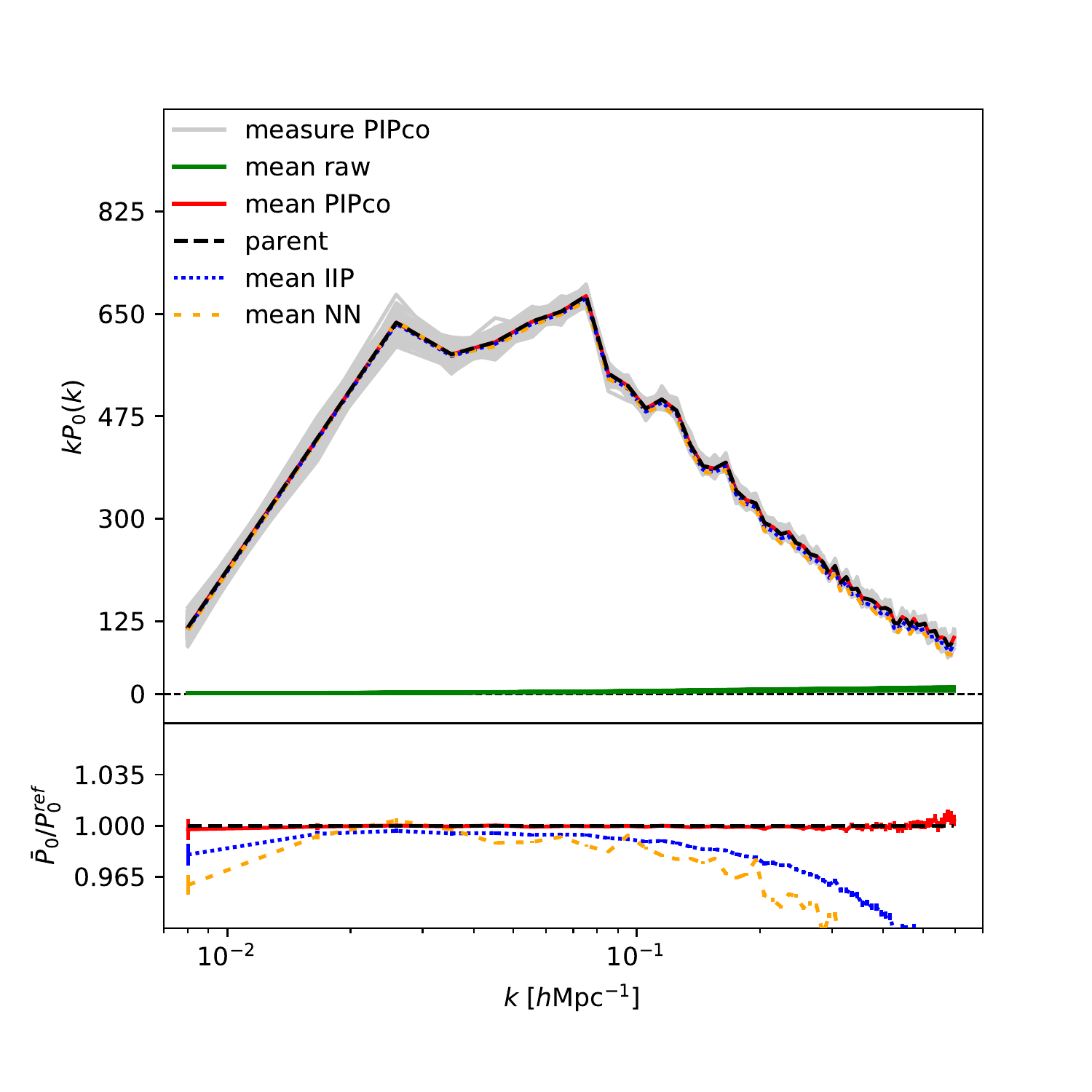} \\
   \vspace{-0.5cm}
   \includegraphics[width=7.3cm]{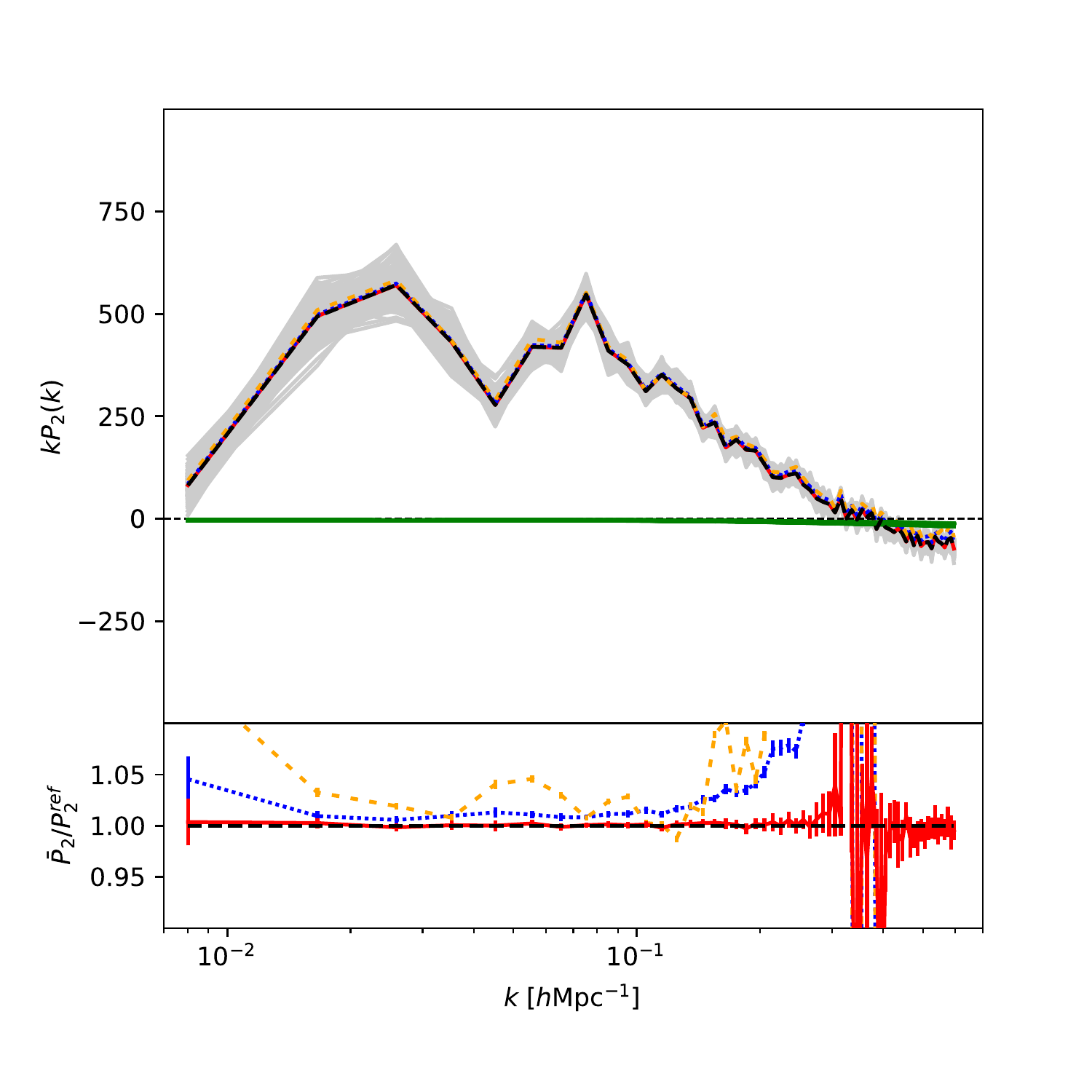} \\
   \vspace{-0.5cm}     
   \includegraphics[width=7.3cm]{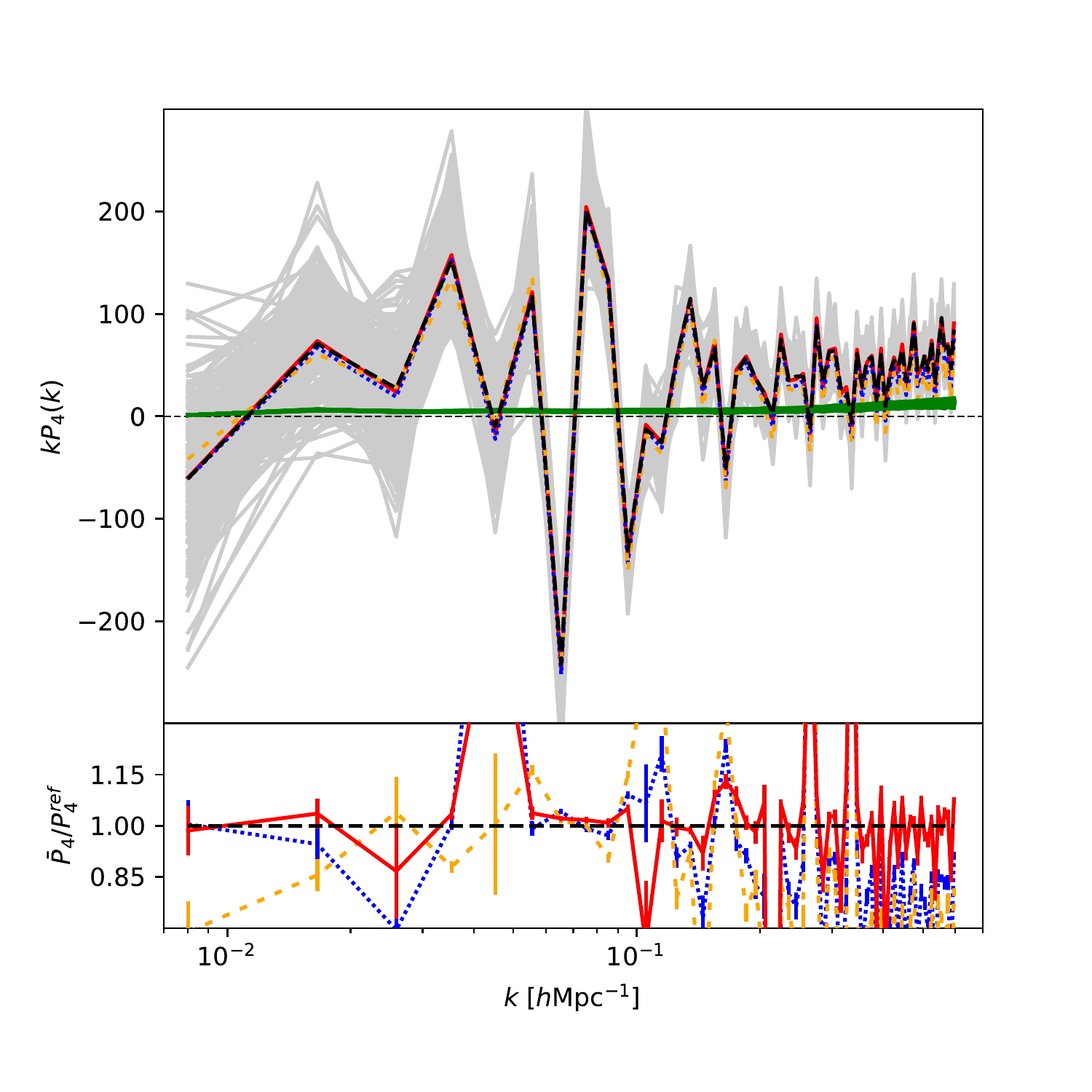} \\
   \vspace{-0.5cm}        
   \caption{Legendre multipoles of the redshift-space power spectrum $P$ (multiplied by the wave number $k$ for visualisation purposes) measured from the simulation and compared to their reference values obtained directly from the parent sample (black dashed). In the upper panel of each of the three plots we show the measurements obtained from $K=217$ realisations of the targeting via PIP weights (solid grey) together with their mean (solid red); the corresponding IIP-to-PIP corrections, as defined in the text (solid green); the mean of the measurements obtained via IIP weights (dotted blue); the mean of the measurements obtained via NN weights (short-dashed orange). In the lower panels we show the ratio between mean and reference, with error bars of the mean.}
   \label{fig:P_comp}
 \end{center}
\end{figure}

In Fig. \ref{fig:P_comp} we compare measurements of $P_\ell$ obtained via IIP (dotted blue) with those obtained via the nearest neighbour correction (NN,  short dashed orange). This latter technique is a traditional countermeasure to the missing observation problem, which consists of assigning the weight of a missing galaxy to its closest (angular) companion (e.g. \citealt{anderson2012}).  
Both curves correspond to the mean from $K=217$ realisations of the targeting.
The reference value is the full parent sample (dashed black).
For all measurements in this section we used FTTW\footnote{Fastest Fourier Transform in the West: \url{http://fftw.org}} with $512^3$ grid nodes and cloud-in-cell grid assignment.
No antialiasing procedure has been applied.
Note that, since we are dealing with a periodic box, the estimator expression, Eq. (\ref{eq:P_estimator}), gets significantly simplified and we actually do not need a random sample.
We will perform tests with more realistic survey geometries in future works.
The inverse probabilities are obtained via the zero-truncated estimator, which, as discussed in Sec.~\ref{sec:sampling}, appears to be the most meaningful and straightforward choice in most of the situations.
All the monopoles are shot-noise subtracted according to
%\begin{equation}
%P_0({\bf k}) = (1+\alpha)\int d^3 r \ \bar{n} ({\bf r})w^2({\bf r}) \ ,
%\end{equation} 
%\begin{equation}
%P_0({\bf k}) = (1+\alpha)\int d^3 r \ \bar{n}_w^2 ({\bf r}) \ ,
%\end{equation} 
%\begin{equation}
%P_0({\bf k}) = (1+\alpha)\int d^3 r \  \sigma_w^2({\bf r}) \ ,
%\end{equation} 
\begin{equation}
S_0({\bf k}) = \int d^3 r \  \sigma_w^2({\bf r}) + \alpha \int d^3 r \ \bar{n}({\bf r})\ ,
\end{equation} %whereas for higher order multipoles the correction is negligible. 
whereas $S_\ell =0$ for $\ell > 0$.
The quantity $\sigma_w^2$ is defined by $\langle w^2 \rangle = \sigma_w^2 \delta V$, where we adopted standard procedure of dividing the survey volume into a grid with cells of volume $\delta V$ containing no more than one particle each, see e.g. \citet{feldman1994}.
In practice we estimate $S_0$ just by summing over all the observed galaxies, $\int d^3 r \  \sigma_w^2({\bf r}) \rightsquigarrow \sum_i w_i^2$.

It is clear from the figure that, not surprisingly, IIP weights perform better than the standard NN but a systematic effect still persists, especially for large $k$.   
To address this undesired residual bias we need to modify Eq.~(\ref{eq:P_estimator}) to account for PIP corrections.
Here we present the final result of the calculations, a more detailed derivation is presented in App. \ref{app:PIPco_derivation}, but it should be clear from the equation itself that the correction consists of a weighted sum over $k$-space modes, as anticipated earlier.
The full PIP power spectrum, conveniently expressed in terms of its IIP counterpart (which includes shot-noise correction) and an additive pure PIP contribution, reads  
\begin{align}\label{eq:P_PIPco}
P^{\rm PIP}_\ell(k) &= P^{\rm IIP}_\ell(k) + \frac{(2\ell+1)}{I} \int \frac{d\Omega_k}{4\pi} \sum_{ij} A_{ij} e^{i{\bf k}\cdot({\bf r}_i - {\bf r}_j)} \mathcal{L}_\ell(\hat{\bf k}\cdot\hat{\bf \eta}_{ij}) \nonumber \\
&= P^{\rm IIP}_\ell(k)+ {(-i)}^\ell \frac{(2\ell+1)}{I} \sum_{ij} A_{ij} j_\ell(ks_{ij}) \mathcal{L}_\ell(\hat{\bf s}_{ij}\cdot\hat{\bf \eta}_{ij}) \ ,
\end{align}
where
\begin{equation}\label{eq:PIPco}
A_{ij} = w_{ij}^{{\rm PIP}} - w_i^{\rm IIP} w_j^{\rm IIP} \ ,
\end{equation}
%\begin{align}\label{eq:P_PIPco}
%\nonumber P_\ell({ k})=& \ \frac{(2\ell+1)}{I}\int \frac{d\Omega_k}{4\pi}\, \left[ \int d{\bf r}_1\,\int d{\bf r}_2\,F({\bf r}_1)F({\bf r}_2)\right.\\
% &\times \left.e^{i{\bf k}\cdot({\bf r}_1-{\bf r}_2)}\mathcal{L}_\ell(\hat{\bf k}\cdot\hat{\bf r}_h)-S({\bf k})\right. \\
%& + \left. \sum_{i \neq j} A_{ij} e^{i{\bf k}\cdot({\bf r}_i - {\bf r}_j)} \mathcal{L}_\ell(\hat{\bf k}\cdot\hat{\bf r}_{ij}^{(h)}) \right] \ ,
%\end{align}
%\begin{align}\label{eq:P_PIPco}
%P^{\rm PIP}_\ell(k) &= P^{\rm IIP}_\ell(k) + \frac{(2\ell+1)}{I}\int \frac{d\Omega_k}{4\pi} \left[ \sum_{i \neq j} A_{ij} e^{i{\bf k}\cdot({\bf r}_i - {\bf r}_j)} \mathcal{L}_\ell(\hat{\bf k}\cdot\hat{\bf \eta}_{ij}) \right] \nonumber \\
%&= P^{\rm IIP}_\ell(k)+ {(-i)}^\ell (2\ell+1) \sum_{i \neq j} A_{ij} j_\ell(ks_{ij}) \mathcal{L}_\ell(\hat{\bf s}_{ij}\cdot\hat{\bf \eta}_{ij}) \ ,
%\end{align}
%where
%\begin{equation}\label{eq:PIPco}
%A_{ij} \propto w_{ij}^{{\rm PIP}} - w_i^{\rm IIP} w_j^{\rm IIP} \ ,
%\end{equation}
%\begin{align}\label{eq:P_PIPco}
%P^{\rm PIP}_\ell(k) = P^{\rm IIP}_\ell(k)+ {(-i)}^\ell (2\ell+1) \sum_{i \neq j} A_{ij} j_\ell(ks) \mathcal{L}_\ell(\hat{\bf s}\cdot\hat{\bf r}_{ij}^{(h)}) \ ,
%\end{align}
and we defined ${\bf s}_{ij} = {\bf r}_i - {\bf r}_j$.
In the second raw of the equation we carried out the $k$-space angular integration ($j_\ell$ are spherical Bessel functions of the first kind, see e.g. \citealt{wilson2017}).
This is computationally convenient when looping over the separations ${\bf s}_{ij}$, since it allows us to update a one dimensional array rather than a three dimensional one.  
Also note that, due to trivial symmetries, it is in practice convenient to restrict the sum to $i>j$ and correct the normalisation accordingly. 
Even with this latter tricks, a full evaluation of the PIP correction can be computationally prohibitive for the typical number of galaxies, $\mathcal{O}(10^7)$, observed by modern spectroscopic surveys.
Luckily, if $A_{ij} = 0$ above some angular separation, $s_c \ll L$, where $L$ is the scale of the survey, the whole process can be speeded up by several order of magnitude through the use of linked lists (our choice) or similar computational shortcuts, which allow us to efficiently exclude from the computational loops all the pairs with separation larger than $s_c$ (or any other arbitrary separation).   
$s_c$ is an input parameter for these algorithms, which can, and should, be predetermined from the sample itself. 
%Due to the form of the PIP correction, Eq. (\ref{eq:PIPco}), the quantity of interest becomes $DD_{\rm PIP} - DD_{\rm IIP}$.
%In Fig. \ref{fig:sampling_correction_P} we show this quantity measured with (solid blue) and without (dashed red) the sampling correction discussed in Sec. \ref{sec:sampling_ind_vs_pair}.
%\begin{figure}
% \begin{center}
%   \includegraphics[width=7.5cm]{{/Users/davide/qua/corfu_pawe/plt/MDR1_62_0.00005_shift_spasst_ms1_w_bw217_IIPauwpc_PIPvsIIP_reg_ang_comp_diff}.pdf}
%   \label{fig:sampling_correction_P}
% \end{center}
%\end{figure}
%As expected we see that without a proper treatment of the sampling effects, Eq. \ref{eq:dist_ind_vs_pair_zt}, we would be misled by the presence of an apparent large-scale selection correlation.
%On the other hand, on scale scales smaller than  $s_c \approx 2 h^{-1}$Mpc the selection correlation is genuine and the PIP correction must be taken into account.   
Thanks to the statistical tools developed in Sec. \ref{sec:sampling}, we can safely say that the scale below which the selection correlation is genuine and the PIP correction must be taken into account corresponds to $s_c \approx 2 h^{-1}$Mpc, see the lower panel of Fig. \ref{fig:corr_eff}. 

In Fig. \ref{fig:P_comp} we show the PIP-corrected measurements of $P_\ell$ obtained from the 217 realisations of the targeting (solid grey) together with their mean (solid red) and the amplitude of each individual correction (solid green).
The PIP approach yields unbiased measurements on all scales and for all the multipoles.    
For the figure we adopted a conservative $s_c = 5 h^{-1}$Mpc, but we have checked that smaller values of the selection-correlation length yield identical results.
Only by going well inside the selection-correlated region, $s_c \lesssim 1 h^{-1}$Mpc, we eventually started seeing some appreciable deviation from the true value.
Even with this conservative choice of $s_c$ the evaluation of the PIP correction is extremely fast\footnote{The computational time scales with the number of pairs with angular separation smaller than $s_c$. Larger and denser samples require more time, but, realistically, the most crucial ingredient will always be $s_c$. Each survey and the correspondent targeting strategy deserve a specific treatment, but it seems clear that an increase of the correlation length of, say, a factor 10 would have a very relevant impact on the total cpu time.}, the amount of cpu time is comparable or even smaller than that required for the raw power spectrum calculation via FFTs.
It is worth noting that all the largest systematic-error fluctuations, lower part of each panel of Fig. \ref{fig:P_comp}, correspond to zero-crossing regions of the reference value, which is at the denominator, and, therefore, they are just plotting artefacts.
Regarding the largest wave numbers, here we are limited by aliasing but the PIP correction is not and there is no obvious reason why it should not work up to arbitrarily large values of $k$.

\begin{figure}
 \begin{center}
   \includegraphics[width=7.5cm]{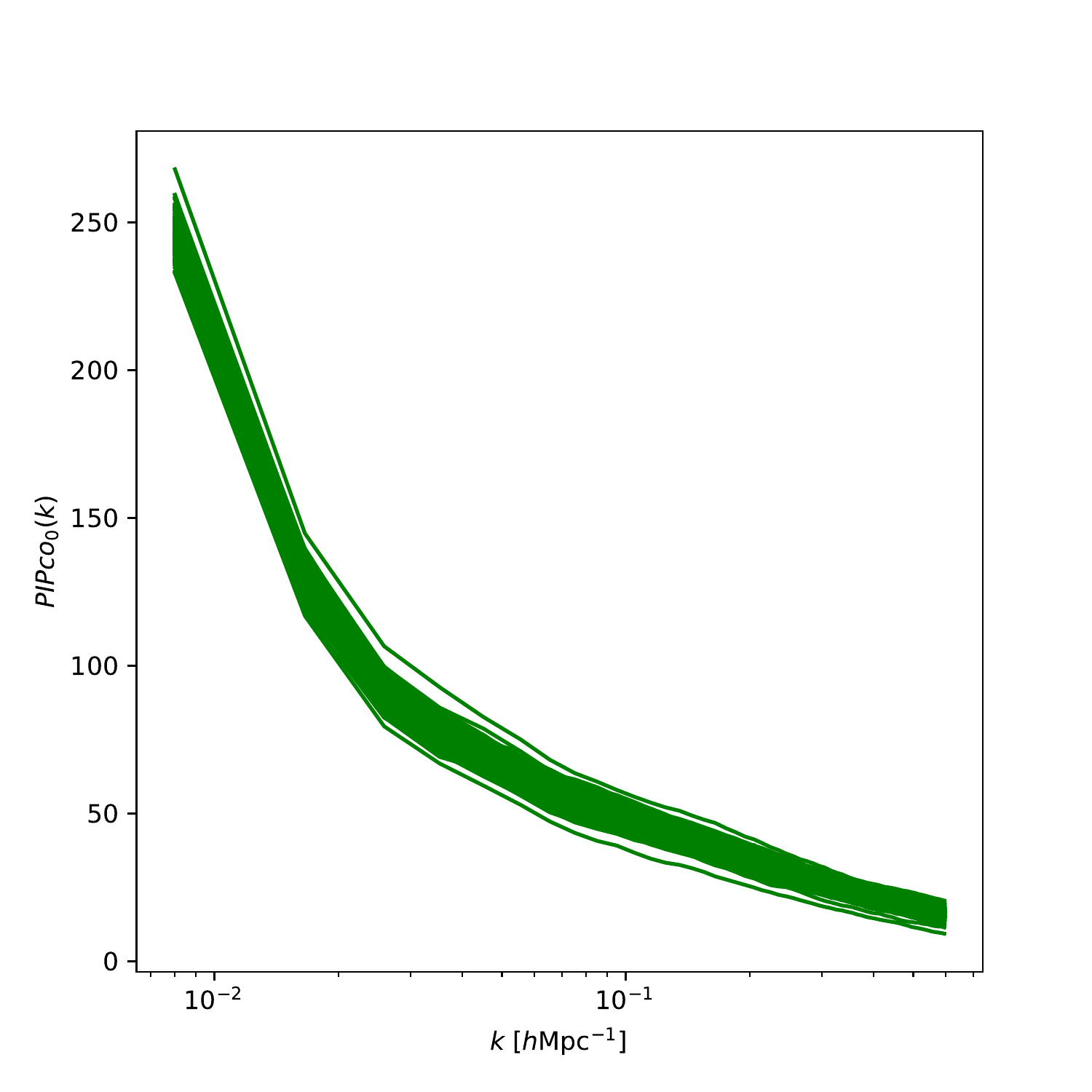} \\
   \vspace{-0.3cm}
   \includegraphics[width=7.5cm]{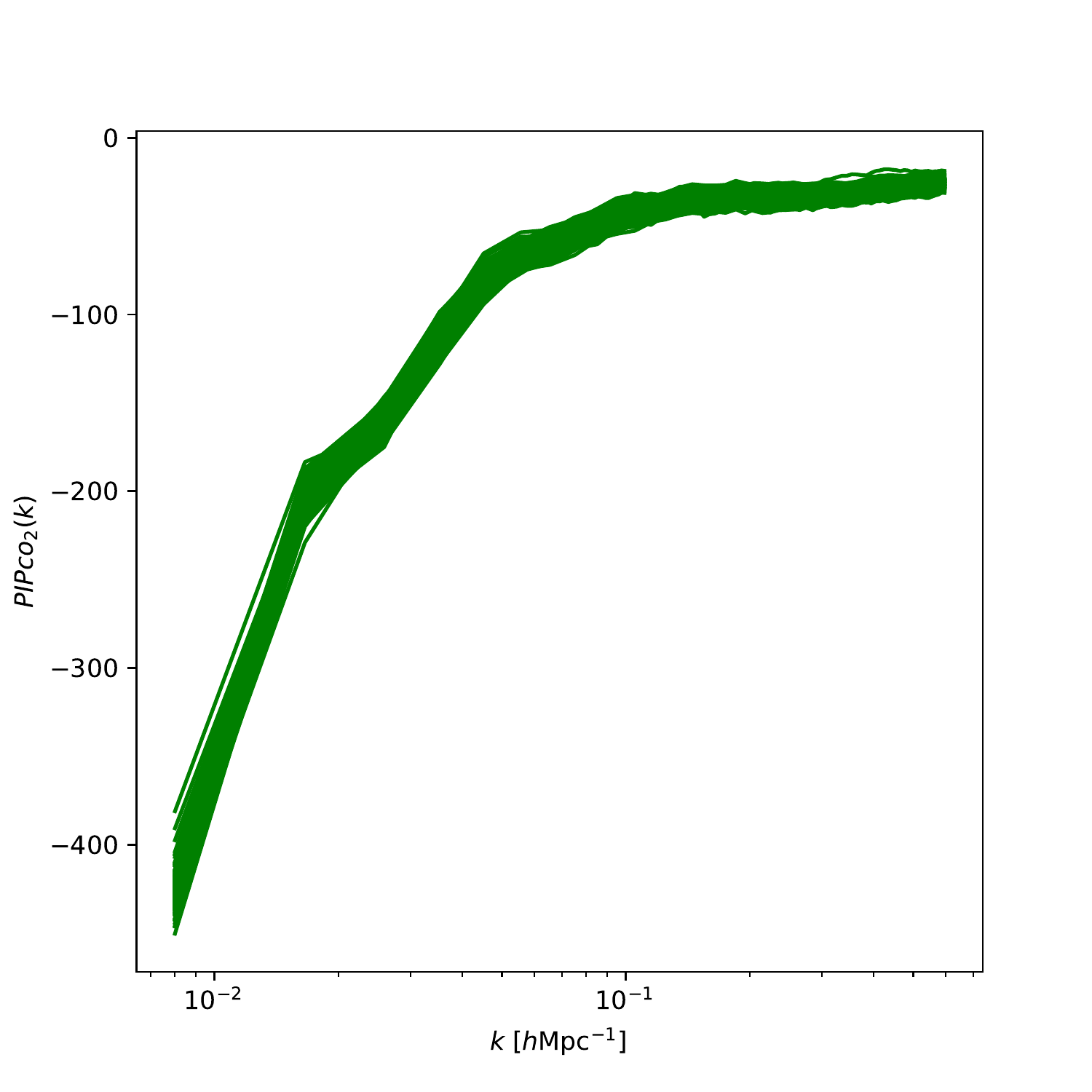} \\
   \vspace{-0.3cm}     
   \includegraphics[width=7.5cm]{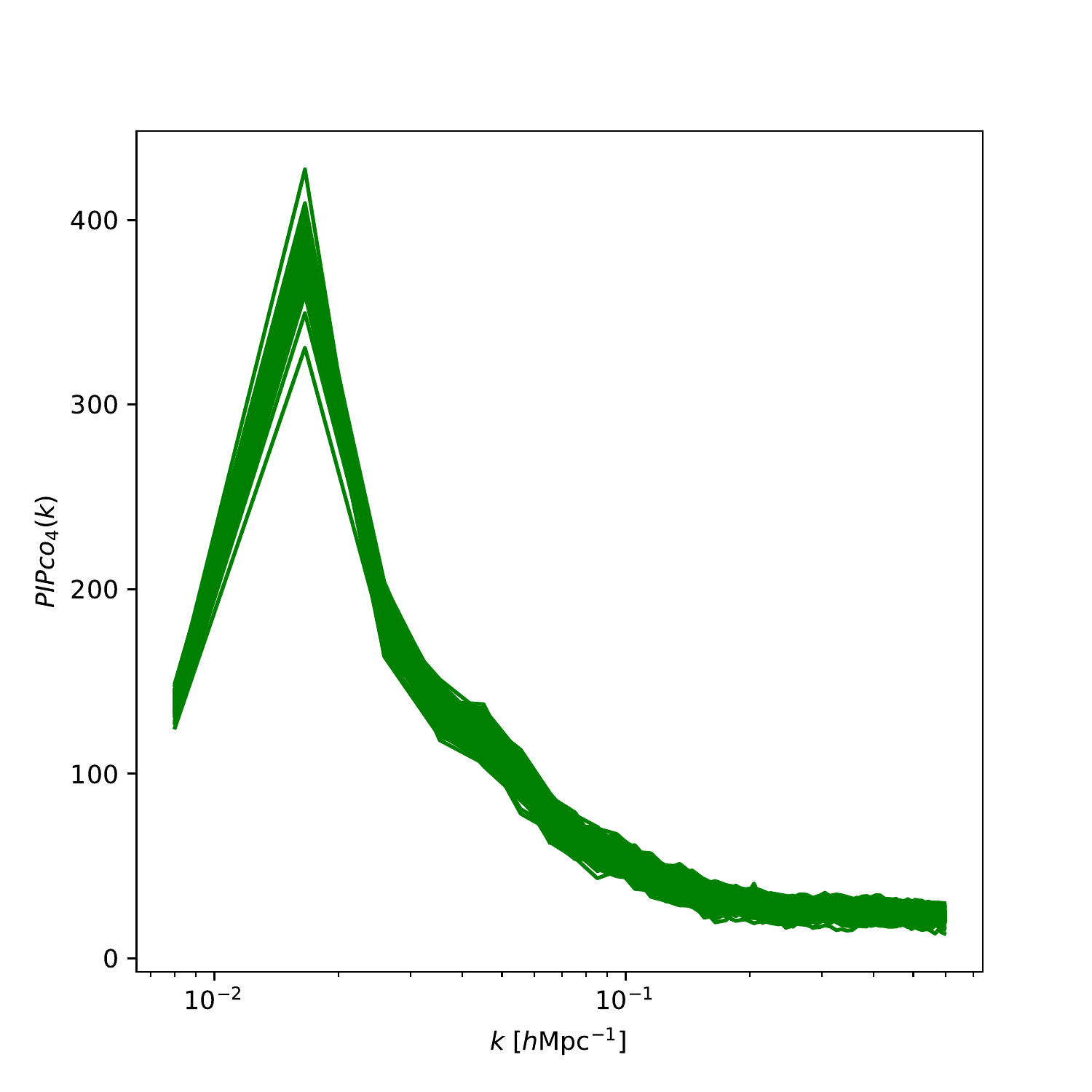} \\  
   \caption{True amplitude of the IIP-to-PIP corrections of Fig. \ref{fig:P_comp}, i.e. without the $k$ factor on the y axis, in order to show their actual $k$ dependence.}
   \label{fig:PIPco}
 \end{center}
\end{figure}

In Fig. \ref{fig:PIPco} we can have a closer look to the raw PIP contributions measured from the 217 samples.
Thanks to the fact that the quantities on the y axis are not multiplied by $k$, at variance with Fig. \ref{fig:P_comp}, we can appreciate their actual scale dependence.  
Quite clearly, even for the monopole there is no easy way to incorporate the missing-observation effect in a constant shot-noise term to be marginalised over.

\section{Generalising the probability weights}\label{sec:generalising} 

As anticipated in Sec. \ref{sec:IP}, there is ample room for generalisations of the whole probability-weights prescription.
The key observation is that $b/p$, where $b$ is a binary variable, can be actually interpreted as $b/\langle b \rangle$.
It then becomes obvious that any weight in the form $\phi / \langle \phi \rangle$, where $\phi \in \mathbb{R}$ and $\langle \phi \rangle \ne 0$, share with the inverse probability the property of yielding unbiased estimates.
In other words, probability weights are just a special case, more precisely the binary limit, of a more general statistical recipe.

\subsection{Generalised nearest neighbour}

For finiteness, we can imagine of using NN weights rather than $b$, which means that the variable of interest is now $\phi_{\rm NN} \in \mathbb{N}$.
More precisely, for any given galaxy in the parent sample we have
\begin{equation}
\phi^{\rm NN}_i =
\begin{cases}
0 & \text{galaxy discarded in the $\eta$-th realisation} \\
a_\eta & \text{galaxy selected in the $\eta$-th realisation} \\
\end{cases} \ ,
\end{equation}
where $a_\eta \in \{1,\dots,N_{gal}\}$ is the number of nearest (unobserved) neighbours of that galaxy in the $\eta$-th realisation. 
If the total number of realisation of the targeting is $K$, the weight of a pair becomes
\begin{equation}
%w_{ij} = K \frac{\phi_{i \eta} \phi_{j \eta}}{\phi_i \cdot \phi_j}
w_{ij \eta} = \frac{\phi_{i \eta} \phi_{j \eta}}{\sum^K_\beta \phi_{i \beta} \phi_{j\beta} / K} \ ,
\end{equation}
where we omitted the superscript NN. 
%where we wrote $\sum^K_n \phi^{(n)}_i \phi^{(n)}_j$ as a scalar product.
Here we are using $K / \sum^K_\beta \phi_{i \beta} \phi_{j\beta}$ as an estimator for $\langle \phi_i \phi_j \rangle^{-1}$, ignoring, for simplicity, any possible sampling effect similar to those discussed in Sec.~\ref{sec:sampling}.
These generalised NN weights have very similar properties to the PIP ones, in the sense that they share the same pairwise nature, which, in both cases, yield unbiased estimates of the clustering\footnote{The generalised NN weights are compatible with fast bitwise-operator-based pair counting algorithms, just slightly more complex than those routinely used for the PIP weights, see \citet{bianchi2017}. In essence, each NN weight can be seen a collection of a few standard bitwise weights. This is actually the strategy adopted for the measurements presented in this work.}.
Analogously to the PIP case, the unbiasedness follows by construction from the fact that each pair is counted once on average.

The main difference is that the new weights can vary from one realisation to one another, i.e. they are no longer in the form $w=bf$, with $f={\rm const}$, but rather $f$ depends on $\eta$.
As a consequence, despite the two estimators having the same expectation value, variance and all higher order moments may differ. 
It is easy to see that, in the idealised case of a single object, the minimum variance is obtained via inverse probability.
Nonetheless in any real-life scenario we have to deal with summary statistics, e.g. correlation functions, measured from multiple objects, e.g. pairs of galaxies, in which case the optimality of the inverse probability may no longer hold. 
This is the topic of the following discussion.

\subsection{Variance of the weights}\label{sec:variance}

In the idealised case of a single object\footnote{Note that here by single versus multiple objects we do not mean a galaxy versus a pair or a triplet but rather a galaxy versus many galaxies or a pair of galaxies versus many pairs of galaxies, etc. In other words, by single versus multiple objects we do not mean IIP versus PIP.} the variance reads
\begin{equation}
\sigma^2 = \left\langle \frac{\phi^2}{\langle \phi \rangle^2} \right\rangle - \left\langle \frac{\phi}{\langle \phi \rangle} \right\rangle^2 =  \frac{\langle \phi^2 \rangle}{\langle \phi \rangle^2} - 1 \ ,
\end{equation}
which in the case of pure inverse probability, $\phi = b$, reduces to
\begin{equation}
\sigma^2_b = \frac{1}{p} - 1 \ .
\end{equation}
When dealing with missing observation it is desirable to have zero weight for an object that is not observed (Sec. \ref{sec:IP}), therefore the variables $b$ and $\phi$ are not independent, precisely $\phi = 0$ if $b=0$ (not necessarily viceversa).
This, together with the fact that both variables are renormalised in order to have expectation value equal to one, creates a lower bound in the variance, which is obtained when the variable is constant on its support, i.e. in the inverse probability case.
In other words, the generalised weight of an object $\phi / \langle \phi \rangle$ can be interpreted as redistributing the probability weight $b/p$ of that object among the realisations in which the object has been observed, while keeping the sum over all the realisations fixed. 
If there is only one object, the process can only add variance to the estimator.

Things change when we consider multiple objects.
If $N$ is the total number of objects, the variance becomes
\begin{align}\label{eq:variance_multi}
\sigma^2 =&  \left\langle \left(\sum_{i=1}^N \frac{\phi_i}{\langle \phi_i \rangle} \right)^2 \right\rangle - \left\langle \sum_{i=1}^N \frac{\phi_i}{\langle \phi_i \rangle} \right\rangle^2 \nonumber \\
= &  \sum_{i=1}^N \frac{\langle \phi_i^2 \rangle}{\langle \phi_i \rangle^2}  + 2 \sum_{i > j}^N \frac{\langle \phi_i \phi_j \rangle}{\langle \phi_i \rangle  \langle \phi_j \rangle} - N^2 \ ,
\end{align}
which, for $\phi = b$, reduces to
\begin{equation}\label{eq:variance_multi_b}
\sigma^2_b = \sum_{i=1}^N \frac{1}{p_i} + 2 \sum_{i > j}^N \frac{p_{ij}}{p_i p_j} - N^2 \ .
\end{equation}
Not surprisingly, now the variance also depends on irreducible cross terms, $\langle \phi_i \phi_j \rangle$, which reduce to $p_{ij}$ for the inverse-probability weights.
As regards the minimum variance, similarly to the single-object case, the first term on the righthand side of Eq.~(\ref{eq:variance_multi}) is minimised by setting $\phi = b$.
This is not necessarily true for the second term.
Indeed a clever choice of $\phi$ can reduce the second term enough to make the entire variance smaller.
We show in App.~\ref{app:toy} a purely pedagogical toy example that proofs the concept, but it should be clear from Eq. (\ref{eq:variance_multi}) that the key ingredient is the anticorrelation of the different $\phi_i$ associated to different objects.
For finiteness, we can particularise the above considerations to the familiar correlation-function case.
In this scenario $N$ is the number of pairs of the parent sample in a given separation bin, $N=DD^{(p)}({\bf s})$.
Eq. (\ref{eq:variance_multi_b}) represents the variance of the PIP weights (ignoring any sampling effect, Sec.~\ref{sec:sampling}), which depends on the 4-point function via the $p_{ij}$ term in the second sum on the righthand side of the equation.

While discussing the variance of the weights it is useful to make a few remarks, in order to avoid confusion.
In our modelling we require each single object to be unbiased on its own, i.e. each single object on average is counted once.
We will refer to this requirement as {\it strong unbiasedness}.
Strictly speaking, strong unbiasedness is not necessary when the quantity of interest is the result of the contribution of many different objects.
Once again, it is convenient to formulate the problem in the familiar correlation-function language.
Clearly, in this scenario what we really need is that the total pair counts are unbiased, i.e. $\langle DD({\bf s}) \rangle = DD^{(p)}({\bf s})$.
It is indeed possible to imagine pair weights capable of recovering $DD^{(p)}$ (which, just for clarity, in the general notation of Eqs. (\ref{eq:variance_multi}) and (\ref{eq:variance_multi_b}) corresponds to the quantity denoted with $N$) while not obeying to the strong-unbiasedness requirement.
Such weights could in principle allow us to reduce the overall variance, thank to the simple fact that they add further degrees of freedom to play with for the modelling. 
The fundamental problem with this approach is that, in practice, in order to derive this kind of weights, an a priori knowledge of the result is required.
For example, if we knew $DD^{(p)}$ in advance, we could weight each pair with separation ${\bf s}$ by $DD^{(p)}({\bf s})/DD({\bf s})$ and recover unbiased correlation function with zero variance, but this is, of course, a circular argument: if we knew $DD^{(p)}({\bf s})$, we would not have a missing-observation problem at all.
Relying on strong unbiasedness appears to be the only robust way to deal with missing observations.
In essence, by enforcing it, we take advantage of the fact that each pair is an irreducible building block, which has to contribute a known-a-priori term $N=1$ to the total budget, irregardless of the separation.

%Clearly, in this scenario it suffices that the total pair count is unbiased, i.e., following the notations adopted for Eq. (\ref{eq:variance_multi}), $\langle DD \rangle = N$.
%Dropping the individual unbiasedness in favour of a global one could potentially help in reducing the variance, since the latter is obviously weaker than the former.
%In other words, one might be tempted to use something along the line of $N \times DD / \langle DD \rangle$ as an estimator. 
%The problem is that in real life $N$ is unknown because, due to missing observation, we do not know how many pairs fall in a specific separation bin.
%Actually, the determination of $N$ as function of the separation is exactly the problem that we need to solve and if we knew its value in advance the issue of missing observation simply would not exist.
%In real life we do not perform sums from 1 to $N$ such as those in Eq. (\ref{eq:variance_multi}), but rather sums over the observed pairs alone.
%The goal is to weight these pairs in such a way that the result of their sum is equal (on average) to $N$, which is actually the quantity we want to measure.
%There is no obvious way to achieve this other than in the limit of infinitely-small bins, for which $N$ is known a priori, i.e. $N=1$, which is just another way to say that we need to have unbiased individual pairs because that 

\subsection{Generalised angular upweighting}

Among the most popular weighting schemes, angular upweighting (AUW, see e.g. \citealt{hawkins2003}) is certainly one of the furthest from the concept of strong unbiasedness.
%Angular upweighting (AUW, see e.g. \citealt{hawkins2003}) represents an interesting example of a weighting scheme that attempts to recover unbiased estimates of the correlation function trough direct manipulation of the total pair counts, therefore without imposing strong unbiasedness.
The technique consists of correcting the 3-dimensional pair counts by weighting each pair according to its angular separation $\theta$.
Specifically, the weights have the form  $DD^{(p)}_a(\theta) / DD_a(\theta)$, where numerator and denominator are the angular pair counts of parent and observed sample, respectively.   
The so obtained 3-dimensional pair counts are designed to match exactly the angular pair counts of the parent sample, but, clearly, the individual contribution of each pair is not forced to be equal to~1.
When used as a missing-observation countermeasure, it has been shown that AUW on its own is not unbiased (e.g. \citealt{guo2012}), at least not by construction, in the sense that its effectiveness in recovering the true 3-dimensional clustering will always depend on the sample under examination.
%The above considerations about individual versus global corrections can be seen as one of the reasons behind this unsatisfactory behaviour. REPHRASE IT
%In order to avoid confusion, it is appropriate to say that it has also been show \citep{percival2017} that, when combined with an already unbiased clustering estimator, %AUW has the beneficial effect of reducing the variance while keeping the measurement unbiased.
%AUW becomes a very effective tool for variance reduction.
On the other hand, when combined with an already unbiased clustering estimator, AUW becomes a very effective tool for variance reduction, as shown by \citet{percival2017}. 
Indeed using PIP plus AUW is the strategy successfully adopted by \citealt{bianchi2017, bianchi2018, mohammad2018, smith2019}.
Essentially, it allows us to optimise the usage of the information available by bringing back into 3-dimensional measurements the full, missing-observation free, angular information.
Consequently, we expect the so obtained variance to represent (or, at least, be very close to) a lower limit.  

%It is interesting to note that AUW can be modified in order to incorporate strong unbiasedness, without PIP weights.
It is interesting to note that strong unbiasedness can be enforced in the AUW-plus-PIP weights. 
Following once again the simple concept $w=\phi/\langle\phi\rangle$, in the same fashion as the generalised NN weights, for the generalised AUW the weight of a pair becomes
\begin{equation}\label{eq:genAUW}
w = b \frac{DD^{(p)}_a}{DD^{\rm PIP}_a} \left\langle b \frac{DD^{(p)}_a}{DD^{\rm PIP}_a} \right\rangle^{-1} \ ,
\end{equation}
which is, by construction, a fully debiased version of the standard AUW ($b$ is the usual binary variable encoding whether the pair has been selected or not and $DD^{(p)}_a / DD^{\rm PIP}_a$ is a function of $\theta$). It is also possible to not use PIP weights at all, $DD^{\rm PIP}_a \rightsquigarrow DD_a$, but this approach requires iterations, such as those described in App.~\ref{app:xipe}, to reach a variance comparable to that obtained via the weights defined in Eq.~(\ref{eq:genAUW}).

\subsection{Comparison}

\begin{figure}
 \begin{center}
   \vspace{-1.0cm}      
   \includegraphics[width=7.4cm]{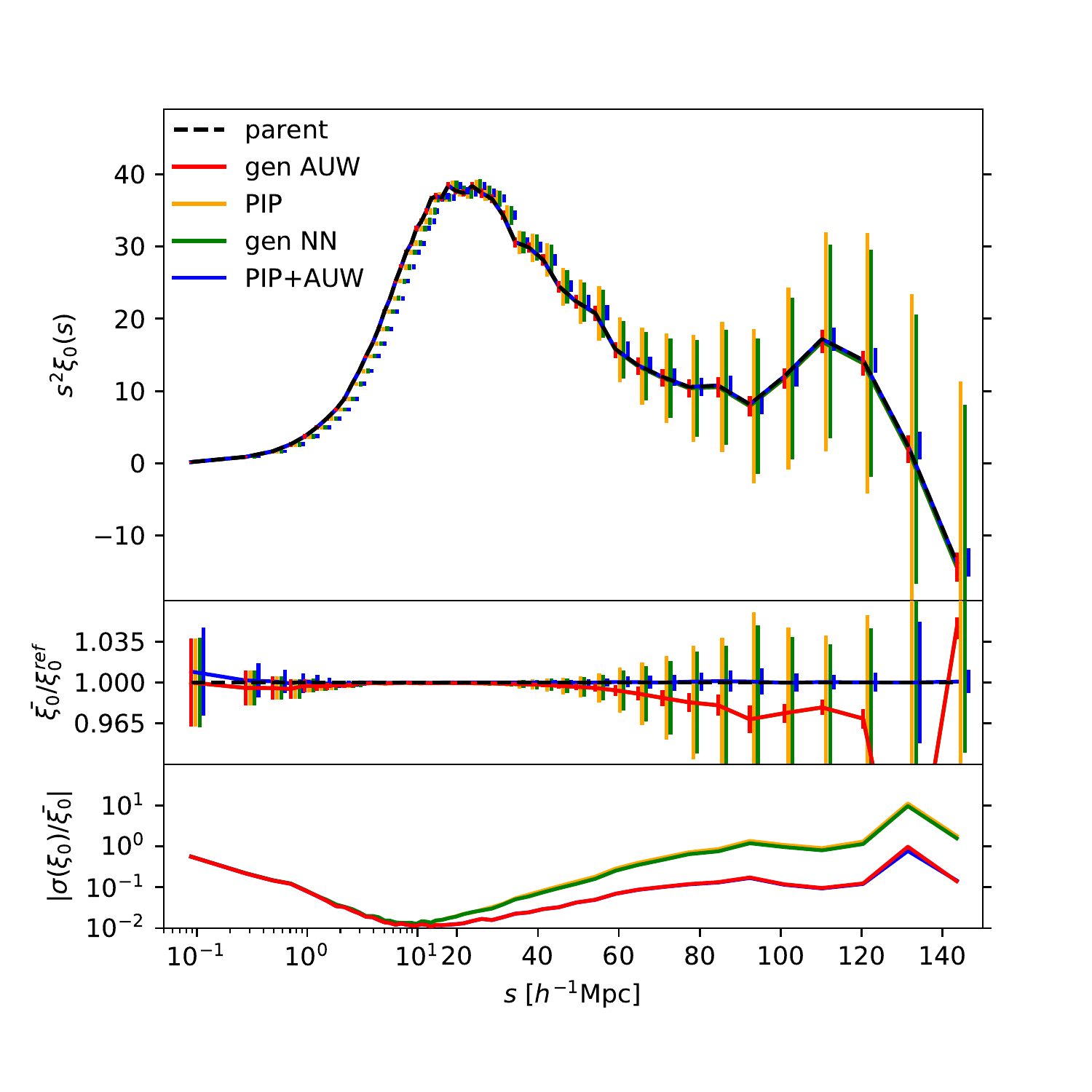} \\
   \vspace{-0.6cm}
    \includegraphics[width=7.4cm]{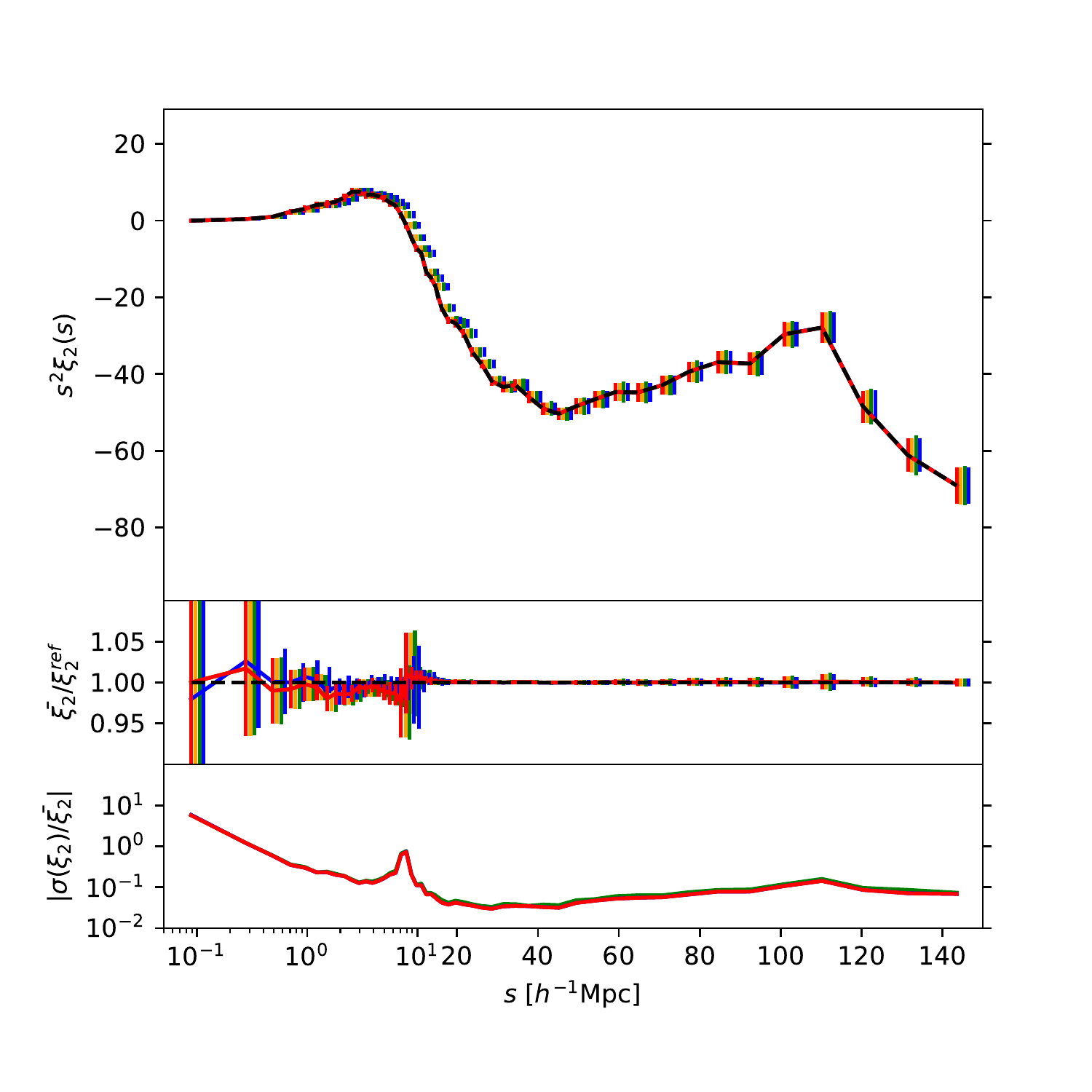} \\
    \vspace{-0.6cm}     
     \includegraphics[width=7.4cm]{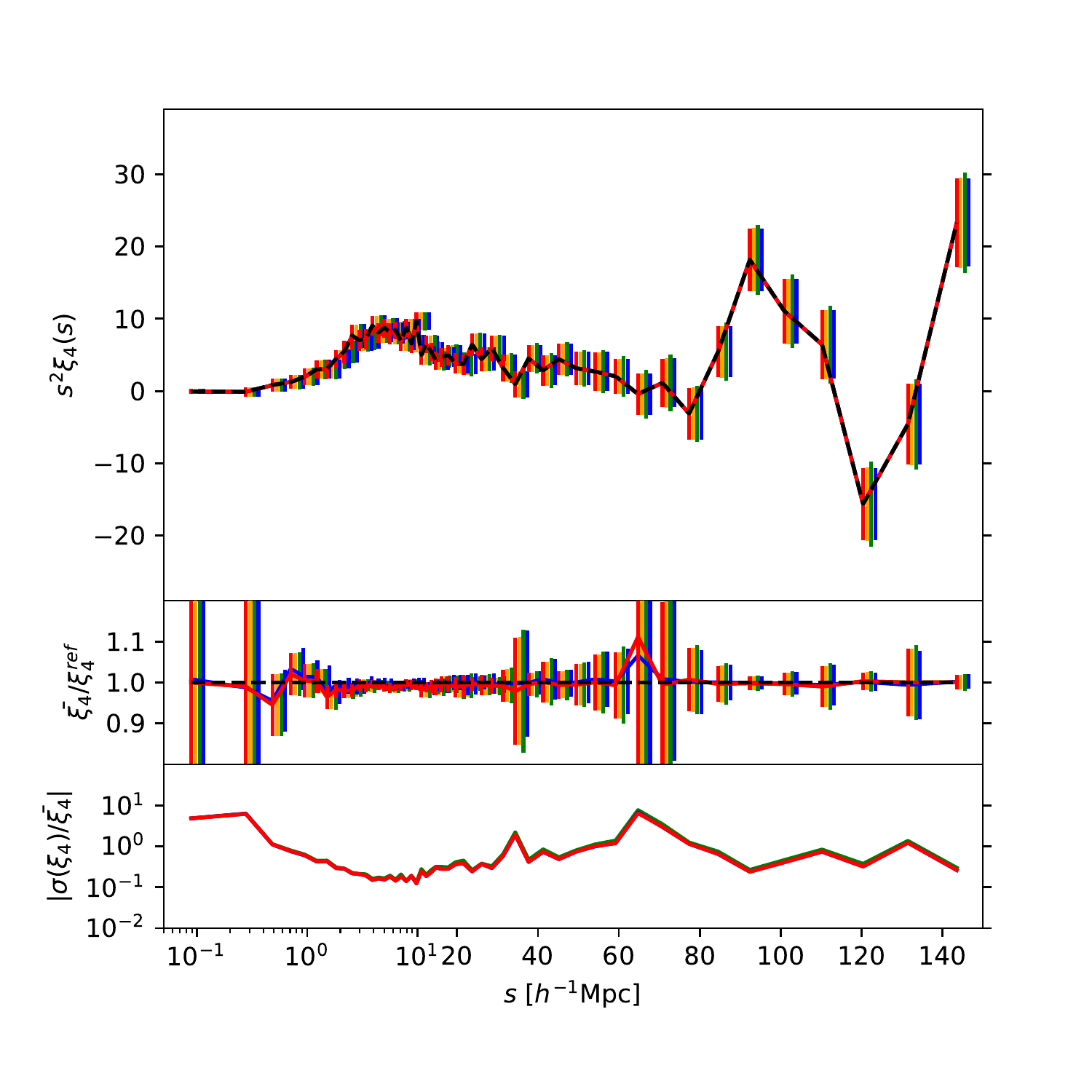} \\
     \vspace{-0.6cm}          
    \caption{Comparison of the Legendre multipoles of the correlation function $\xi$ measured from the simulation via different weights: generalised AUW (solid red); pure PIP (solid orange); generalised NN (solid green); PIP plus AUW (solid blue). The reference values are obtained directly from the parent sample (black dashed). All the weighted measurements are the mean from $K=217$ realisation of the targeting. For illustrative purposes the normalisation of all the pair counts is kept fixed, $n_{DD} = N_{gal} (N_{gal}-1) / 2$. In the top panel of each of the three plot we report the amplitude of the multipoles (multiplied by the separation squared $s^2$ for visualisation purposes) with error bars corresponding to the standard deviation. Central panels: ratio between the weighted multipoles and the reference with error bars of the mean. Lower panels: standard deviation of the multipoles divided by their mean.}
   \label{fig:gen_w_comp}
 \end{center}
\end{figure}

In Fig. \ref{fig:gen_w_comp} we show a comparison between this generalised AUW and the other different weighting schemes just discussed.
More examples and further discussions about advantages/drawbacks and implementation of the different approaches are presented in Apps. \ref{app:norm}, \ref{app:generalised_comparison} and \ref{app:xipe}. 

%The conclusion of our tests is that the iterative weighting scheme yield results identical to those obtained via PIP plus AUW.
In the figure we report the correlation function measured via generalised AUW (red solid), PIP weights alone (solid orange), generalised NN (solid green), PIP plus AUW (solid blue).
It is important to stress that the purpose of the figure is to illustrate the different properties of the different estimators, rather than determining which one performs better.
In order to emphasise such differences, we adopt a fixed normalisation for all pair counts, corresponding to the standard normalisation of the parent sample, $n_{DD} = N_{gal}(N_{gal}-1)/2$.
As discussed in the following, this effectively disfavours some of the measurements as, e.g. pure PIP weights.
Indeed, as shown in App. \ref{app:norm}, when all the estimators are properly normalised, they yield almost identical results.
On the other hand, we show in App.~\ref{app:generalised_comparison} that it is possible to imagine more extreme targeting algorithms for which, even when properly normalised, different estimators give different results.

First we note that, with the only exception of PIP plus AUW, all different approaches yield the same identical expectation value.
In other words, red, orange and green curves in the central panels are perfectly superimposed and indistinguishable (the error bars are displaced for clarity).
This is expected as a direct consequence of having imposed individual unbiasedness.
On the largest scales these methods show a small systematic lack of power in the monopole.
This is also expected since there are pairs that have not been targeted in any of the $K=217$ realisations.
The effect disappears when the proper normalisation factor is adopted (see App. \ref{app:norm}), but as anticipated above, here we prefer not to do that, in order to help the reader visualise the conceptual differences between the different approaches\footnote{Obviously, since by construction there are no zero-probability pairs in our sample, the systematic effect can also be removed by just increasing the number of realisations, irregardless of the normalisation.}. 
By construction, the optimal normalisation for PIP plus AUW (solid blue) coincides with that of the parent sample and, as a consequence, we do not observe any systematic effect. 

As regards the variance, PIP plus AUW and generalised AUW yield the smallest variance, at least for the monopole (for the quadrupole and hexadecapole we do not see any appreciable behaviour difference).
Generalised NN seems to have a slightly smaller variance compared pure PIP weights, but keep in mind that with a proper normalisation all the estimators, including those based on AUW, yield similar variance, at least for the relatively regular targeting algorithm considered here.
We show in App. \ref{app:generalised_comparison} that the gap between pure PIP and generalised NN can be much larger under more extreme (but less realistic) targeting criteria.
On the other hand, AUW, either coupled to PIP or generalised, seems to define an effective lower bound for the variance, under any circumstance.
In the following we briefly discuss the usefulness of having different possible approaches to choose from. 
    
Compared to PIP plus AUW, the main advantage of the generalised AUW is that it yields unbiased estimates even if the angular correlation function is wrong, in which case the only consequence is that the variance is not fully optimised.
Indeed, a quick inspection of Eq. (\ref{eq:genAUW}) reveals that the amplitude of the angular correlation function gets canceled out in the weighting process, the only relevant information is contained in the fluctuations from one realisation to one another. 
This can be useful,  e.g., when dealing with volume limited subsamples.
In such circumstances it is impossible to establish if a missing galaxy belongs to the sample or not because to do so we would need to know its redshift.
As a consequence, the true angular correlation function of the subsample is unknown, we only know the angular correlation of the full parent sample and, in principle, the two may differ.  
With the generalised approach it is possible to improve the variance of the measurements without the risk of biasing the results on account of improper angular upweighting.

On the other hand, when the angular correlation of the sample is known, it is probably preferable to use PIP plus AUW, which is simpler and computationally faster\footnote{At variance with PIP and generalised NN weights, for the generalised AUW only a fraction the bitwise-operator machinery can be used, as we no longer have to deal with binary variables or small integers but rather with real numbers. This makes the pair counts slower.}. 
It is also worth reminding that, as first discussed in \citet{bianchi2018}, beside the beneficial effect on the variance, coupling traditional AUW to PIP weights also corrects for pairs with zero or too-low-to-be-sampled probability, i.e. the unobservable ones.
This is not rigorous, since it means assuming that the behaviour of the unobservable pairs can be inferred from that of the observable ones, which is not necessarily true.  
Despite this formal issue, the correction was shown to be very effective \citep{bianchi2018, mohammad2018, smith2019}.
It is important to realise that the generalised weights do not have this property.
By construction the expectation value is exactly the same of PIP weights alone, which, depending on the situation, may or may not be a desirable property.  

Beside their pedagogical purpose, it is not clear if there is any realistic situation in which is convenient to use the generalised NN weights.
One possible scenario is that of reconstruction, see Sec. \ref{sec:IP}.
As briefly discussed in that section, the general rule is that anytime we try to infer $n$-point functions directly from the galaxy density field, i.e. weighting galaxies rather than $n$-plets of galaxies, we are exposed to the risk of interpreting correlation signals purely coming from the targeting as real cosmological information.
There are scenarios, like reconstruction, for which accepting this approximation seems reasonable, in which case, in order to minimise the systematic error on the final result, it is desirable to have a small-variance weighted density field.
This may be achieved by weighting each galaxy with generalised NN, or similar approaches.

%\section{Reconstruction}

%THE PLAN IS TO SKIP THIS UNLESS WE WANT TO SPLIT THE PAPER IN TWO, IN WHICH CASE IT MIGHT BE USEFUL TO HAVE MORE STUFF 

\section{Conclusions}\label{sec:conclusions}

Inverse probability weights are an advanced, unbiased by construction, countermeasure to the problem of missing observations.
%In this work we have expanded on their original formulation, with the purpose of exploiting their full potential and answering important questions that arose after they were first introduced by \citep{bianchi2017}.
%With this work we have answered important questions that arose after they were first introduced by \citep{bianchi2017}, and expanded on their original formulation, with the purpose of exploiting their full potential
With this work we expanded on their original formulation \citep{bianchi2017}, with the purpose of achieving a deeper understanding of their statistical properties, which we then used to discuss new possible applications of the weights, including the derivation of an estimator for the anisotropic power spectrum.  
We summarise our main results as follows. 

\begin{itemize}
\item
We developed a more mature formalism that allows us to discuss about probability weights with the traditional statistical language, i.e. in terms of well defined random variables, expectation values, probability distributions, etc.
\item
Within this formalism, we defined three possible estimators for the inverse probability, namely the inverse-count, the efficient and the zero-truncated estimator.
All the three estimators are asymptotically unbiased, i.e. they converge to the true $1/p$ value when the number of realisations of the targeting $K$ tends to infinity.   
Hence we investigated the convergence properties (with respect to $K$) of the three estimators, showed analytically that the inverse-count estimator is outperformed by its efficient and zero-truncated counterparts and confirmed the result via correlation-function measurements from simulations.  
We therefore recommend to always use either the efficient or the zero-truncated estimator, which share the same convergence properties.
The main difference between the two is that with the zero-truncated estimator it is possible to have perfect cancellation of the fluctuations even when averaging over a finite number of realisations, which may and may not be a desirable property, according to what one wants to test. 
Since, in practice, the zero-truncated is the estimator that minimises the computational effort (all the targeting realisations can be used to derive the weights), it seem reasonable to adopt it as the default choice.
\item
We identified subtle but important sampling effects that arise when comparing IIP and PIP weights.
In brief, even when the selection probabilities of two galaxies are independent, due to the finiteness of the number of targeting realisations, estimating the inverse probability of a pair directly and as the product of the two individual inverse probabilities gives different results.
We showed how to model exactly such effect and how this new insight into the inverse-probability mechanisms can be used (i) to compute very accurate normalisation factors in virtually no time and (ii) to properly determine the selection-correlation length (a crucial ingredient in the Fourier space formalism that follows).
\item
We developed an unbiased-by-construction estimator for the anisotropic power spectrum in the presence of missing observation.
The derivation follows the same inverse-probability strategy adopted for its configuration-space counterpart, but, rather than pairs, we weight Fourier-space modes, for which the configuration-space separation ${\bf s}$ effectively acts a wave number and the usual wave number ${\bf k}$ as the variable. 
The estimator is conveniently split into a IIP component, which can be evaluated via the current standard algorithms, and an irreducible PIP component, which requires looping over the different pair separations.
For fast evaluation we rely on the fact that this latter term, by constructions, solely depends on pairs with separation smaller than the (angular) selection-correlation scale.
Such scale varies from survey to survey and can be safely determined via the statistical tools summarised above.    
\item
We proposed a very natural generalisation of the inverse probability weights, from $b/p$, where $b$ is a binary random variable (selected/discarded) and $p$ the associated selection probability, to $\phi/\langle\phi\rangle$, where $\phi$ is a generic random variable.
We provided two examples of practical implementation of the concept by defining the generalised nearest neighbour weights and the generalised angular upweighting.
It is not clear yet in which occasion these two new weighting schemes should be adopted, as, at least when tested on our main simulation, they do not seem to improve over the more traditional PIP plus angular upweighting or just PIP weights with realisation-dependent normalisation.
We can nonetheless think of a couple of scenarios in which they could prove useful, namely when dealing with volume-limited samples and reconstruction. 
We leave this topic to further investigations.
\item
Beside the one summarised above, there are more avenues of generalisation that we identified in this paper.
Such ideas can be seen as a first step in the exploration of the applicability of the inverse-probability perspective to more general statistical problems, beyond missing observations. 
In brief, we showed that the inverse probability $b/p$, or similarly $\phi/\langle\phi\rangle$, is not the only possible approach when trying to recover the true spatial distribution of a set of test particles that have been distorted by a stochastic process.
Additive or more complex functions of these quantities can indeed accomplish the same goal. 
%We also discussed how inverse probability weights can be seen as a special case in general formalism for which, rather than focusing on the probability distribution of a field $\mathcal{P}(\varphi)$, we consider weight functions $\mathcal{W}$, directly associated to test particles.
From an even larger perspective, we can interpret the inverse-probability weights as a special case of a general statistical formalism in which, rather than focusing on the probability distribution of a generic field $\mathcal{P}(\varphi)$, we consider weight functions $\mathcal{W}(\eta)$, directly associated to test particles (as usual $\eta$ identifies one possible realisation of the field).
These functions do not have a specific meaning on their own, in the sense that there is no natural order with respect to $\eta$ (i.e. it does not matter how the realisations are sorted), but the $n$-point statistics of $\varphi$ can be obtained by taking their scalar products.
%These functions contain the same information as the original distributions and all the $n$-pt statistics of $\varphi$ can be obtained by taking their scalar products.
A rigorous formalisation of the concept goes beyond the scope of this work, but the idea might prove helpful in situations in which (at least part of) the stochastic process can be simulated.
For example, we are exploring whether it can help in modelling the stochasticity of galaxy bias and redshift-space distortions.
%TOO MUCH?
\end{itemize}

%$IN PRINCIPLE, USEFUL TO FORMULATE MULTIVERSE GRAVITY MODELS.  

From a theoretical perspective, the Fourier-space formalism developed for the power spectrum can easily be applied to higher order $n$-point functions.
One practical complication comes from the fact that the loop over all the irreducible (in terms of selection correlation) quantities, becomes a loop over triangles for the bispectrum, tetrahedrons for the trispectrum, etc. 
If, for finiteness, we focus on the bispectrum, such loop is orders of magnitude faster than a brute-force loop over all the triangles, but, still, much slower than the simple loop over pair separations that we performed in this paper.
Testing whether a straightforward extension of the power-spectrum approach to the bispectrum yields computationally-acceptable results or some clever approximation is required, goes beyond the purpose of this work.     

The pair counting problem is tractable in configuration space, nonetheless, it is worth noting that the same reasoning followed in the derivation of the PIP-corrected power spectrum directly applies to FFT-based estimators of the correlation function, such as that proposed by \citet{slepian2016}, and similarly for the, much less tractable, $n$-plets counts of higher order configuration-space functions.

Finally, we note that relying on a finite selection-correlation length is also a key aspect of the Fourier-space approach proposed by \citet{hahn2017} as a fibre-collision countermeasure for the BOSS survey.
The main difference compared to our strategy is that, rather than focusing on recovering unbiased power-spectrum estimates, \citet{hahn2017} choose to incorporate the effect of missing observations directly into the model used for cosmological inference.
More precisely, the authors measure the power spectrum via standard NN weights and assume that the residual bias can be modelled as a top-hat function of the perpendicular separation convolved with the power-spectrum model.
Since the convolution requires the knowledge of the small-scale nonlinear power spectrum, which is not well predicted by theory, the authors introduce two nuisance parameters that approximately compensate for this issue.
For a direct comparison with our approach it is convenient to substitute NN with IIP weights, which is a perfectly legitimate way to extend the original \citet{hahn2017} derivation (since NN weights can be seen as an approximation of IIP weights, \citealt{bianchi2015b}, even this small modification should make the model more robust).
From this perspective it becomes clear that the quantity that the authors model as a top-hat function is the selection correlation itself, i.e. the curves in Fig. \ref{fig:corr_eff}. 
One important advantage of our approach is that we do not assume anything about the selection correlation other the fact than it becomes negligible above some separation, in accordance with the behaviour observed in the figure.
Strictly speaking, even the selection-correlation length itself is not a real parameter of the model as we just use it for computational purposes.
We can, e.g., arbitrarily set the selection-correlation length at a value larger than the true one without changing the results.    
%One important advantage of our approach, beside using using IIP rather than NN weights (the latter can be seen as an approximation of the former, \citealt{bianchi2015b}), is that we adopt the weaker assumption that the selection correlation is zero above some angular separation but completely free below that separation, in accordance with the behaviour observed in the figure.
Furthermore, by using PIP weights to recover the small-scale power spectrum, we do not need to add any nuisance parameter, as everything is measured directly from the data themselves.

\section*{Acknowledgements}

We thank Will Percival and Mike Wilson for their helpful feedback and H\'ector Gil-Mar\'in for useful discussions. 
DB and LV acknowledge support of European Union’s Horizon 2020 research and innovation programme ERC (BePreSySe, grant agreement 725327).
DB acknowledges support of MDM-2014-0369 of ICCUB (Unidad de Excelencia Maria de Maeztu).
LV acknowledges support of MINECO grant PGC2018-098866-B-I00.

%%%%%%%%%%%%%%%%%%%%%%%%%%%%%%%%%%%%%%%%%%%%%%%%%%



%\bibliographystyle{mnras}
\bibliographystyle{mnras}
\bibliography{./biblio_db}

%\begin{thebibliography}{99}
%\end{thebibliography}

%%%%%%%%%%%%%%%%%%%%%%%%%%%%%%%%%%%%%%%%%%%%%%%%%%

%%%%%%%%%%%%%%%%% APPENDICES %%%%%%%%%%%%%%%%%%%%%

\appendix

\section{Evaluation of the expectation values of the inverse-probability estimators}\label{app:evaluation_expect}

The expectation value of the efficient estimator, Eq. (\ref{eq:eff_expect}), can be derived by substituting $f_K=\frac{K+1}{c+1}$ into
\begin{equation}
\langle w_K \rangle = p \sum_{c=0}^K f_K(c) \binom{K}{c} p^c (1-p)^{K-c} \ ,
\end{equation}
which gives
\begin{equation}
\langle w_K \rangle = p \sum_{c=0}^K \binom{K+1}{c+1} p^c (1-p)^{K-c} = \sum_{c=1}^{K'} \binom{K'}{c} p^{c} (1-p)^{K'-c} \ ,
\end{equation}
where we defined $K'=K+1$.
The final result is obtained by completing the zero-th order moment of the binomial distribution, which we know that must be equal to 1,
\begin{align}
\langle w_K \rangle &= \sum_{c=0}^{K'} \binom{K'}{c} p^c (1-p)^{K'-c} - (1-p)^{K'} = 1 - (1-p)^{K+1} \ .
\end{align}
Similarly for the zero-truncated estimator, Eq. (\ref{eq:zt_expect}), we substitute $f_K=\frac{K}{c}$ into
\begin{equation}
\langle w_K \rangle = p \sum_{c=1}^K f_K(c) \binom{K-1}{c-1} p^{c-1} (1-p)^{K-c} \ ,
\end{equation}
which gives
\begin{align}
\langle w_K \rangle &= p \sum_{c=1}^K \binom{K}{c} p^{c-1} (1-p)^{K-c} \nonumber \\
&= p \sum_{c=0}^K \binom{K}{c} p^{c-1} (1-p)^{K-c} - (1-p)^K \nonumber \\
&= \sum_{c=0}^K \binom{K}{c} p^c (1-p)^{K-c} - (1-p)^K = 1 - (1-p)^K \ .
\end{align}

\section{Derivation of the probability distributions of the recurrence}\label{app:distributions}

In this appendix we describe the reasoning behind the derivation of the distributions that link pair and individual (sampled) probabilities, Eqs.~(\ref{eq:dist_ind_vs_pair})~and~(\ref{eq:dist_ind_vs_pair_zt}), which we rewrite here, for convenience,
\begin{equation}\label{eq:expect_app}
\mathcal{P}(c_{12} | c_1, c_2) = \binom{c_1}{c_{12}} \binom{K - c_1}{c_2 -c_{12}} {\binom{K}{c_2}}^{-1}  \ ,
\end{equation}
where $0 \le c_{12} \le \min(c_1,c_2)$ and 
\begin{equation}\label{eq:zt_expect_app}
\mathcal{P}_{\rm zt}(c_{12} | c_1, c_2) = \binom{c_1 - 1}{c_{12} - 1} \binom{K - c_1}{c_2 -c_{12}} {\binom{K - 1}{c_2 - 1}}^{-1}  \ ,
\end{equation}
where $1 \le c_{12} \le \min(c_1,c_2)$.
For simplicity, we just focus on the first of the two distributions, i.e. the one associated to the inverse-count and the efficient estimators (the derivation of the zero-truncated distribution, $\mathcal{P}_{\rm zt}$, is very similar and briefly discussed at the end of this appendix).
In the following, it is important to keep in mind that the distribution is conditional and, as a consequence, $c_1$ and $c_2$ (and, obviously, the number of realisations $K$) act as fixed parameters. 

\begin{figure}
 \begin{center}
   \includegraphics[width=7.5cm]{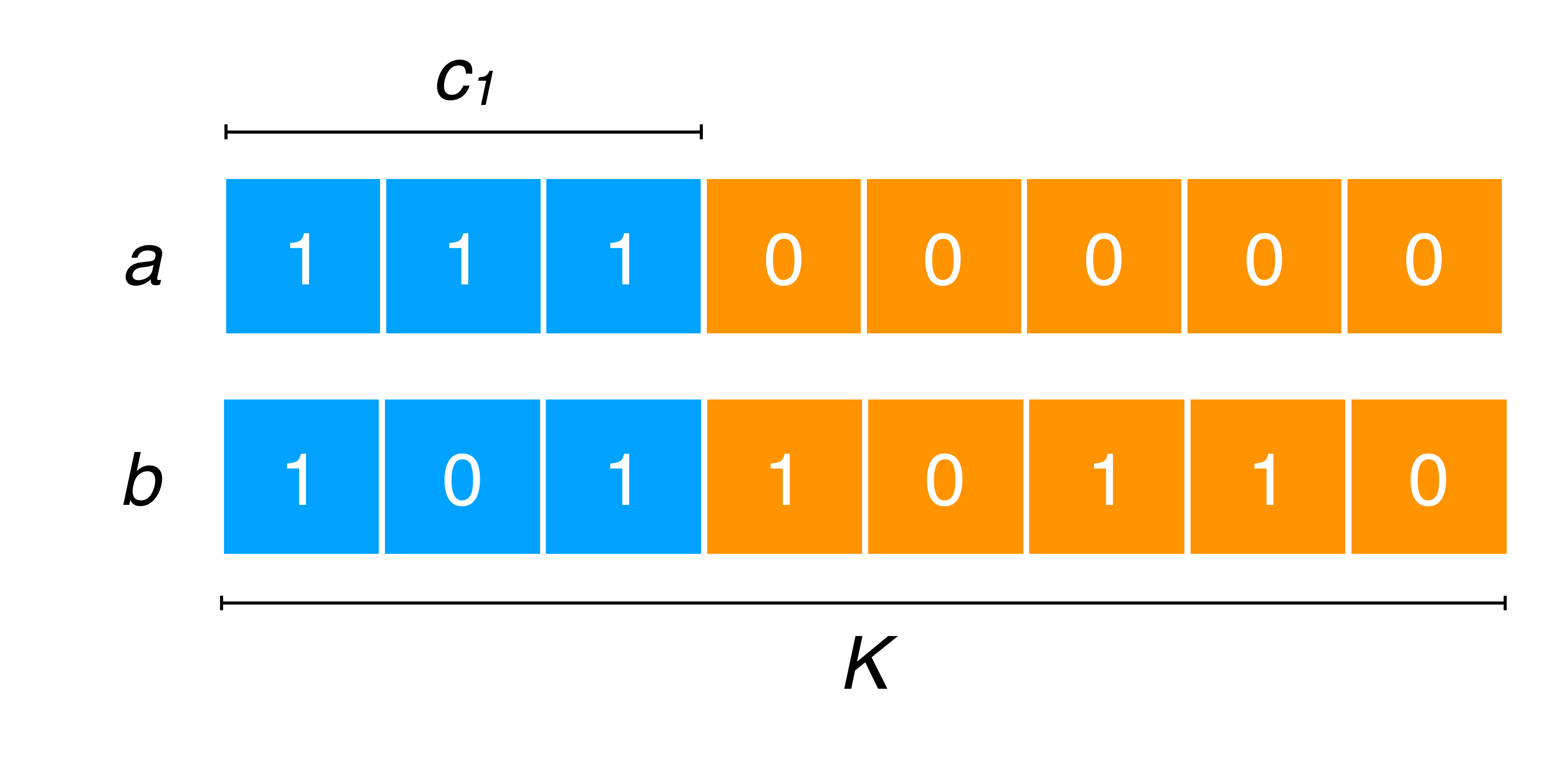}
   \caption{Example of $K=8$ possible outcomes of the selection process for two arbitrary objects $a$ and $b$ conveniently sorted in order to have the array associated to object $a$ in a two-block form.}
   \label{fig:selection_arrays}
 \end{center}
\end{figure}

The basic idea consists of counting all possible states of a two-object system, with their multiplicity, which depends on $c_{12}$.
More explicitly, we can imagine of having two arrays with $K$ binary elements each, $a = \{a_1,\dots,a_K\}$ and $b = \{b_1,\dots,b_K\}$, such that the individual recurrences are $c_1=\sum_i a_i$ and $c_2=\sum_i b_i$. 
In the case of independent variables, the corresponding pairwise recurrence, $c_{12} = \sum_i a_ib_i$, is the result of ``chance alignments''  of the two arrays. 
Since the order with respect to the index $i$ is irrelevant, with no loss of generality, we can always think of sorting the two arrays in such a way that array $a$ is in a two-block form, with $a_i=1$ for $1 \le i \le c_1$ and $a_i=0$ for $c_1 < i \le K$, see Fig. \ref{fig:selection_arrays}.
In this scenario, $c_{12}$ corresponds to the number of ones in the second array with $i < c_1$, i.e. the number of ones in the first block of array~$b$.  
%The quantity of interest is then the multiplicity associated to any given value of $c_{12}$, i.e. how many states of the system give the same $c_{12}$. 
As anticipated, our goal is to evaluate the $c_{12}$ multiplicity, i.e. we whish to count how many possible configurations of the system yield the same value of $c_{12}$. 
Thanks to the two-block interpretation, it is now easy to see that the quantity we are looking for is the number of possible permutations of the first block of array $b$ times the number of possible permutations of the second block of array $b$, which correspond to $\binom{c_1}{c_{12}}$ and $\binom{K - c_1}{c_2 - c_{12}}$, respectively.
In conclusion, for any given value of $c_{12}$, the multiplicity is $\binom{c_1}{c_{12}} \binom{K - c_1}{c_2 -c_{12}}$, which means
\begin{equation}
\mathcal{P}(c_{12} | c_1, c_2) \propto \binom{c_1}{c_{12}} \binom{K - c_1}{c_2 -c_{12}}  \ .
\end{equation}
The remaining term ${\binom{K}{c_2}}^{-1}$ is a normalisation factor.
One instructive way to derive it consists of reasoning in terms of the actual number of possible configurations of the system. 
So far we focused on relative multiplicity, which is essentially the reason why we had the freedom to rearrange array $a$ in a two-block form.
If instead we want the true number of all possible states associated to a given $c_{12}$ we need to account for all possible permutation of array $a$, which yields a further $\binom{K}{c_1}$ factor.
Now we can build the normalised distribution as the number of these states, $\binom{K}{c_1} \binom{c_1}{c_{12}} \binom{K - c_1}{c_2 -c_{12}}$,  divided by the number of all possible states of the system $\binom{K}{c_1} \binom{K}{c_2}$, which finally gives Eq.~(\ref{eq:expect_app})

Eq. (\ref{eq:zt_expect_app}) can be derived following the same reasoning, with the only difference that $c_{12}$ is forced to be larger or equal than 1.
Roughly speaking, this implies solving the same stochastic problem but for the variable $x$ in $K-1$ dimensions, where $c_{12}=x+1$.

\section{Derivation of the PIP correction to the power spectrum}\label{app:PIPco_derivation}

As discussed in the text, it is convenient to rewrite the estimator in terms of sum over $k$-space modes.
Since $n({\bf r}) = \sum_i \delta_D({\bf r} - {\bf r}_i)$ and $n_s({\bf r}) = \sum_i \delta_D({\bf r} - \tilde{\bf r}_i)$, Eq. (\ref{eq:P_estimator}) can be expressed as
\begin{align}
P_\ell (k )=& - \frac{(2\ell+1)}{I} \int \frac{d\Omega_k}{4\pi} \left[ \sum_{ij} \mathcal{Y}_\ell({\bf r}_i,{\bf r}_j,{\bf k}) - \alpha \sum_{ij} \mathcal{Y}_\ell({\bf r}_i,\tilde{\bf r}_j,{\bf k}) \nonumber \right. \\
&\left. - \alpha \sum_{ij} \mathcal{Y}_\ell(\tilde{\bf r}_i,{\bf r}_j,{\bf k}) + \alpha^2 \sum_{ij} \mathcal{Y}_\ell(\tilde{\bf r}_i,\tilde{\bf r}_j,{\bf k}) + S_\ell ({\bf k}) \right] \ ,
\end{align}
where we defined $\mathcal{Y}_\ell ({\bf r}_1,{\bf r}_2,{\bf k}) = e^{i{\bf k}\cdot({\bf r}_1-{\bf r}_2)}\mathcal{L}_\ell \left[ \hat{\bf k}\cdot\hat{\bf \eta}({\bf r}_1, {\bf r}_2) \right]$.
The full PIP extension can be obtained by defining
\begin{align}
P^{\rm PIP}_\ell (k )=& - \frac{(2\ell+1)}{I} \int \frac{d\Omega_k}{4\pi} \left[ \sum_{ij} w^{\rm PIP}_{ij}  \mathcal{Y}_\ell({\bf r}_i,{\bf r}_j,{\bf k}) \nonumber \right. \\
&\left. - \alpha \sum_{ij} w^{\rm IIP}_i \mathcal{Y}_\ell({\bf r}_i,\tilde{\bf r}_j,{\bf k}) - \alpha \sum_{ij} w^{\rm IIP}_j \mathcal{Y}_\ell(\tilde{\bf r}_i,{\bf r}_j,{\bf k})\nonumber \right. \\
&\left. + \alpha^2 \sum_{ij} \mathcal{Y}_\ell(\tilde{\bf r}_i,\tilde{\bf r}_j,{\bf k}) + S_\ell ({\bf k})  \right] \ .
\end{align}
By adding and subtracting to the integrand the quantity $\sum_{ij} w^{\rm IIP}_i w^{\rm IIP}_j \mathcal{Y}_\ell({\bf r}_i,{\bf r}_j,{\bf k}) $ we can conveniently isolate the IIP power spectrum, 
\begin{align}\label{eq:P_PIPco}
P^{\rm PIP}_\ell(k) &= \frac{(2\ell+1)}{I} \int \frac{d\Omega_k}{4\pi} \sum_{ij} \left( w^{\rm PIP}_{ij}  - w^{\rm IIP}_i w^{\rm IIP}_j \right) \mathcal{Y}_\ell({\bf r}_i,{\bf r}_j,{\bf k}) \nonumber \\
+& \ P^{\rm IIP}_\ell(k) \ .
\end{align}

\section{Minimising the variance through generalised weights: a simple proof of concept.}\label{app:toy}

Here we present a simple toy example whose purpose is to provide an explicit illustration of how the generalised weights defined in Sec.~\ref{sec:generalising} work.
Beside the instructional function, this appendix is effectively an analytic proof of the fact that the variance obtained via inverse probability can be reduced by adopting this more general perspective, as argued in Sec. \ref{sec:generalising}.

Imagine we have two objects $a$ and $b$, i.e. the smallest possible non-trivial set (as discussed in Sec. \ref{sec:generalising}, in the single-object case it is not possible to improve on pure inverse probability).
For finiteness, these objects can be two pairs of galaxies for which we are performing pair counts, i.e. we want to recover $DD=2$ with the minimum possible variance.
Imagine a selection algorithm such that there are three possible equally likely outcomes, $\{ b_a, b_b\} = \{0,1\}, \{1,1\}, \{1,0\}$, where, as usual, 1 means selected and 0 discarded.  
Since the three configurations are equally likely, the correspondent inverse probability weights are $\{b_a/p_a, b_b/p_b\} = \{0,3/2\}, \{3/2,3/2\}, \{3/2,0\}$, which results in the following pair counts, $DD= 3/2, 3, 3/2$.
The average is $\langle DD \rangle = 2$, obviously unbiased, and the variance is  $\langle DD^2 \rangle - \langle DD \rangle^2 = 1/2$.

It is trivial to see that the set of weights $\{\phi_a/\langle \phi_a \rangle, \phi_b/\langle \phi_b \rangle\} = \{0,2\}, \{1,1\}, \{2,0\}$ yields zero-variance unbiased pair counts.
Note that these weights verify two requirements that are of fundamental importance when modelling missing observations: (i) discarded objects have zero weight; (ii) each individual pair yields unbiased pair counts, i.e.on average is counted once.
Obviously, the result is invariant with respect to any transformation $\phi \rightsquigarrow \phi \times {\rm const}$.

It is important to note that zero-variance weights do not always exist.
If, e.g., we repeat all the above considerations replacing the selection algorithm with one for which the possible equally likely outcomes are $\{ b_a, b_b\} = \{0,1\}, \{0,1\}, \{1,0\}$, it is easy to see that the minimum variance is obtained through pure inverse probability and differs from zero. 
Similarly, it is possible to imagine situations in which the zero-variance solution is not unique.

\section{Impact of the different normalisation factors}\label{app:norm}

\begin{figure}
 \begin{center}
   \includegraphics[width=7.5cm]{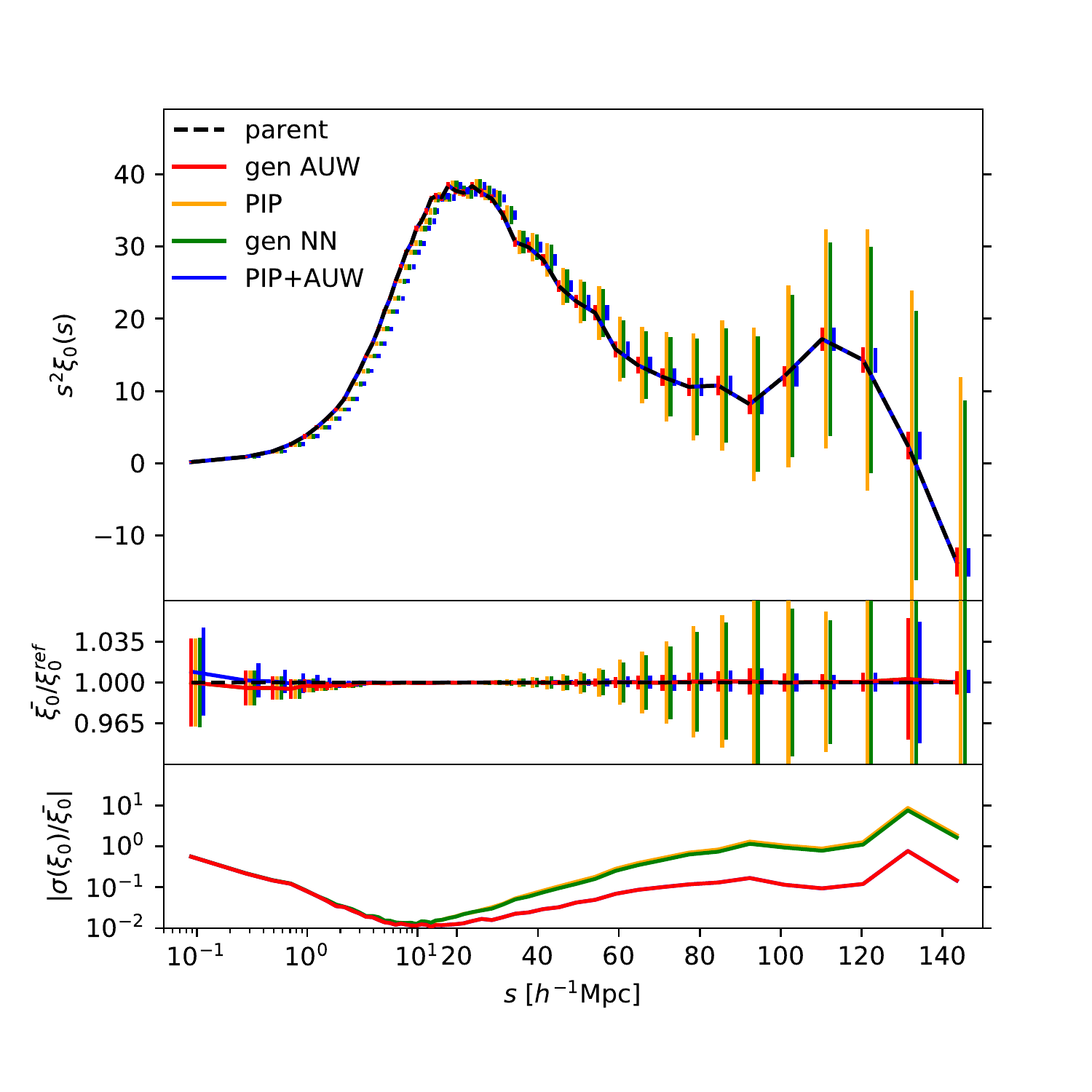} \\
    \caption{Same as the top plot (i.e. the plot for the monopole) of Fig. \ref{fig:gen_w_comp}, but with a different pair-count normalisation for pure PIP, generalised AUW and generalised NN, as described in the text.}
   \label{fig:arenorm}
 \end{center}
\end{figure}

\begin{figure}
 \begin{center}
   \includegraphics[width=7.5cm]{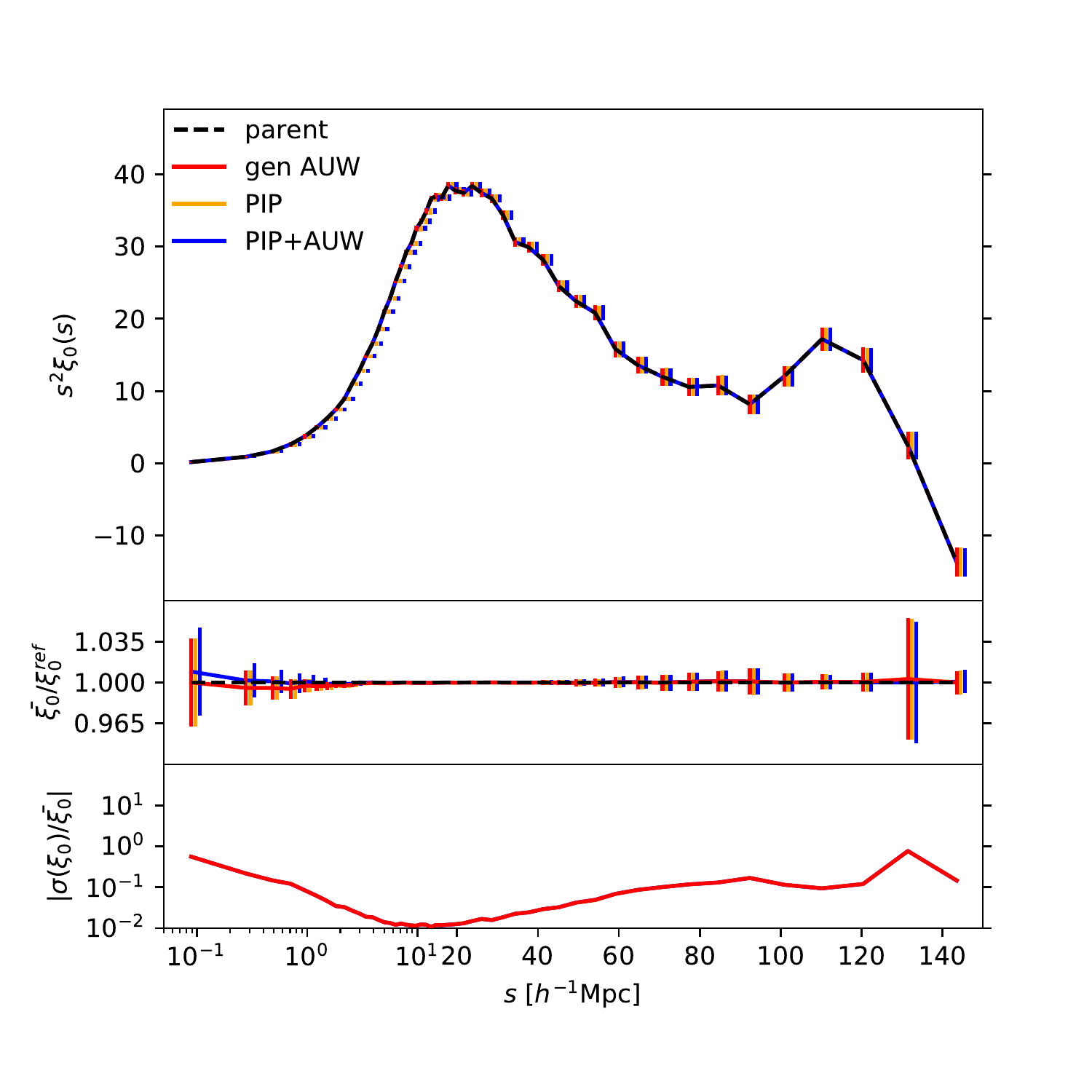} \\
    \caption{Same as Fig. \ref{fig:arenorm}, but with a different (realisation-dependent) pair-count normalisation for the pure PIP measurements, as described in the text. NN measurements are not shown.}
   \label{fig:renorm}
 \end{center}
\end{figure}

Fig. \ref{fig:arenorm} is the same as the upper panel of Fig. \ref{fig:gen_w_comp} but with renormalised measurements, in order to remove the systematic bias induced by the pairs that have not been observed in any of the $K=217$ realisations.
Specifically, rather than normalising all $DD$ counts by the total number of unweighted pairs, $n_{DD} = N_{gal} (N_{gal} - 1) / 2$, we infer the debiased normalisation factor as follows. 
Eq. \ref{eq:norm} provide us with a fast way compute the correct normalisation factor when the $DD$ counts are performed via PIP weights.
Such factor varies from one realisation to one another (i.e. we have $K$ normalisation factors), thus allowing us manipulate mean and variance at once (see Fig. \ref{fig:renorm} and discussion below).
These factors cannot be directly applied to the other weighting schemes, as they are specific for pure PIP weights.
On the other hand, since pure PIP, generalised NN and generalised AUW share by construction the same expectation value, it is possible to infer a fixed normalisation factor, which does not depend on realisation or weighting scheme and accounts for systematic effect without affecting the variance.
This factor is simply obtained by averaging over the $K$, realisation dependent, PIP factors obtained via Eq. \ref{eq:norm}.
Note that, since in our sample, by construction, there are no zero-probability pairs, it would suffice to increase the number of realisations to remove any systematic effect.
Nonetheless, keeping the number of realisations low is in practice desirable and it is important to know if and when the too-few-realisations artefacts can be compensated by a simple, fast-to-compute renormalisation.
It is clear from the figure that, at least for the sample under examination, the debiased normalisation works very effectively.
Form a quantitative point of view, the ratio between debiased and biased normalisation factor is just about $0.99997$.
Due to the relation between correlation function and pair counts, $\xi  = DD/RR -1$, even such a small deviation from one can translate into a significant fractional error when the correlation tends to zero.    

In Fig. \ref{fig:renorm} we show a comparison between generalised AUW (renormalised as above), PIP plus AUW and pure PIP with the full realisation-dependent normalisation, Eq. (\ref{eq:norm}).
We did not include generalised NN in this comparison because we do not have a fast-to-compute expression for their realisation-dependent normalisation.
Obviously, for any estimator the exact realisation-dependent normalisation can be computed by directly counting all the (weighted) pairs in the sample.
Despite this being doable for the number of objects considered in this paper, it becomes extremely computational expensive for the numbers expected from next generation surveys.
We therefore prefer not to follow this avenue, in order to avoid confusion on what is realistically feasible or not. 
As anticipated, when pure PIP weights are normalised via Eq. (\ref{eq:norm}), not only the expectation value, but also the variance is largely improved.
Indeed, form the figure it is clear that overall behaviour of renormalised PIP weights is almost identical to that of its angular-upweighted counterparts.
It is nonetheless worth saying that such results depends on the sample under examination, and it is not necessarily true in more extreme (and less realistic) situations, as, e.g., the targeting strategy presented in App. \ref{app:generalised_comparison}.

\section{Behaviour of the generalised weights in more extreme conditions}\label{app:generalised_comparison}

\begin{figure}
 \begin{center}
   \includegraphics[width=7.5cm]{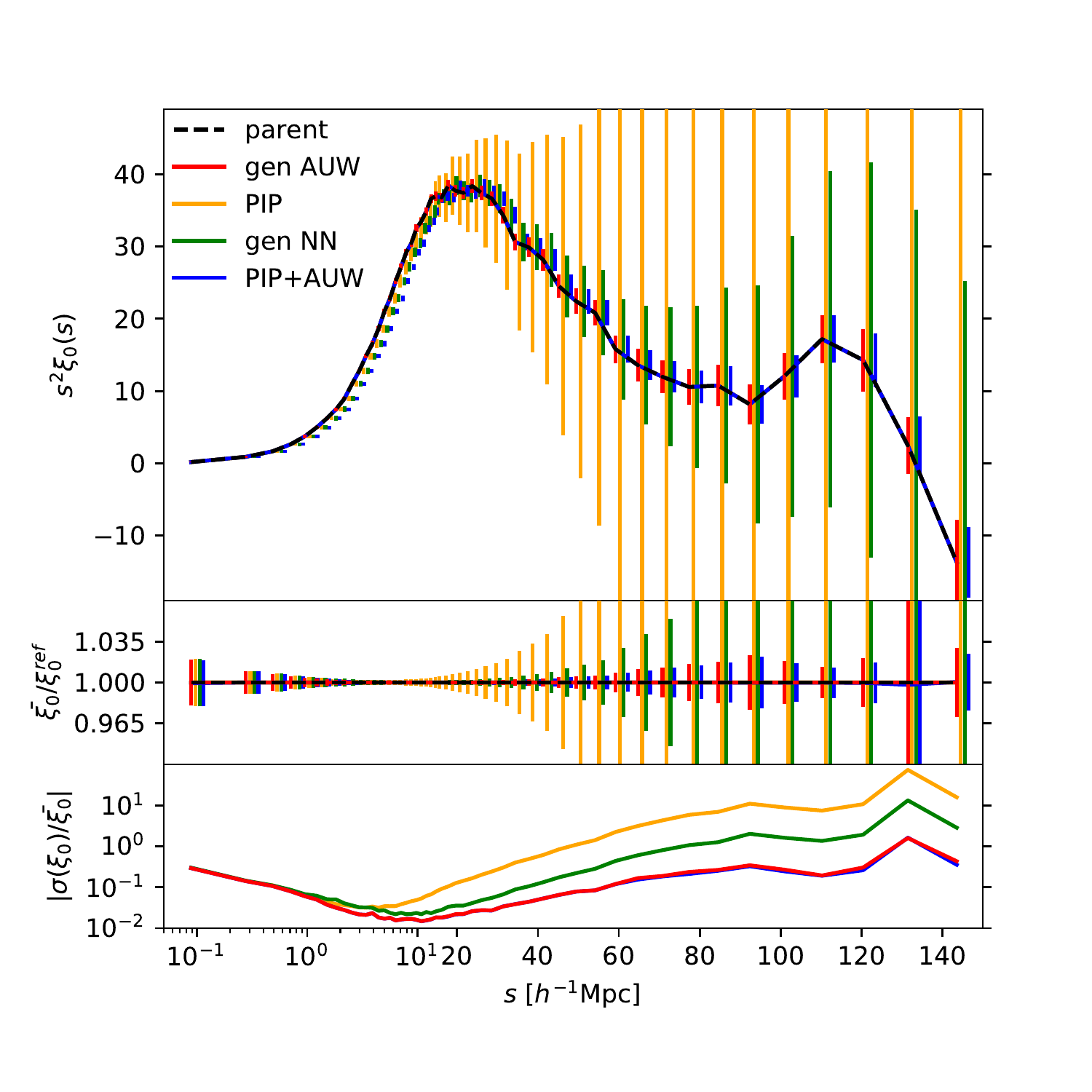} \\
   \vspace{-0.5cm}
    \includegraphics[width=7.5cm]{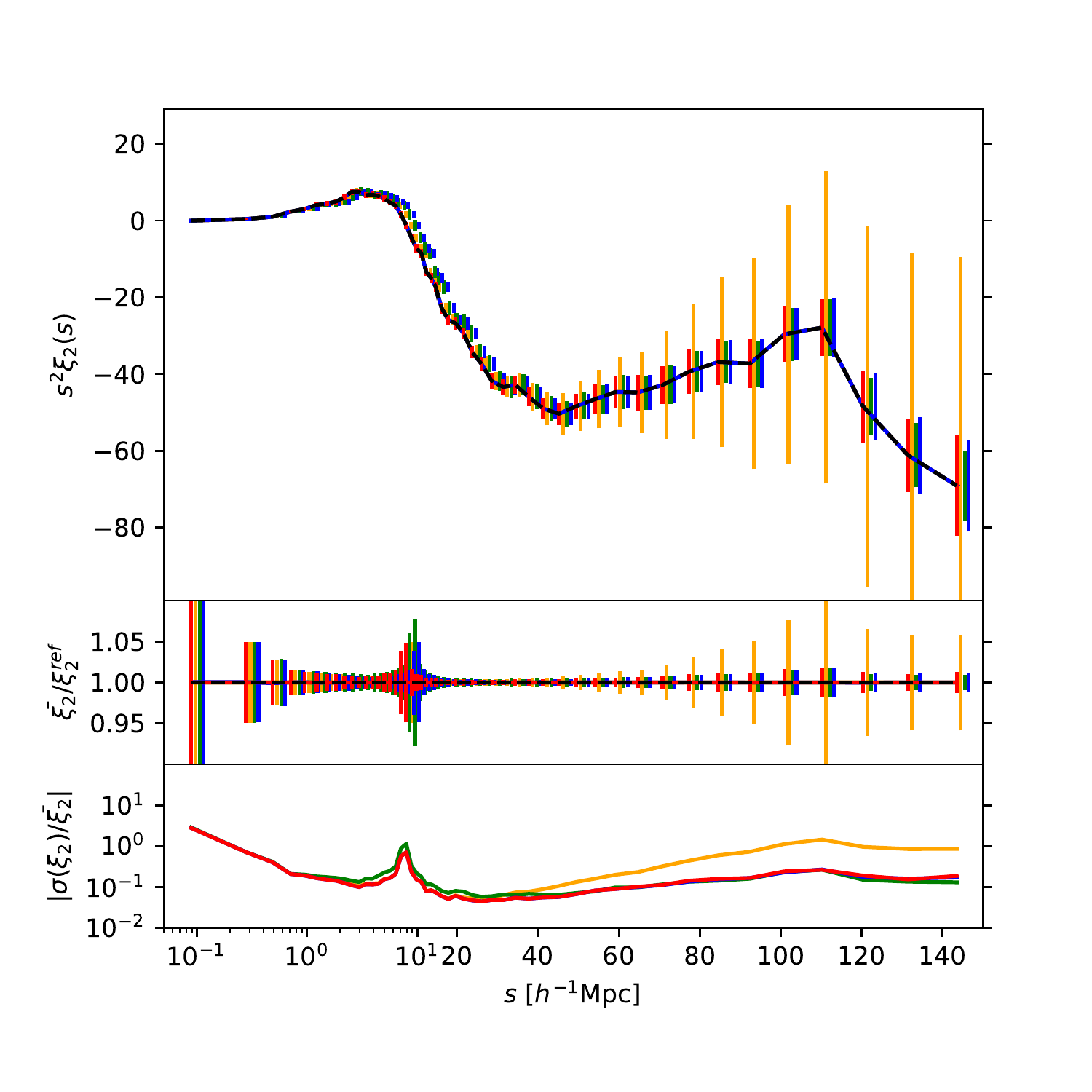} \\
    \vspace{-0.5cm}     
     \includegraphics[width=7.5cm]{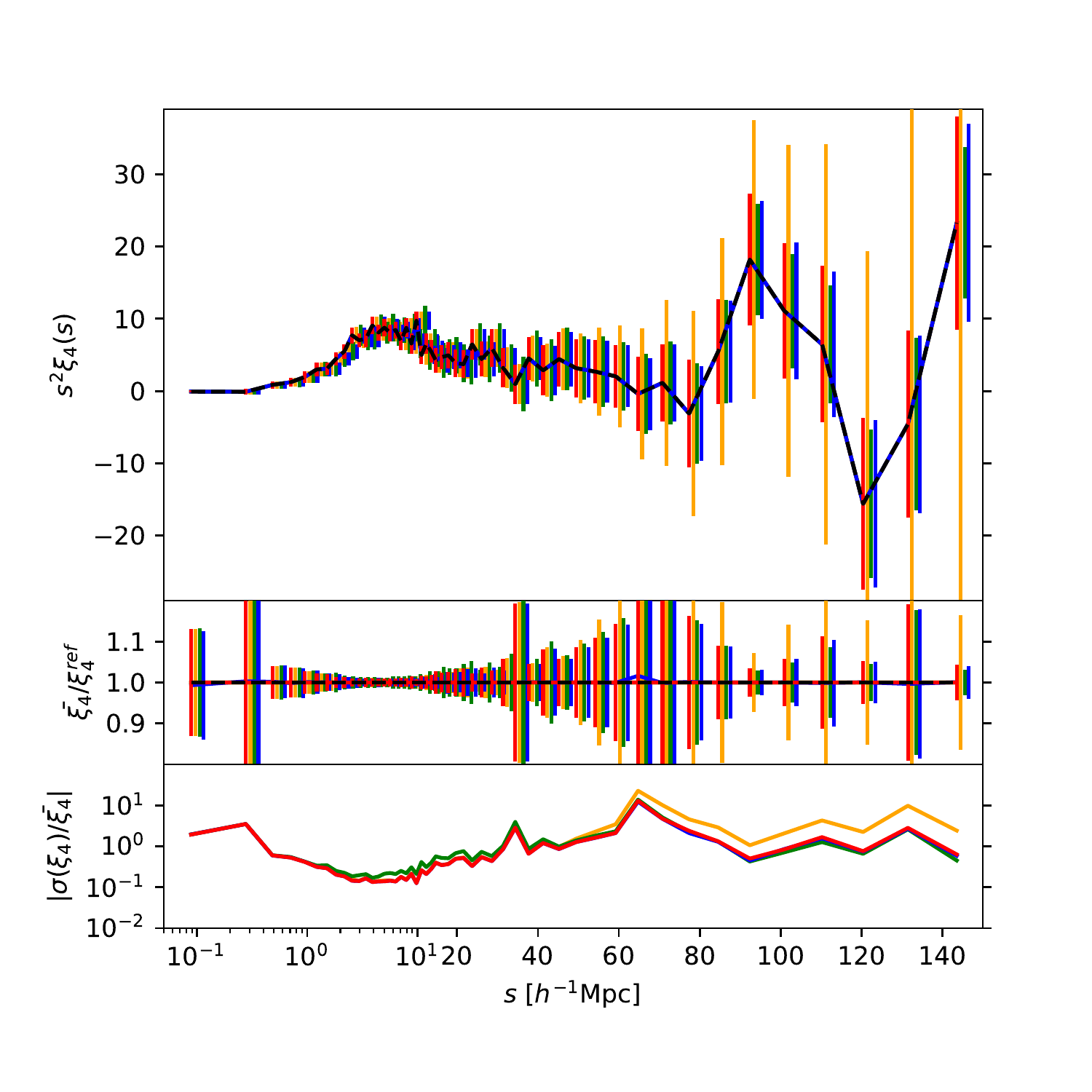}
    \caption{Same as Fig. \ref{fig:gen_w_comp}, but for a more extreme targeting strategy, as described in the text.}
   \label{fig:gen_w_comp_50-100}
 \end{center}
\end{figure}

Fig. \ref{fig:gen_w_comp_50-100} is the same as Fig. \ref{fig:gen_w_comp} but for a different targeting strategy in which only a subset of the full survey area is observed twice. 
Specifically the second pass of the algorithm is performed on 100 squares of $50 h^{-1}$Mpc side randomly distributed over the whole survey area (observing strategy OSmulti in \citealt{bianchi2017}), which yields an average galaxy completeness of about $0.66$. 
Since the squares are allowed to overlap, there are regions observed more than twice.
The resulting increased selection probability for the pairs that belong to triplets and more complex structures has beneficial effect on the missing-observations corrections, especially on small scales.
What is not beneficial and makes this strategy more extreme compared to the one considered throughout the rest of this paper, is that the second-pass region varies significantly from one realisation to one another creating an effective selection correlation up to scales roughly comparable to those of the squares.
This is not how galaxy surveys usually work, but helps illustrating some interesting points.

Compared to Fig.~\ref{fig:gen_w_comp}, here we observe two main differences.
First, despite having adopted the same normalisation, in Fig.~\ref{fig:gen_w_comp_50-100} there are no signs of systematic effects.
This is direct consequence of the higher selection probability of close pairs mentioned above.
In other words, with this targeting strategy, $K=217$ realisations are enough to target (almost) all pairs at least once.
Second, the variance of the generalised NN weights is significantly smaller than that of the pure PIP weights.
On the other hand, analogously to Fig.~\ref{fig:gen_w_comp}, PIP plus AUW and generalised AUW perform almost identically and always yield the lowest variance.

\section{Estimating the angular correlation function via generalised weights}\label{app:xipe}

In this appendix we show measurements of the angular correlation function, or, more properly, the perpendicular correlation function, as we are in plane parallel approximation.
This might sound odd because, since we have a parent sample, in principle, we do not need any sophisticated missing-observation correction to measure the angular correlation function.
Nonetheless, there are at least two reasons why it is still interesting to discuss this topic.
First, as anticipated in Sec. \ref{sec:generalising}, when using volume-limited samples it is not obvious how to define the parent sample.
Second, it gives us the opportunity to show more in detail how the generalised AUW defined in Sec. \ref{sec:generalising} works.

\begin{figure*}
 \begin{center}
   \includegraphics[width=7.5cm]{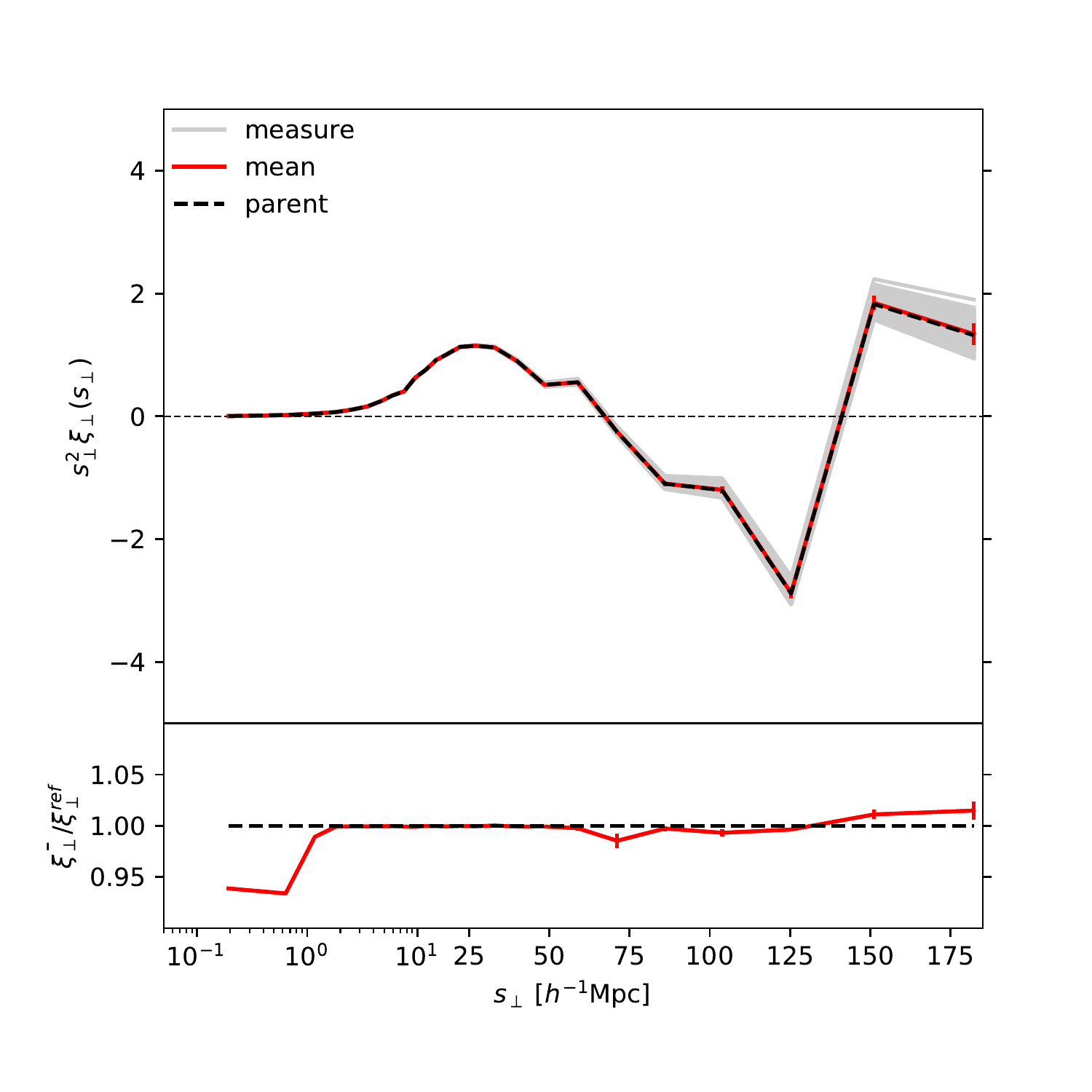}
   \includegraphics[width=7.5cm]{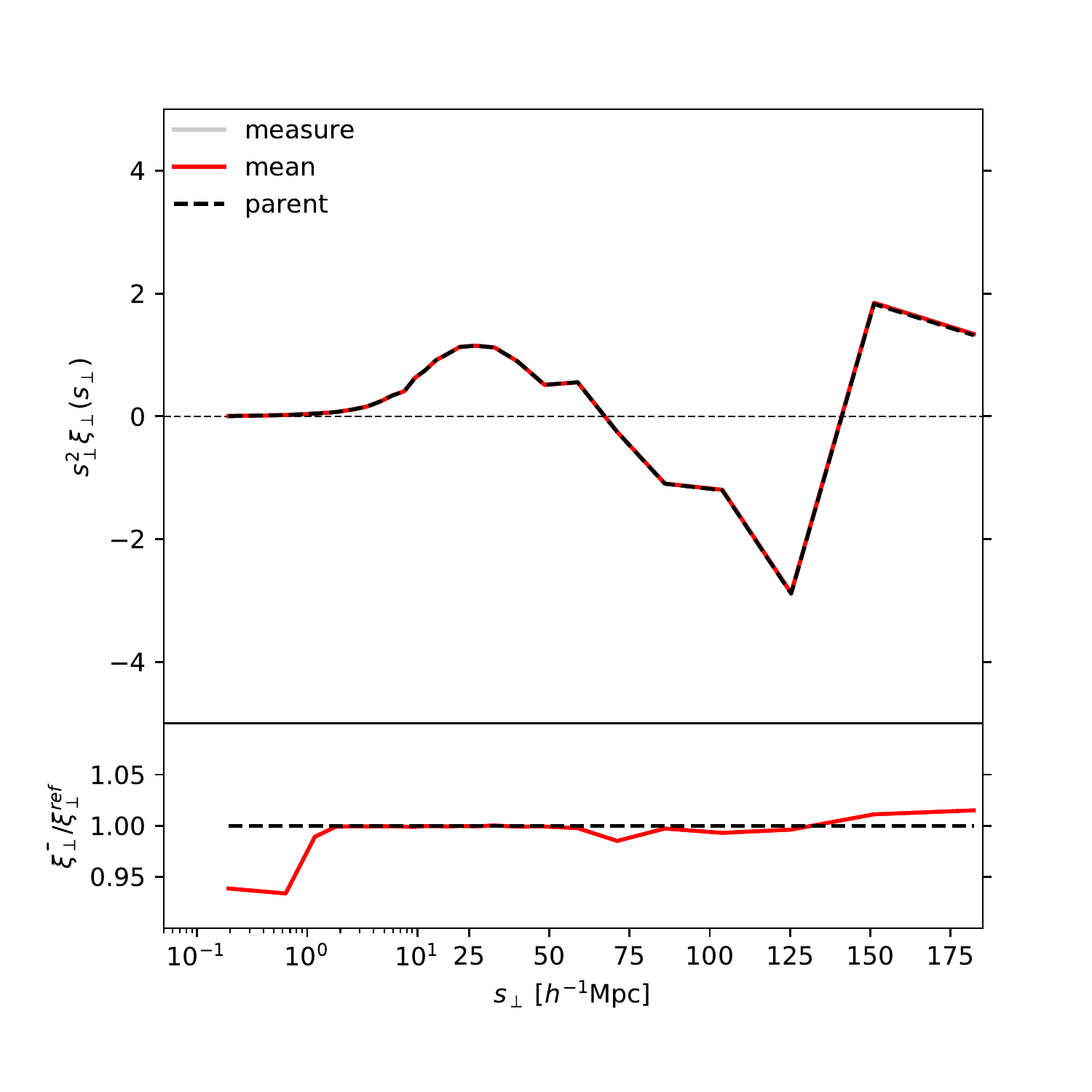}
    \caption{Measurement of the perpendicular correlation function $\xi_\perp$, which, since we are in plane parallel approximation, ideally corresponds to the angular correlation function. The measurements (solid grey) and their mean (solid red) are obtained from $K=217$ realisations of the targeting via generalised AUW and compared to the reference value obtained directly from the parent sample (dashed black). Right and left correspond to first and second iteration, respectively, as described in the text. Top panels: amplitude of the correlation function multiplied by the perpendicular separation squared $s_\perp^2$, for visualisation purposes. Bottom panels: ratio between the mean of the measurements and the reference, with error bars of the mean.}
   \label{fig:xipe}
 \end{center}
\end{figure*}

Imagine to measure the angular correlation function via standard AUW.
The weight of each pair is $w = b DD^{(p)}_a / DD_a$, where numerator and denominator are respectively the pair counts of parent and observed sample (as a function of of the angular separation $\theta$) and $b$ the usual binary variable.
By construction we always obtain the correlation of the parent sample.
If we replace AUW with its generalised counterpart,
\begin{equation}\label{eq:genAUW_app}
w = b \frac{DD^{(p)}_a}{DD^{\rm PIP}_a} \left\langle b \frac{DD^{(p)}_a}{DD^{\rm PIP}_a} \right\rangle^{-1} \ ,
\end{equation}
due the factor $\langle b DD^{(p)}_a / DD^{\rm PIP}_a \rangle^{-1}$, the angular correlation of different realisations of the targeting fluctuates around that of the parent sample, which is recovered only on average, analogously to its 3-dimensional counterpart, see left panel of Fig. \ref{fig:xipe}. 
%Actually, the unbiasedness of the generalised AUW comes at the cost of increased variance compared to standard AUW.
%This effect turns out to be completely negligible for the three-dimensional measurements (Figs. \ref{}), but not for the angular correlation function, as shown in the upper panel of Fig. \ref{fig:xipe}.  
%In essence, the reason behind this behaviour is that with the generalised AUW also the angular part has variance, whereas in the standard scenario the angular variance is forced to be zero.  
%In other words, due the factor $\langle DD^{(p)}_a / DD_a \rangle^{-1}$, the angular correlation of different realisations of the targeting fluctuates around that of the parent sample, which is recovered only on average, analogously to its 3-dimensional counterpart.  
It seems clear anyway that it formally exists a solution (in general more than one) which combines strong unbiasedness (defined in Sec. \ref{sec:generalising}) and zero (or at least negligible) angular variance.
Unfortunately a direct evaluation of such solution is prohibitively computationally expensive, as it involves solving a large set of algebraic equations, roughly as large as the total number of pairs.
On the other hand we can get arbitrarily close to that solution with an iterative approach.
%We discuss the details of the procedure to App. \ref{app:generalised_comparison}, where we also present a comparison of the results obtained with the different approaches just discussed.
Essentially, we can repeat the process by defining a new set of weights,
\begin{equation}
w' = b \frac{DD^{(p)}_a}{DD^{\rm PIP}_a} \frac{DD^{(p)}_a}{DD'_a} \left\langle b \frac{DD^{(p)}_a}{DD^{\rm PIP}_a} \frac{DD^{(p)}_a}{DD'_a} \right\rangle^{-1} \ ,
\end{equation}
%\begin{equation}\label{eq:genAUW_app}
%w = b \prod^{N_{\rm iter}}_{i=1} \frac{DD^{(p)}_a}{DD_{a,i}} \left\langle b \prod^{N_{\rm iter}}_{i=1} \frac{DD^{(p)}_a}{DD_{a,i}} \right\rangle^{-1} \ .
%\end{equation}
where $DD'_a$ are the angular pair counts obtained with the previous iteration, i.e. via Eq. (\ref{eq:genAUW_app}).
In the right panel of Fig. \ref{fig:xipe} we show that this approach allows us to reach virtually zero variance with just one more iteration.
The pair counts have been normalised via the averaging procedure described in App. \ref{app:norm}, any residual systematic effect is evidence of an intrinsically scale-dependent distortion coming from the pairs that have not been selected at least once in the $K=217$ realisation.
Since in our sample there are no pairs with zero probability, such effects can be removed by increasing the number of realisations. 

Interestingly, for the whole technique to work, the knowledge of the parent-sample pair counts is not required, as it gets cancelled out by the expectation value.
We can, e.g., get the same identical result by setting $DD^{(p)}_a$ to any arbitrary constant value, which is what we actually did for the measurements in the figure.
Similarly, we do not need to know the exact value of the various $DD^{\rm PIP}_a$ and $DD'_a$ but only how they fluctuate from one realisation to one another.
If we get these pair counts wrong, e.g. because we are dealing with a volume limited sample, the only resulting effect is a non optimal reduction of the variance.

As discussed in Sec. \ref{sec:generalising}, one could think of using $DD_a$, rather than $DD^{\rm PIP}_a$.
We have checked that with this approach the convergence get slower, in the sense that one additional iteration is required to reach a comparable variance.
In other words, not surprisingly, PIP weights provide us with an excellent initial guess for the iterative process, to the extent that for 3-dimensional measurements no iteration is required at all, see Sec. \ref{sec:generalising}.

%%%%%%%%%%%%%%%%%%%%%%%%%%%%%%%%%%%%%%%%%%%%%%%%%%

% Don't change these lines
\bsp	% typesetting comment
\label{lastpage}
\end{document}